\documentclass[12pt,twoside]{article}
\usepackage{inputenc}
\usepackage[backend=biber,style=numeric-comp,autocite=plain,sorting=none,%
giveninits=true]{biblatex}
\usepackage{amsmath,amsfonts,amssymb,amsthm,mathabx,bbold}
\usepackage{esint}
\usepackage{times}
\usepackage{hyperref}
\hypersetup{
  breaklinks=true,   %
  colorlinks=true,   %
  pdfusetitle=true,  %
}
\usepackage{xurl}
\usepackage{tensor}
\usepackage{color}
\usepackage{caption}
\usepackage{enumitem}
\usepackage{graphicx}
\usepackage{array}
\usepackage{amsbsy}
\usepackage{appendix}
\usepackage{mathrsfs}
\usepackage{physics}
\usepackage{cancel}
\usepackage{yfonts}
\usepackage{chngcntr}
\usepackage{xcolor}
\usepackage{soul}
\usepackage{cancel}
\usepackage{subfigure}
\usepackage{wrapfig}
\usepackage[leqno,fleqn,intlimits,newmultline]{empheq}

\advance\textheight\topskip%
\numberwithin{equation}{section} \makeatletter
\@addtoreset{equation}{section}

\newcommand{\D}{\text{d}}
\newcommand{\sign}{\text{sgn}}

\begin{document}

\title{Model spaces as constrained Hamiltonian systems: I.~Application to $\mathrm{SU}(2)$}

\author{Glenn Barnich, Thomas Smoes}

\def\mytitle{Model spaces as constrained Hamiltonian systems: \\ I.~Application
  to $\mathrm{SU}(2)$}

\pagestyle{myheadings} \markboth{\textsc{\small G.~Barnich, T.~Smoes}}
{\textsc{\small Model spaces: I. $\mathrm{SU}(2)$}}

\addtolength{\headsep}{4pt}

\begin{centering}

  \vspace{1cm}

  \textbf{\Large{\mytitle}}

    \vspace{1.5cm}

    {\large Glenn Barnich, Thomas Smoes}

\vspace{1cm}

\vspace{0.5cm}

\begin{minipage}{.9\textwidth}\small \it \begin{center}
    Physique Th\'eorique et Math\'ematique \\ Universit\'e libre de Bruxelles
  and International Solvay Institutes\\ Campus Plaine C.P. 231, B-1050
  Bruxelles, Belgium \\
  E-mail:
  \href{mailto:Glenn.Barnich@ulb.be}{Glenn.Barnich@ulb.be}, \href{mailto:Thomas.Smoes@ulb.be}{Thomas.Smoes@ulb.be}
\end{center}
\end{minipage}

\end{centering}

\vspace{0.5cm}

\vspace{.5cm}

\begin{center}
  \begin{minipage}{.9\textwidth} \textsc{Abstract}. Motivated by group-theoretical
    questions that arise in the context of asymptotic symmetries in gravity, we
    study model spaces and their quantization from the viewpoint of constrained
    Hamiltonian systems. More precisely, we propose that a central building
    block in the construction of the model space for a generic Lie group $G$ is
    the symplectic submanifold of $T^*G$ that one obtains when one imposes only
    the second class constraints in the construction of the coadjoint orbit as a
    symplectic quotient. Before turning to the non-compact infinite-dimensional
    groups relevant in the gravitational setting, we work out all details in the
    simplest case of $\mathrm{SU}(2)$. Besides recovering well-known results on
    the quantum theory of angular momentum from a unified perspective, the
    analysis sheds some light on the definition and properties of
    spin-weighted/monopole spherical harmonics.
 \end{minipage}
\end{center}

\vfill
\thispagestyle{empty}

\newpage

\tableofcontents

\section{Introduction}\label{sec:introduction-1}

The solution space of asymptotically anti-de Sitter gravity in three dimensions
in Fefferman-Graham gauge is known in closed from~\cite{Banados1999,Skenderis:1999nb}. It can be
identified with the coadjoint representation of its asymptotic symmetry group
given by two copies of the Virasoro group~\cite{Brown:1986nw}. The same goes for
asymptotically flat spacetimes in three dimensions, where the solution space~\cite{Barnich:2010eb}
corresponds to the coadjoint representation of the centrally extended BMS3
group~\cite{Bicak1989,Barnich:2006avcorr}.

The coadjoint representation admits a partition into orbits.
Each of these coadjoint orbits is a symplectic manifold that can in principle be
quantized~\cite{kirillov1976elements,Kostant1970,Souriau1970,A.A.Kirillov897,DavidA.Vogan1997}.
What is needed in the context of three dimensional gravity is not really a
quantization of each of the coadjoint orbits separately, but of all of them at
once in a consistent fashion. This is where model spaces come in.

{\em The model space of a Lie group $G$ is a classical $G$-invariant system whose
quantization yields a Hilbert space that carries all unitary irreducible
representations of $G$ with multiplicity one.}

We start here with geometric actions, i.e., Lagrangian particle actions
associated to individual coadjoint orbits. When quantized through path integral
methods, they have been used in~\cite{Alekseev:1988vx,Alekseev1989} to produce group characters. The
study of what the model space should correspond to in the case of the Virasoro
group has been initiated in~\cite{Alekseev:1990mp}. The case of compact Lie groups has been
worked out in detail in terms of Darboux coordinates related to the
Gelfand-Zetlin basis. The relation to earlier work in~\cite{bernstein1976models} had been left open.
Their proposal has been discussed further in~\cite{La:1990ty} from the viewpoint of the
Hilbert space and operator quantization.

In order to relate path integral and operator quantization (see
e.g.~\cite{Henneaux:1985kr}), it is useful to reformulate geometric actions as
constrained Hamiltonian systems~\cite{Barnich2022}, even if these actions are already of first
order. As summarized in the first section below, the geometric action for a
fixed coadjoint vector is related to the one of the cotangent bundle $T^*G$ by
natural primary constraints. Those associated with the little algebra of the
coadjoint vector are first class, while the other ones are second class. From
this viewpoint, the heuristic proposal is the following:

{\em A necessary step in the construction of the model space from $T^*G$ is to
  drop the first class constraints while imposing only those of second class.
  Equivalently, when starting from the geometric action for a single coadjoint
  orbit, it is necessary to replace the components of the coadjoint vector along
  the complement of the annihilator of the little algebra by additional
  dynamical variables.}

When formulated in these terms, one can use the full flexibility of constrained
Hamiltonian systems, such as introducing additional spurious degrees of freedom
in the form of (generalized) auxiliary fields, and conversion of second into
first class systems. While in field theory applications, this allows for a local
formulation with manifest Lorentz invariance, the aim here is manifest
covariance under the Lie group symmetries.

On the quantum level, in addition to geometric quantization, the whole arsenal
of operator and path integral methods for quantization of constrained
Hamiltonian systems may be used to construct the unitary irreducible
representations associated to the model space. Again, the advantage with respect
to reduced phase space quantization is manifest covariance. Furthermore, there
is no need to find Darboux coordinates in order to evaluate the path integral.

Conversely, these systems can serve as completely tractable, non-trivial
applications of constrained Hamiltonian systems.

After fixing notations and conventions for the description of Lie groups and
algebras in Section~\ref{sec:notat-gener}, geometric actions together with their global and gauge
symmetries as well as their relation to coadjoint orbits are reviewed in
Section~\ref{sec:geometric-actions}. The analysis~\cite{Barnich2022} of geometric actions as constrained Hamiltonian
system of the cotangent bundle $T^{*}G$ is reviewed in Section~\ref{sec:phase-space-primary}. Generalities
on geometric actions associated to the cotangent bundle $T^{*}G$ are briefly
discussed in Section~\ref{sec:unconstrained-model-1}, while the proposal for the symplectic submanifold that
serves as a building block for the model space is presented in~\ref{sec:model-space}.

The remainder of the current paper is devoted to illustrating the proposal in
detail in the simplest non-trivial case, the group ${\rm SU}(2)$, where all
results are perfectly well-known. We start by reviewing the coadjoint orbits,
which correspond to the foliation of $\mathbb R^{3}$ by two-spheres together
with the associated Hopf fibration of $\mathrm{SU}(2)$, in Section~\ref{sec:coadjoint-orbits}.
Section~\ref{sec:reduced-phase-space} is devoted to the appropriate global description needed to
avoid the Gribov obstruction. The associated model space, with an additional
dynamical variable directly related to the radius of the sphere, is described
next in Section~\ref{sec:su2-model-space}. This formulation of the model space is connected to
Schwinger's construction in terms of an isotropic two dimensional harmonic
oscillator~\cite{osti_4389568} by embedding $\mathrm{SU}(2)$ as a subgroup of the group of
non-zero quaternions of modulus one in Sections~\ref{sec:unconstr-model-as} and~\ref{sec:su2-model-space-2}.

In the last part of the paper, we review reduced phase space quantization of single
coadjoint orbits (Section~\ref{sec:quant-singke-coadj}), the quantization of $T^{*}\mathrm{SU}(2)$
(Section~\ref{sec:quantization-tsu2}) in terms of Wigner functions, the Dirac quantization of both the
model space and single orbits from Dirac quantization of $T^{*}\mathrm{SU}(2)$
(Section~\ref{sec:reduct-after-quan}), and the quantization of the second class
system in the cotangent bundle of quaternions that leads to Schwinger's
description of the quantized model space (Section~\ref{sec:quant-extend-space}). Finally, in the
quantized model space, we compute the partition function associated to the
Casimir operator.

Several technical appendices make the paper more self-contained.
Appendix~\ref{sec:path-integr-meas} addresses subtleties related to the path
integral measure. In Appendix~\ref{sec:parametrizations-su2}, left and right invariant Maurer-Cartan forms
and vector fields are computed in the various parametrizations of
$\mathrm{SU}(2)$ needed for the different applications while Appendix~\ref{sec:non-unim-quat} is
devoted to quaternions.

\section{Geometric actions as constrained Hamiltonian systems}\label{sec:generalities}

\subsection{Notation and generalities}\label{sec:notat-gener}

The following elements of Lie group and algebra theory (see
e.g.~\cite{Spindel1985,Gibbons2006}) are useful, both for the Hamiltonian
formulation of coset spaces~\cite{Nicolai:1992xx,Matschull:1994vi} and of geometric actions~\cite{Barnich2022}.

Let $g^i$ be (arbitrary) local coordinates on a Lie group $G$ and $e_\alpha$,
$\alpha=1,\dots,n$ be a basis of its Lie algebra $\mathfrak g$, with
$[e_{\alpha},e_{\beta}]=f\indices{^{\gamma}_{\alpha\beta}}e_{\gamma}$. The
generators of right/left translations are the left/right invariant vector fields
\begin{equation}
  \label{eq:1}
  \vec L_{\alpha}=L\indices{_\alpha^i}\frac{\partial}{\partial g^i}\quad/\quad \vec R_{\alpha}=R\indices{_\alpha^i}\frac{\partial}{\partial g^i},
\end{equation}
which satisfy
\begin{equation}
  \label{eq:213}
  [\vec L_{\alpha},\vec L_\beta]=f\indices{^{\gamma}_{\alpha\beta}}\vec L_\gamma,\quad [\vec R_{\alpha},\vec R_\beta]
  =-f\indices{^{\gamma}_{\alpha\beta}}\vec R_\gamma,\quad [\vec L_{\alpha},\vec R_\beta]=0.
\end{equation}
Equivalently, if $\xi,\eta\in \mathfrak g$ and
$\vec L_\xi=\xi^\alpha \vec L_\alpha$, $\vec R_\xi=\xi^\alpha\vec R_\alpha$, these relations may be written as $[\vec L_\xi,\vec L_\eta]=\vec L_{[\xi,\eta]}$, $[\vec R_\xi,\vec R_\eta]=-\vec R_{[\xi,\eta]}$, $[\vec L_\xi,\vec R_\eta]=0$.

The left/right invariant Maurer-Cartan forms $L=g^{-1}dg/R=d g g^{-1}$ are given
by
\begin{equation}
  \label{eq:2}
  L=e_{\alpha}L^{\alpha}=L_{i}dg^{i}=e_\alpha  L\indices{^\alpha_i}dg^i\quad/\quad R=e_{\alpha}R^{\alpha}=R_{i}dg^{i}=e_\alpha R\indices{^\alpha_i}dg^i,
\end{equation}
where
\begin{equation}
  \label{eq:3}
  L\indices{_\alpha^i}L\indices{^\alpha_j}=\delta^i_j=R\indices{_\alpha^i}R\indices{^\alpha_j},\quad
  L\indices{_\alpha^i}L\indices{^\beta_i}=\delta_\alpha^\beta=  R\indices{_\alpha^i}R\indices{^\beta_i}.
\end{equation}
and
\begin{equation}
  \label{eq:214}
   dL^{\alpha}=-\frac{1}{2}f\indices{^{\alpha}_{\beta\gamma}}L^{\beta}L^{\gamma},\quad
dR^{\alpha}=\frac 12 f\indices{^{\alpha}_{\beta\gamma}}R^{\beta}R^{\gamma},
\end{equation}
or equivalently, $dL+\frac 12[L,L]=0$, $dR-\frac 12[R,R]=0$. In these terms, the
adjoint and coadjoint actions are given by
\begin{equation}
  \label{eq:31}
  {\rm Ad}_{g}\xi=e_{\beta}R\indices{^{\beta}_{i}}L\indices{_{\alpha}^{i}}\xi^\alpha,\quad
  {\rm Ad}^{*}_{g}\mu=\mu_\gamma L\indices{^{\gamma}_{j}}R\indices{_{\alpha}^{j}}e^{\alpha}_{*},
\end{equation}
where $\mu=\mu_\gamma e^\gamma_*\in \mathfrak g^*$

If there is a Lie algebra metric $g_{\alpha\beta}$ invariant under the coadjoint
representation, the associated bi-invariant metric on the Lie group is
\begin{equation}
  \label{eq:34} g_{ij}=g_{\alpha\beta}R\indices{^{\alpha}_{i}}R\indices{^{\beta}_{j}}=
  g_{\alpha\beta}L\indices{^{\alpha}_{i}}L\indices{^{\beta}_{j}}.
\end{equation}
It follows from the last of~\eqref{eq:213} that
\begin{equation}
  \label{eq:78}
  \mathcal L_{\vec L_{\xi}}R^{\beta}=0=\mathcal L_{\vec R_{\xi}}L^{\beta},
\end{equation}
so that left and right invariant vector fields are Killing vectors,
\begin{equation}
  \label{eq:15}
\mathcal L_{\vec R_{\xi}}g_{ij}=0=\mathcal L_{\vec L_{\xi}}g_{ij}.
\end{equation}
Furthermore, the Christoffel symbols are given by
\begin{equation}
  \label{eq:134}
  \Gamma\indices{^{i}_{jk}}=\frac 12 R\indices{_{\alpha}^{i}}f\indices{^{\alpha}_{\beta\gamma}}
  R\indices{^{\beta}_{j}}R\indices{^{\gamma}_{k}}-R\indices{^{\alpha}_{j}}\partial_{k}R\indices{_{\alpha}^{i}}
  =-\frac 12 L\indices{_{\alpha}^{i}}f\indices{^{\alpha}_{\beta\gamma}}
  L\indices{^{\beta}_{j}}L\indices{^{\gamma}_{k}}-L\indices{^{\alpha}_{j}}\partial_{k}L\indices{_{\alpha}^{i}},
\end{equation}
while the scalar Laplacian is given by
\begin{equation}
  \label{eq:137}
  \Delta_{G}f=g^{ij}D_{i}\partial_{j}f=|g|^{-\frac 12}\partial_{i}(|g|^{\frac 12} g^{ij}\partial_{j}f)=g^{\alpha\beta}\vec L_{\alpha}\vec L_{\beta}f=g^{\alpha\beta}\vec R_{\alpha}\vec R_{\beta}f.
\end{equation}

\subsection{Geometric actions}\label{sec:geometric-actions}

Let $g(t)$ denote maps from $\mathbb R$ to $G$, or in other words, consider a particle that moves on $G$. For a fixed nonzero covector,
$0\neq \mu\in \mathfrak g^*$, let $G_{\mu}$ be its little group,
\begin{equation}
  \label{eq:240}
  G_{\mu}\ni h, {\rm Ad}^*_h \mu=\mu.
\end{equation}
and consider left translations by elements $h^{-1}\in G_{\mu}$, and also right
translations by elements $k\in G$,
\begin{equation}
  \label{eq:216}
  g(t)\to h^{-1}(t)g(t),\quad g(t)\to g(t)k(t),
\end{equation}
with associated infinitesimal transformations
\begin{equation}
  \label{eq:220}
  \delta^R_\epsilon g^i=-\epsilon^\alpha (t)R\indices{_\alpha^i},\ \epsilon(t)\in\mathfrak g_{\mu}\subset \mathfrak g,\quad \delta^L_{\xi}g^{i}=\xi^{\alpha}(t)L\indices{_{\alpha}^{i}},\ \xi(t)\in \mathfrak g.
\end{equation}

We consider here geometric actions of the type
\begin{equation}
  \label{eq:5}
  S^{\mathcal O_{\mu}}[g;\mu]=\int dt\, [\langle \mu,{\dot g} g^{-1}\rangle-V]
  =\int dt\, \big[\mu_{\alpha} R\indices{^\alpha_i}\dot g^i-V\big],
\end{equation}
where $V(g)$ is required to be gauge invariant, so that the geometric action is suitably gauge invariant,
\begin{equation}
  \label{eq:241}
  \delta^R_{\epsilon}V=0\Longrightarrow \delta^R_{\epsilon} S^{{\mathcal O_{\mu}}}=\int dt\, \frac{d}{dt} \langle \mu,\epsilon\rangle.
\end{equation}
Accordingly, the geometric action is associated with the set of right cosets
$G_{\mu}\backslash G$, rather than with $G$ itself, and the little group $G_{\mu}$
is the gauge group of the model\footnote{At this stage, it has not yet been
  shown that the model has no other gauge symmetries, which would be the case if
  $G_{\mu}$ were only part of the gauge group. That there are no other gauge
  symmetries and that $G_{\mu}$ is indeed the full gauge group follows for
  instance from the analysis in~\ref{sec:phase-space-primary}.}. In turn,
$G_{\mu}\backslash G$ is isomorphic to the coadjoint orbit $\mathcal O_{\mu}$
containing $\mu$.

Note that, under finite gauge transformations $g'=hg$ with $h\in G_\mu$, the right
invariant Maurer-Cartan form and the one form $a_g=\langle \mu,dg g^{-1} \rangle$ transform as
\begin{equation}
  \label{eq:111}
  dg'g^{\prime -1}=hdgg^{-1}h^{{-1}}+dhh^{{-1}},\quad a_{g'}=a_g+\langle \mu,dh h^{-1}\rangle,
\end{equation}
while the two form $\sigma_g=d a_g=\langle  \mu,dg g^{-1}\wedge dg g^{-1} \rangle$ is invariant
\begin{equation}
  \label{eq:112}
  \sigma_{g'}=\sigma_g,
\end{equation}
because $d(dh h^{-1})=\frac 12 [dhh^{-1},dhh^{-1}]$ and $dhh^{{-1}}$ belongs to
the little algebra $\mathfrak g_{\mu}$. Accordingly, if $\gamma$ is a closed loop on
$G$, the gauge invariant functional\footnote{See~\cite{Wiegmann:1989hn} in a closely related context with a $U(1)$ gauge group.} related to the (kinetic term of the)
geometric action is a Wilson loop~\cite{Wilson:1974sk}
\begin{equation}
  \label{eq:115}
  W_{\gamma}=P \exp \oint_{\gamma} a_{g},
\end{equation}
which may be expressed in terms of the two form by using the non-abelian Stokes'
theorem~\cite{Halpern:1978ik,Arefeva1980}
\begin{equation}
  \label{eq:117}
  W_{\gamma}=\mathcal P \exp \int_{\Sigma} \sigma_{g},\quad \partial \Sigma=\gamma.
\end{equation}

Under the infinitesimal transformations $\delta^L_{\xi}$ associated to right
translations, the kinetic term transforms as
\begin{equation}
  \label{eq:244}
  \delta^L_{\xi} \langle \mu, \dot g g^{{-1}}\rangle =Q_{\dot \xi},\quad Q_{\xi}=\langle \mu,{\rm Ad}_{g}\xi\rangle =\mu_{\alpha}
  R\indices{^{\alpha}_{i}}L\indices{_{\beta}^{i}}\xi^{\beta}.
\end{equation}
We furthermore require here that the time dependence of $\xi(t)$ may be determined
through the equation
\begin{equation}
  \label{eq:243}
  Q_{\dot \xi}-\delta^L_{\xi}V=0 .
\end{equation}
It then follows that the geometric action is invariant under the associated
infinitesimal global symmetries, $\delta^L_{\xi}S^{\mathcal O_{\mu}}=0$ with
associated Noether charges given by $Q^L_{\xi}$.

For instance, if $V=Q^L_{\eta}$ for some fixed $\eta\in\mathfrak g$, it is indeed gauge invariant. The corresponding geometric action, on which we concentrate below, is denoted by
\begin{equation}
  \label{eq:246}
  \boxed{S^{{\mathcal O_{\mu}}}[g;\mu,\eta]=\int dt\, \langle \mu,\dot gg^{-1}-{\rm Ad}_{g}\eta\rangle
  =\int dt\, \big[\mu_{\alpha} R\indices{^\alpha_i}\dot g^i-\mu_{\alpha} R\indices{^\alpha_i}L\indices{_{\beta}^{i}}\eta^{\beta}\big]}.
\end{equation}
In this case, the time-dependence of $\xi$ may be fixed through $\dot \xi=[\xi,\eta]$. Equivalently,
if $Q^L_{\alpha}=\mu_{\beta}R\indices{^{\beta}_{i}}L\indices{_{\alpha}^{i}}$ denote generators for the Noether charges at $t=0$,
\begin{equation}
  \label{eq:46}
  Q^L_{\alpha}(t)=Q_{\alpha}+t Q^L_{\gamma}f\indices{^{\gamma}_{\alpha\beta}}\eta^{\beta}.
\end{equation}
Similarly, for the associated finite transformations by right translations, the
geometric action is invariant provided the time-dependence is suitably fixed,
\begin{equation}
  \label{eq:217}
  g(t)\to g(t)k(t),\quad  {\dot k} k^{-1}={\rm Ad}_k \eta-\eta,\quad
  S^{{\mathcal O_{\mu}}}[gk;\mu,\eta]=S^{{\mathcal O_{\mu}}}[g;\mu,\eta].
\end{equation}

Consider partitions of $\mathfrak g^{*}$ and $\mathfrak g$ into (co)-adjoint orbits,
$\mu'\sim \mu\iff \mu^{'}={\rm Ad}^{*}_{g_1}\mu$, $\eta^{\prime}\sim \eta\iff \eta^{\prime}={\rm Ad}_{g_2}\eta$;
and let $\mu^{\Xi},\eta_{\Sigma}$ denote a set of orbit representatives for these
partitions,
\begin{equation}
  \label{eq:14}
  \mathfrak g^{*}\simeq \bigcup_{\Xi}{\rm Ad}^{*}_{G}\mu^{\Xi},\quad \mathfrak g\simeq \bigcup_{\Sigma}{\rm Ad}_{G}\eta_{\Sigma}.
\end{equation}
Geometric actions associated to Lie algebra covectors or vectors that belong to
the same equivalence classes are (quantum)-mechanically equivalent in the sense
that they are related by field redefinitions,
\begin{equation}
  \label{eq:18}
  S^{{\mathcal O_{\mu'}}}[g;\mu',\eta']=S^{{\mathcal O_{\mu}}}[g';\mu,\eta],\quad g'(t)=g_1^{-1}g(t)g_2.
\end{equation}
It follows that it is enough to study the actions $S^{\mathcal O_{\mu^{\Xi}}}[g;\mu^{\Xi},\eta_{\Sigma}]$
associated to the different orbit representatives.

\subsection{Cotangent bundle and constraints}\label{sec:phase-space-primary}

In this subsection, we provide an alternative way to obtain the geometric
actions associated to individual coadjoint orbits by imposing suitable
constraints on the cotangent bundle of the group.

The cotangent bundle $T^{*}G$ can either be described through
canonical coordinates $(g^{i},p_{i})$ with canonical Poisson brackets
\begin{equation}
  \label{eq:32}
  \{g^{i},p_{j}\}=\delta^{i}_{j},\quad \{g^{i},g^{j}\}=0=\{p_{i},p_{j}\},
\end{equation}
or in terms of (non-Darboux) coordinates $(g^{i},\pi_{\alpha})$ adapted to the globally
well-defined left trivialization $T^*G\simeq G\times \mathfrak g^*$,
\begin{equation}
  \label{eq:80}
  \pi_{\alpha}=R\indices{_{\alpha}^{i}}p_{i},
\end{equation}
for which the fundamental Poisson brackets contain the Kirillov-Kostant-Souriau
brackets,
\begin{equation}
  \label{eq:215} \{\pi_{\alpha},\pi_{\beta}\}=f\indices{^{\gamma}_{\alpha\beta}}\pi_{\gamma},\quad\{g^{i},\pi_{\alpha}\}=R\indices{_{\alpha}^{i}},\quad \{g^{i},g^{j}\}=0.
\end{equation}

When performing a standard Hamiltonian analysis of the geometric
action~\eqref{eq:246}, one gets the primary constraints
$p_{i}- \mu_{\alpha}R\indices{^{\alpha}_{i}}= 0$, because the Lagrangian is
linear in time-derivatives. When expressed in the variables $(g^i,\pi_\alpha)$,
these primary constraints become $\pi_{\alpha}-\mu_{\alpha}=0$. Imposing these
primary constraints through Lagrange multipliers, gives a description of the
geometric action for $\mathcal O_\mu$ as a constrained system on $T^*G$,
\begin{equation}
  \label{eq:19}
  \boxed{S^{{\mathcal O_{\mu}}}[g,\pi,u;\mu,\eta]=\int dt\, \big[\pi_{\alpha} R\indices{^\alpha_i}\dot g^i-\pi_{\alpha} R\indices{^\alpha_i} L\indices{_\beta^i}\eta^\beta-u^{\alpha}\phi^{\mu}_{\alpha}\big],\ \phi^{\mu}_{\alpha}=\pi_{\alpha}-\mu_{\alpha}}.
\end{equation}

The Noether charges for right translations are described through\footnote{There is a sign mistake in equation (3.27) of~\cite{Barnich2022}.}
\begin{equation}
  \label{eq:59}
  Q^L_{\xi}=\pi_{\beta}R\indices{^{\beta}_{i}}L\indices{_{\alpha}^{i}}\xi^{\alpha},\quad \{Q^L_{\xi_{1}},Q^L_{\xi_{2}}\}=-Q^L_{[\xi_{1},\xi_{2}]}.
\end{equation}
They generate the symmetries in the Poisson bracket,
\begin{equation}
  \label{eq:75}
  \delta^L_{\xi}g^{i}=\{g^{i},Q^L_{\xi}\}=L\indices{_{\alpha}^{i}}\xi^{\alpha},\quad \delta^L_{\xi}\pi_{\alpha}=\{\pi_{\alpha},Q_{\xi}\}=0.
\end{equation}
If the Hamiltonian vector fields associated to a phase space function $f$ are
defined as $X_{f}=-\{f,\cdot \}$, with $[X_{f},X_{g}]=-X_{\{f,g\}}$, those
associated to the Noether charges $Q^L_{\xi}$ and to the $\pi_\xi=\pi_{\alpha}\xi^\alpha$ are
\begin{equation}
  \label{eq:79}
  X_{Q^L_{\xi}}=\xi^\alpha\vec L_{\alpha},\quad X_{\pi_{\xi}}= \xi^\alpha [\vec R_{\alpha}+f\indices{^{\gamma}_{\beta\alpha}}\pi_{\gamma}\frac{\partial}{\partial \pi_{\beta}}].
\end{equation}

Applying the algorithm by Dirac to this system is straightforward. There are no
secondary constraints. In particular, it then follows from
$\{Q^L_{\xi},\phi^{\mu}_{\alpha}\}=0$, that the Noether charges are first class functions and
thus also gauge invariant. The nature of the constraints is determined by the
matrix
\begin{equation}
  \label{eq:221}
  C_{\alpha\beta}=\mu_{\gamma}f\indices{^{\gamma}_{\alpha\beta}}.
\end{equation}
Its eigenvectors with eigenvalue zero correspond to the little algebra
$\mathfrak g_{\mu}$ of $\mathfrak g$. The associated constraints are first
class. The constraints associated to the supplementary space in $\mathfrak g$
are second class. More explicitly, let
$\mathfrak g=\mathfrak g_\mu\oplus\mathfrak m$, and consider an adapted basis
$(e_a,e_A)$ of $\mathfrak g$ together with an adapted basis $(e^A,e^a)$ of
$\mathfrak g^*=\mathfrak g^\circ_\mu\oplus\mathfrak n^*$, where
$\mathfrak g^\circ_\mu\in \mathfrak g^*$ is the annihilator of
$\mathfrak g_\mu$. In terms of components,
\begin{equation}
  \label{eq:222}
  e\indices{_{a}^{\alpha}},\quad e\indices{_{A}^{\alpha}},\quad e\indices{^{a}_{\alpha}},\quad e\indices{^{A}_{\alpha}},
\end{equation}
\begin{equation}
  \label{eq:223}
e\indices{_a^\alpha}e\indices{^B_\alpha}=0=e\indices{_A^\alpha}e\indices{^b_\alpha},\quad  e\indices{_{a}^{\alpha}}e\indices{^{b}_{\alpha}}=\delta^{a}_{b},\quad
  e\indices{_{A}^{\alpha}}e\indices{^{B}_{\alpha}}=\delta^{A}_{B},\quad
  e\indices{_{a}^{\alpha}}e\indices{^{a}_{\beta}}+e\indices{_{A}^{\alpha}}e\indices{^{A}_{\beta}}=\delta^{\alpha}_{\beta}.
\end{equation}
When using these matrices to convert indices in the usual way, it follows that
\begin{equation}
  \label{eq:74}
  f\indices{^{C}_{{ab}}}=0,\quad C_{ab}=0=C_{Ab},\quad C_{AB}=\mu_{c}f\indices{^{c}_{AB}}+\mu_{C}f\indices{^{C}_{AB}}\ \mathrm{invertible}.
\end{equation}
The first and second class constraints are then, respectively,
\begin{equation}
  \label{eq:81}
  \phi^{\mu}_{a}=0,\quad \phi^{\mu}_{A}=0,
\end{equation}
while the Dirac brackets are
\begin{equation}
  \label{eq:84}
  {\{g^{i},g^{j}\}}^{*}=R\indices{_{A}^{i}}C^{-1 AB}R\indices{_{B}^{j}},\
  {\{g^{i},\pi_{a}\}}^{*}=R\indices{_{a}^{i}},\ {\{\pi_{a},\pi_{b}\}}^{*}=f^{c}_{ab}\pi_{c}\approx0,\ {\{\pi_{A},\cdot \}}^{*}=0,
\end{equation}
where $\approx 0$ means here ``vanishing on the first class constraint surface'',
and
\begin{equation}
  \label{eq:20}
  {\{\phi^{\mu}_{a},\phi_{b}^{\mu}\}}^{*}=f\indices{^{c}_{ab}}\phi_{c}^{\mu}.
\end{equation}

As usual for second class constraints, a more economic, but less covariant,
description is achieved by using the fact that the coordinates associated to the
second class constraints, here $\pi_{A}$, and the associated Lagrange
multipliers $u^{A}$ are auxiliary fields whose equations of motion,
\begin{equation}
  \label{eq:27}
  u^{A}=R\indices{^{A}_{i}}\dot g^{i}+R\indices{^{A}_{i}}L\indices{_{\beta}^{i}}\eta^{\beta},\quad \pi_{A}=\mu_{A},
\end{equation}
may be solved in the action~\eqref{eq:19}. In this reduced description, the geometric action becomes
\begin{multline}
  \label{eq:22}
  S^{{\mathcal O_{\mu}}}_{R}[g^{i},\pi_{a},u^{a};\mu_{A},\eta^{\beta}]=\int dt\, [
  \pi_{a}R\indices{^{a}_{i}}\dot g^{i}+\mu_{A}R\indices{^{A}_{i}}\dot g^{i}\\-
  \pi_{a}R\indices{^{a}_{i}}L\indices{_{\beta}^{i}}\eta^{\beta}-\mu_{A}R\indices{^{A}_{i}}
  L\indices{_{\beta}^{i}}\eta^{\beta}-u^{a}\phi^{\mu}_{a}],
\end{multline}
and only involves the first class constraints.

Elimination of the first class constraints by the Lagrange multiplier method then gives back the starting point~\eqref{eq:5}. In that case, the potential 1-form is pre-symplectic,
\begin{equation}
  \label{eq:35}
 a^{\mu}=\mu_{\alpha} R\indices{^{\alpha}_{i}}dg^{i},\quad \sigma^{\mu}=da^{\mu}=\frac 12 C_{AB}R\indices{^{A}_{i}}R\indices{^{B}_{j}}dg^{i}\wedge dg^{j},
\end{equation}
with null vectors of the associated $2$-form given by $R\indices{_{\alpha}^{i}}\frac{\partial}{\partial g^{i}}$, whereas by
construction, the potential one-form associated to~\eqref{eq:22} gives rise to a symplectic two-form,
\begin{equation}
  \label{eq:35b}
  a^{R}=\pi_{a}R\indices{^a_i}dg^i+\mu_A R \indices{^A_i}dg^i,\quad
  \sigma^R=da^R=\frac 12 C_{AB}R\indices{^A_i}R\indices{^{B}_{j}}dg^i\wedge dg^{j}-R\indices{^{b}_{i}}dg^{i}\wedge d\pi_{b},
\end{equation}
and, on the variables of the reduced theory, the Poisson structure determined by
the inverse matrix coincides with the Dirac brackets in~\eqref{eq:84}.

\subsection{Geometric action for the cotangent bundle}\label{sec:unconstrained-model-1}

When dropping all constraints $\phi_{\alpha}^{\mu}$, one gets an action associated to $T^{*}G\simeq G\times \mathfrak g^*$, described in the variables adapted to the left trivialization,
\begin{equation}
  \label{eq:6}
  \boxed{S^{T^{*}G}[g,\pi;\eta]=\int dt\, \langle \pi,\dot g g^{-1} -{\rm Ad}_g \eta \rangle=\int dt\, \big[\pi_\alpha R\indices{^\alpha_i}\dot g^i-\pi_\alpha R\indices{^\alpha_i} L\indices{_\beta^i}\eta^\beta\big]}.
\end{equation}
The associated potential one-form
gives rise to a symplectic two-form,
\begin{equation}
  \label{eq:40}
  a=\pi_{\alpha}R\indices{^{\alpha}_{i}}dg^{i},\quad \sigma=da=\frac 12 \pi_{\alpha}f\indices{^{\alpha}_{\beta\gamma}}R\indices{^{\beta}_{i}}
  R\indices{^{\gamma}_{j}}dg^{i}\wedge dg^{j}-R\indices{^{\beta}_{i}}dg^{i}\wedge d\pi_{\beta},
\end{equation}
whose inverse matrix determines the Poisson brackets~\eqref{eq:215}.

The equations of motion for~\eqref{eq:6} are
\begin{equation}
  \label{eq:224}
  \dot g^{i}=\{g^{i},Q^L_{\eta}\}=L\indices{_{\alpha}^{i}}\eta^{\alpha},\quad \dot \pi_{\alpha}=\{\pi_{\alpha},Q^L_{\eta}\}=0.
\end{equation}
It follows that, besides the $Q^L_{\xi}$, the $\pi_\alpha$ (or the $\pi_\xi=\xi^\alpha\pi_\alpha$) are also
constants of the motion. The associated global symmetries correspond to
invariance of the action under global left translations, $g'=hg$,
$\pi'=h\pi h^{-1}$, with $\pi=\pi_{\alpha}e_{*}^{\alpha}$ and $h$ constant.

Studying Hamiltonian reductions on the level sets of $\pi_{\xi}$ amounts to
performing the previous analysis in reverse.

On $T^{*}G$,  the kinetic term can be written in terms of canonical coordinates or the ones adapted to the left or the right trivialization as
\begin{equation}
  \label{eq:254}
  \int dt\ p_{i}\dot g^{i}=\int dt\ \pi_{\alpha} R\indices{^{\alpha}_{i}}\dot g^{i}=\int dt\ \psi_{\alpha}L\indices{^{\alpha}_{i}}\dot g^{i},
\end{equation}
where $\psi_{\alpha}=L\indices{_{\alpha}^{i}}p_{i}$. Its global symmetries
include both right and left translations: in Darboux coordinates
$(g^{i},p_{j})$, they act as
\begin{equation}
  \label{eq:265}
  \delta^{L}_{\xi} g^{i}=\xi^\alpha L\indices{_{\alpha}^{i}},\
  \delta^{L}_{\xi}p_{i}=-\xi^\alpha p_{k}\partial_{i} L\indices{_{\alpha}^{k}},
  \quad \delta^{R}_{\xi} g^{i}=\xi^\alpha R\indices{_{\alpha}^{i}},\
  \delta^{R}_{\alpha}p_{i}=-p_{k}\partial_{i} R\indices{_{\alpha}^{k}},
\end{equation}
with associated Noether charges
\begin{equation}
  \label{eq:266}
  Q^{L}_{\xi}=p_{i}L\indices{_{\alpha}^{i}}\xi^\alpha,\quad Q^{R}_{\xi}=p_{i}R\indices{_{\alpha}^{i}}\xi^\alpha,
\end{equation}
while in the $(g^{i},\pi_{\alpha})/ (g^{i},\psi_{\alpha})$ parametrizations of phase
space, they act as
\begin{equation}
  \label{eq:267}
  \delta^{L}_{\xi} \pi_{\beta} =0,\ \delta^{R}_{\xi}\pi_{\beta}=-\pi_{\gamma}f\indices{^{\gamma}_{\alpha\beta}}\xi^\alpha\ /\
  \delta^{L}_{\xi} \psi_{\beta} =\psi_{\gamma}f\indices{^{\gamma}_{\alpha\beta}}\xi^\alpha,\ \delta^{R}_{\xi}\psi_{\beta}=0,
\end{equation}
with associated Noether charges
\begin{equation}
  \label{eq:268}
  Q^{L}_{\xi}=\pi_{\beta} R\indices{^{\beta}_{i}}L\indices{_{\alpha}^{i}}\xi^\alpha,\ Q^{R}_{\xi}=\pi_{\alpha}\xi^\alpha,\ / \
  Q^{L}_{\xi}=\psi_{\alpha}\xi^\alpha,\  Q^{R}_{\xi}=\psi_{\beta}R\indices{^{\beta}_{i}}L\indices{_{\alpha}^{i}}\xi^\alpha.
\end{equation}

Let $\pi=\pi_\alpha e_*^\alpha$, $\psi=\psi_\alpha e_*^\alpha$. In terms of the left trivialization of
$T^*G \simeq G\times \mathfrak g^*\ni (g,\pi)$, the momentum maps
\begin{equation}
  \label{eq:13}
  Q^L:T^*G\to \mathfrak g^*,\quad Q^R:T^*G\to \mathfrak g^*,
\end{equation}
for the lift of the right and left translations to $T^*G$ are explicitly given
by
\begin{equation}
  \label{eq:9}
  Q^L={\rm Ad^*}_{g^-1}\pi,\quad Q^R=\pi,
\end{equation}
whereas for the right trivialization $T^*G \simeq G\times \mathfrak g^*\ni (g,\psi)$, they are given by
\begin{equation}
  \label{eq:10}
  Q^L=\psi,\quad Q^R={\rm Ad^*}_{g^-1}\psi.
\end{equation}

In the case where there is an invariant metric $g_{\alpha\beta}$ on the Lie algebra (giving rise to a bi-invariant metric $g_{ij}$ on the group), the Hamiltonian
\begin{equation}
  \label{eq:48}
  H=\frac 12 g^{ij}p_{i}p_{j}=\frac 12 g^{\alpha\beta}\pi_{\alpha}\pi_{\beta}=\frac 12 g^{\alpha\beta}\psi_{\alpha}\psi_{\beta},
\end{equation}
preserves all these symmetries. When eliminating the momenta by their own equations of motion from
\begin{equation}
  \label{eq:87}
  S^{T^{*}G}[g;\pi;H]=\int dt\, \big[\pi_\alpha R\indices{^\alpha_i}\dot g^i-H\big],
\end{equation}
one finds the action that gives rise to geodesic motion on $G$,
\begin{equation}
  \label{eq:248}
  S[g]=\frac 12 \int dt\,  g_{ij}\dot g^{i}\dot g^{j}.
\end{equation}

Less symmetric choices are also interesting in the context of integrable
systems. When there is a metric $g'_{\alpha\beta}$ on the Lie algebra that is not
necessarily invariant under the coadjoint representation, it may be extended to a
right invariant metric $g'_{ij}=g'_{\alpha\beta}R\indices{^{\alpha}_{i}}R\indices{^{\beta}_{j}}$
on the group. In this case, the associated geodesic flow is still invariant
under right translations because the left invariant vector fields are still
Killing vectors of this metric.

\subsection{Heuristic proposal for the model space}\label{sec:model-space}

When starting from $T^*G$ of dimension $2n$, with symplectic actions provided by
the lift of the left and right multiplications, and a given coadjoint vector
$\mu\in \mathfrak g^*$, with little algebra $\mathfrak g_\mu$ of dimension $k$,
one may associate three symplectic submanifolds that inherit symplectic actions
of G that descend from the lift of right multiplication:

\begin{itemize}

\item the orbit $\mathcal O_\mu$ of dimensions $2n-n-k=n-k$ obtained by imposing all the
  primary constraints $\phi^\mu_\alpha=0$ and quotienting by the gauge orbits associated to
  the first constraints $\phi^\mu_a=0$,

\item $\mathcal F_\mu$ of dimension $2n-2k$ obtained by imposing only the first class
  constraints $\phi^\mu_a=0$ and quotienting by their gauge orbits,

\item $\mathcal M_\mu$ of dimension $2n-(n-k)=n+k$ obtained by imposing only the second class
  constraints $\phi^\mu_A=0$.

\end{itemize}

Our main contention here is that a central ingredient in the construction of a
model space for a generic Lie group $G$ is the symplectic submanifold
$\mathcal M_{\mu}$. More explicitly, the geometric action associated to
$\mathcal M_{\mu}$ is
\begin{equation}
  \label{eq:225}
  \boxed{S^{\mathcal M_{\mu}}[g,\pi,u^{A};\mu_{A},\eta^{\beta}]=\int dt\,[\pi_{\alpha}R\indices{^{\alpha}_{i}}\dot g^{i}
    -\pi_{\alpha}R\indices{^{\alpha}_{i}}
    L\indices{_{\beta}^{i}}\eta^{\beta}-u^{A}(\pi_{A}-\mu_{A})].}
\end{equation}
A reduced description, in this case without any constraints, is again achieved
by solving the second class constraints in the action, which yields
\begin{equation}
  \label{eq:218}
  S^{\mathcal M_{\mu}}_{R}[g,\pi_{a};\mu_{A},\eta^{\beta}]=\int dt\,[\pi_{a}R\indices{^{a}_{i}}\dot g^{i}
  +\mu_{A}R\indices{^{A}_{i}}\dot g^{i}
    -\pi_{a}R\indices{^{a}_{i}}
    L\indices{_{\beta}^{i}}\eta^{\beta}-\mu_{A}R\indices{^{A}_{i}}
    L\indices{_{\beta}^{i}}\eta^{\beta}\big].
\end{equation}
This description may also be directly obtained from the starting point
description of the geometric action for $\mathcal O_{\mu}$ in~\eqref{eq:246} by replacing the
components along the supplementary space $\mathfrak n^*$ to the annihilator of
the little algebra of the coadjoint vector $\mu$ by new dynamical variables,
$\mu_{a}\to \pi_{a}$. The associated symplectic and bracket structures are given in~\eqref{eq:35b}
and~\eqref{eq:84}. Adding these new variables in terms of the globally well-defined
trivialization of $T^*G$ rather than in terms of Darboux coordinates goes some
way to take into account some of the remarks of~\cite{La:1990ty} on the original
proposal in~\cite{Alekseev:1990mp}.

For completeness, let us also provide a description of the symplectic space
$\mathcal M_{\mu}$ in terms of Hamiltonian reduction, rather than through Dirac's
language of constraints and their classification. It is well-known~\cite{Marsden1974} that the
coadjoint orbit $\mathcal O_{\mu}$ of $\mu\in \mathfrak g^*$ may be described in terms
of the momentum map $Q^R$ associated to the lift of left multiplication to
$T^*G$ through
\begin{equation}
  \label{eq:11}
  \mathcal O_{\mu}= {(Q^R)}^{-1}(\mu)/G_{\mu}.
\end{equation}
In our previous discussion, the restriction to ${(Q^R)}^{-1}(\mu)$ corresponds to
imposing both the first and the second class constraints $\phi^{\mu}_a=0$,
$\phi_A^{\mu}=0$.

Consider the projection $p^*_{\mu}:\mathfrak g^*\to \mathfrak g_{\mu}^\circ $ on
the annihilator $\mathfrak g_{\mu}^\circ$ of the little algebra $\mathfrak g_{\mu}$. The
symplectic submanifold $\mathcal M_{\mu}\subset T^*G$ is obtained from $T^*G$ and the
momentum map $Q^R$ associated to the lift of left multiplication to $T^*G$, and
an element $\mu^\circ\in \mathfrak g_{\mu}^\circ$ through
\begin{equation}
  \label{eq:12}
 \mathcal M_\mu= {[p^*_{\mu}\circ Q^R]}^{-1}(\mu^\circ).
\end{equation}
This is the equivalent of imposing only the second class constraints
$\phi_A^{\mu}=0$. As explicitly shown through the Dirac analysis above, there are no gauge orbits in this case. In terms of the left trivialization
$T^*G\simeq G\times \mathfrak g^*\ni (g,\pi)$, the associated symplectic and
bracket structures are given in~\eqref{eq:35b} and~\eqref{eq:84}.

What matters for the construction of the representations of $G$ is that both
$\mathcal O_{\mu}$ and $\mathcal M_{\mu}$ inherit Hamiltonian actions of $G$ that
descend from the lift of right multiplication to $T^*G$.

Coadjoint orbits are classified in terms of continuous and discrete parameters
$c^i,d^I$. If a complete set of orbit representatives $\mu^{c,d}$ has been
identified, one needs to consider the $\mathcal M_{\mu^{c,d}}$ for each of these
orbit representatives. In the particular case of $\mathrm{SU}(2)$ discussed in
detail below, there is only one continuous parameter, the norm
$\boldsymbol{\mu}=||\mu||$ of the coadjoint vector $\mu$, that completely
classifies the orbits. For $\boldsymbol{\mu}>0$, the orbits are 2-dimensional.
For $\boldsymbol{\mu}=0$, the orbit reduces to the origin $\mu=0$, while the
little group $G_{0}=\mathrm{SU}(2)$. This case is naturally
associated with the trivial representation.

\subsection{Variety of descriptions and quantization schemes}\label{sec:descriptions}

Other descriptions for the coadjoint orbits and the model space could be imagined. For example, it might be useful to describe the system
associated to $T^{*}G$ itself as a constrained system of a suitable ``embedding''
phase space $ET^{*}G$, even before discussing additional constraints that bring
one down successively to the model space and the individual coadjoint orbits.
This will in particular be the case for $\mathrm{SU}(2)$ below: as a manifold
$\mathrm{SU}(2)$ corresponds to a three-sphere $\mathbb S^{3}$, which is best
understood as a submanifold of the embedding space $\mathbb R^{4}$. On the level
of the group, this corresponds to going from unimodular to general (nonzero)
quaternions.

Furthermore, first-class constraints $\gamma_{a}=0$ may be turned into second-class ones through
gauge fixing conditions $\chi_{b}=0$, which should be reachable, and fix the gauge
completely: $\{\gamma_{a},\chi_{b}\}$ should be invertible on the constraint surface.
Conversely, second-class constraints may be converted into half their number of
first-class ones. Depending on the starting point, a variety of descriptions are
thus possible. For instance,
\begin{itemize}
  \item the cotangent bundle $T^{*}G$ is described as first or second class
        constrained system of $ET^{*}G$,
  \item the space $\mathcal M_{\mu}$ is described as first or a second class
        constrained system of $ET^{*}G$ or $T^{*}G$,
  \item individual coadjoint orbits $\mathcal O_{\mu}$ are described as first
        or second class constrained systems of $ET^{*}G$, $T^{*}G$, or $\mathcal M_{\mu}$.
\end{itemize}

The quantization of these different systems may also be performed in various ways. The most direct, but not necessarily the most transparent, is reduced phase space quantization. It consists of transforming first-class into second-class constraints by a choice of canonical gauge-fixing conditions. One then solves the second-class constraints $\chi_{A}(q,p)=0$ in terms of independent coordinates $y^{i}$ and quantizes the associated Dirac brackets ${\{y^{i},y^{j}\}}^*$ in a suitable Hilbert space. When the geometry of the constraint surface is such that there does not exist a global gauge condition that intersects the gauge orbits once and only once, one has to deal with what is commonly called the Gribov obstruction.

On the other hand, first-class constraints, which may have been obtained after
conversion of a second-class constrained system into a first-class one, may be
imposed after quantization on the Hilbert space of the unconstrained system.
This method is known as Dirac quantization. In the compact case that we consider
below, there are no issues with a divergent inner product and no need for
additional delta functions in the measure.

Even though not strictly necessary in this context of Lie algebras where there
are no structure functions, open algebras or reducible constraints, Hamiltonian
BRST-BFV operator techniques should prove useful for more complicated groups
than $\mathrm{SU}(2)$. Finally, one may also choose to quantize through path integral
methods. How subtleties related to the path integral measure should be handled
is briefly discussed in Appendix~\ref{sec:path-integr-meas}.

\section{Application to $\mathrm{SU}(2)$: Classical theory}\label{sec:classical-theory}

Our notations and conventions for $\mathrm{SU}(2)$ can be found in Appendix~\ref{sec:gener-notat}.

\subsection{(Co) adjoint orbits and Hopf fibration}\label{sec:coadjoint-orbits}

The adjoint and coadjoint representations of $\mathrm{SU}(2)$ are isomorphic to the vector representation of $\mathrm{SO}(3)$.
Indeed, when using that $g^{-1}=g^{\dagger}$, it follows that the matrix of the adjoint representation can be written as
\begin{equation}
  \label{eq:210}
  \mathcal R\indices{^{\alpha}_{\beta}}\equiv R\indices{^{\alpha}_{i}}L\indices{_{\beta}^{i}}=\frac 12{\rm Tr}(\sigma^{\alpha} g\sigma_{\beta} g^{\dagger}),
\end{equation}
which coincides with the standard $2$ to $1$ homomorphism of $\mathrm{SU}(2)$
into $\mathrm{SO}(3)$.

For fixed (co) vector $\mu\in \mathfrak{su}^{(*)}(2)$, the subgroup of
rotations that leave this vector invariant is given by $\mathcal R(\hat \mu,\psi)$.
The (co) adjoint orbits are spheres $\mathbb S^{2}_{\boldsymbol{\mu}}$ of radius
$\boldsymbol{\mu}\geq 0$. On the level of $\mathrm{SU}(2)$, the little group
$G_{\mu}$ corresponds to the subgroup of matrices of the form
\begin{equation}
  \label{eq:228}
  h=e^{-\psi\frac{\hat \mu^{\alpha}}{2}\imath\sigma_{\alpha}},\quad 0\leq \psi <4\pi,
\end{equation}
and is isomorphic to $\mathrm{U}(1)$ for $\mu\neq 0$.

Representatives for the (co)~adjoint orbits may be chosen along the $z$-axis,
\begin{equation}
  \label{eq:238}
  \mu_\alpha^{\epsilon\boldsymbol{\mu}}=\epsilon\boldsymbol{\mu} \delta_{\alpha}^{3},\quad \xi^{\alpha}_{\boldsymbol{\xi}}=\boldsymbol{\xi}\delta^{\alpha}_{3},
\end{equation}
where $\epsilon=\pm1$ is introduced for later convenience and keeps track of
whether the coadjoint representative is chosen along the positive or negative
$z$-axis,
\begin{equation}
  \label{eq:211}
  \mathfrak{su}^{(*)}(2)\simeq \mathbb R^{3}=\bigcup_{{\boldsymbol{\mu}}}\mathcal R_{SO(3)} \mu^{\epsilon{\boldsymbol{\mu}}}.
\end{equation}
In this case, the little group $G_{\mu^{\epsilon{\boldsymbol{\mu}}}}$ is
explicitly described by
\begin{equation}
  \label{eq:239}
  h=\begin{pmatrix} e^{-\imath \epsilon \frac{\psi}{2}} & 0 \\ 0 & e^{\imath\epsilon \frac{\psi}{2}}
  \end{pmatrix}, \quad 0\leq \psi <4\pi.
\end{equation}
for $\boldsymbol{\mu}>0$.

The set of right cosets, $G_{\mu^{\epsilon{\boldsymbol{\mu}}}}\backslash \mathrm{SU}(2)\simeq \mathbb S^{2}$, is directly related to the Hopf fibration. For an explicit description, the following parametrization of $\mathrm{SU}(2)$ through Euler angles is useful,
\begin{equation}
  \label{eq:99x}
  g=e^{-\frac{\psi}{2}\imath\sigma_{3}}e^{-\frac{\theta}{2}\imath\sigma_{2}}e^{-\frac{\phi}{2}\imath\sigma_{3}}
  =\begin{pmatrix} \cos\frac{\theta}{2}e^{-\imath\frac{\psi+\phi}{2}} & -\sin\frac{\theta}{2}e^{-\imath\frac{\psi-\phi}{2}} \\
    \sin\frac{\theta}{2}e^{\imath\frac{\psi-\phi}{2}} & \cos\frac{\theta}{2}e^{\imath\frac{\psi+\phi}{2}}
\end{pmatrix},
\end{equation}
with $0\leq \psi<4\pi$, $0\leq \phi<2\pi$, $0< \theta< \pi$. Explicit
expressions for the left and right invariant vector fields and Maurer-Cartan
forms, as well as the adjoint and coadjoint representations are provided in
Appendix~\ref{sec:adapted-euler-angles-1}.

In this case, a gauge transformation by left multiplication with
$h^{-\epsilon}(t)$ may be used to reach the Borel gauge,
\begin{equation}
  \label{eq:245}
  \psi=0,\quad g_{B}=e^{-\frac{\theta}{2}\imath\sigma_{2}}e^{-\frac{\phi}{2}\imath\sigma_{3}}
  =\begin{pmatrix} \cos\frac{\theta}{2}e^{-\imath\frac{\phi}{2}} & -\sin\frac{\theta}{2}e^{\imath\frac{\phi}{2}} \\
    \sin\frac{\theta}{2}e^{-\imath\frac{\phi}{2}} & \cos\frac{\theta}{2}e^{\imath\frac{\phi}{2}}
\end{pmatrix}.
\end{equation}
As discussed in more details below, there is a Gribov obstruction: the geometry
of the constraint surface and the gauge orbits is such that it forbids the
existence of global gauge conditions. In other words, neither the Borel, nor any
other canonical gauge condition, is globally valid because the Hopf bundle is
non-trivial.

\subsection{Coadjoint orbits: Local and global descriptions}\label{sec:reduced-phase-space}

With the above choice of nonzero covector representative $\mu^{\epsilon{\boldsymbol{\mu}}}, \boldsymbol{\mu}\neq 0$, the
nature of the constraints is determined by $C_{AB}=\epsilon\boldsymbol{\mu}\epsilon_{AB}$, with
$A,B,\dots =1,2$, $\epsilon_{AB},\epsilon^{AB}$ completely skew-symmetric and $\epsilon_{12}=\epsilon^{21}=1$,
so that ${(C^{-1})}^{{AB}}=\frac{1}{\epsilon \boldsymbol{\mu}}\epsilon^{AB}$.
For convenience, let us define
\begin{equation}
  \label{eq:105}
  q=-\epsilon\boldsymbol{\mu}.
\end{equation}
The second class constraints are thus $\pi_{1}=0=\pi_{2}$, while the first class
constraint is $\gamma=\pi_{3}+q=0$, which generates arbitrary shifts in $\psi$.

Locally, a reduced phase space description is achieved by turning this first
class constraint into two second class ones by a suitable gauge fixing
condition. This is done in the parametrization in terms of adapted Euler angles
through the Borel gauge, $\chi=\psi$ which is reachable. It also follows
from~\eqref{eq:47Bx} that $\{\chi,\gamma\}=R\indices{_{3}^{1}}=1$, so that the gauge is
fixed completely.

The non-vanishing Dirac brackets are given by
\begin{equation}
  \label{eq:53}
  {\{\theta,\phi\}}^{*}=R\indices{_{A}^{2}}{(C^{-1})}^{AB}R\indices{_{B}^{3}}=
  -\frac{1}{q\sin{\theta}}.
\end{equation}
When using in addition~\eqref{eq:49Bx}, a local expression for the completely
reduced action is
\begin{equation}
  \label{eq:52}
  \boxed{S^{\mathbb S^{2}_{\boldsymbol{\mu}}}[\theta,\phi;\mu,\xi]=-\int dt\, q[\cos{\theta}\frac{d\phi}{dt}-\cos \theta \boldsymbol{\xi}]}.
\end{equation}
Local Darboux coordinates are obtained by introducing $\nu=\cos\theta$,
$1> \nu> -1$. The associated symplectic potential 1-form and 2-forms are
\begin{equation}
  \label{eq:26}
  a=-q\nu d\phi,\quad \sigma=da=-q d\nu\wedge d\phi=q\sin\theta d\theta\wedge d\phi ,\quad {\{\nu,\phi\}}^{*}=\frac{1}{q}.
\end{equation}

Whereas $\sigma=da$ locally when $0<\theta<\pi$, this cannot be true globally on
$\mathbb S^{2}$ with a smooth $a$. Indeed, if it were, one would have
$\int_{\mathbb S^{2}}\sigma =\int_{\partial \mathbb S^{2}}a =0$, but instead
\begin{equation}
  \label{eq:71}
  \int_{\mathbb S^{2}}\sigma=4\pi q.
\end{equation}

A standard way to get a globally well-defined description\footnote{This approach
  has been used in the case of Dirac's monopole~\cite{Dirac:1931kp}
  in~\cite{Wu:1975es,Wu:1976ge}, which is directly related to the current problem
  when reducing from $\mathbb R^{3}$ to $\mathbb S^{2}$.} is to consider two
overlapping coordinate neighborhoods of $\mathbb S^{2}$, the first one
containing the north pole $U^{\delta}_{+}:\{0\leq \theta<\frac{\pi}{2}+\delta\}$ with
$0<\delta\leq\frac{\pi}{2}$ and the second one containing the south pole,
$U^{\delta}_{-}:\{\frac{\pi}{2}-\delta< \theta\leq \pi\}$. The associated potentials in the respective
patches are given by
\begin{equation}
    \label{eq:97}
    a_{+}=q[1-\cos\theta]d\phi,\quad a_{-}=q[-1-\cos\theta]d\phi,
\end{equation}
while their difference on the overlap $U^{\delta}_{+-}=U^{\delta}_{+}\cap U^{\delta}_{-}:\frac{\pi}{2}-\delta<\theta<
\frac{\pi}{2}+\delta$ is
\begin{equation}
  \label{eq:106}
  a_{+}-a_{-}=d\Phi,\quad \Phi=2q\phi.
\end{equation}
If $\mathbb S^{+},\mathbb S^{-},E$ denote the upper hemisphere, lower hemisphere
and the equator, respectively,
\begin{equation}
  \label{eq:98}
  \int_{\mathbb S^{2}}\sigma=\int_{\mathbb S^{+}}da_{+}+\int_{\mathbb S^{-}}da_{-}=\oint_{E}(a_{+}-a_{-})=4\pi q,
\end{equation}
as it should.

In order to deal with the Gribov obstruction, one may use the implicit
Euler-Rodrigues parametrization of $\mathrm{SU}(2)$ (see
e.g.~\cite{biedenharn1981angular} section 2.5) that consists in describing
$\mathrm{SU}(2)\simeq \mathbb S^{3}$ in terms of coordinates on $\mathbb R^{4}$ by using
four real variables or two complex variables,
\begin{equation}
  \label{eq:138}
  a=\alpha^{0}+\imath\alpha^{3},\ b=\imath(\alpha^{1}+\imath\alpha^{2}),\quad
  g=\begin{pmatrix} \bar a & \bar b \\ -b  & a
  \end{pmatrix}=\alpha^{0}\sigma_{0}+\alpha^{\beta}(-\imath \sigma_{\beta}).
\end{equation}
with parameters constrained to lie on the unit sphere,
\begin{equation}
  \label{eq:127}
  {(\alpha^{0})}^{2}+{(\alpha^{1})}^{2}+{(\alpha^{2})}^{2}+{(\alpha^{3})}^{2}
  =\alpha_{A}\alpha^{A}=1=|a|^{2}+|b|^{2}.
\end{equation}
In terms of adapted Euler-angles
\begin{equation}
  \label{eq:114}
  (a,b)=(\cos\frac{\theta}{2}e^{\imath\frac{\psi+\phi}{2}},
  -\sin\frac{\theta}{2}e^{\imath\frac{\psi-\phi}{2}}).
\end{equation}
In the overlap $U^{\delta}_{+-}$, the gauge condition $\psi=0$ intersects each
orbit once:
$(\cos\frac{\theta}{2}e^{\imath\frac{\phi}{2}}, -\sin\frac{\theta}{2}e^{-\imath\frac{\phi}{2}})$ is
associated with a single element of~\eqref{eq:114}. This is no longer the case in the
neighborhoods $U^{\delta}_{\pm}$. For $U^{\delta}_{+}$, at $\theta=0$,
$(a,b)=(e^{\imath\frac{\psi+\phi}{2}},0)$, which are all in the same orbit. The condition
$\psi=0$ gives $(e^{\imath\frac{\phi}{2}},0)$ and does not fix a single element of this
orbit but leaves an $\mathbb S^{1}$ worth of group elements that belong to the
same orbit\footnote{It thus follows that the Gribov obstruction here is directly related to ``gimbal lock''.}. The only gauge choice that fixes a single group element,
$(a,b)=(1,0)$ is $\chi_{+}=\psi+\phi=0$. In the other neighborhood, $U^{\delta}_{-}$, a
similar reasoning shows that the appropriate gauge choice is $\chi_{-}=\psi-\phi$.

For later use, instead of spherical coordinates, one may also consider the
complex coordinates on $U_{-}=U_{-}^{\frac{\pi}{2}}$ provided by stereographic
projection from the north pole on the complex plane through the origin, and on
$U_+=U^{\frac{\pi}{2}}_{+}$ those provided by the complex conjugate of the
stereographic projection from the south pole on the complex plane through the
origin,
\begin{equation}
  \label{eq:77}
  \zeta_-=\frac{x+\imath y}{1-z}=e^{\imath\phi}\cot\frac{\theta}{2},\quad
  \zeta_+=\frac{x-\imath y}{1+z}=e^{-\imath\phi}\tan\frac{\theta}{2},\quad \zeta_+\zeta_-=1\quad {\rm on}\quad U_{+-}. 
\end{equation}
Expressions of the Maurer-Cartan forms in these coordinates as well as useful identities are given in Appendix \ref{sec:compl-param}. In these terms, if $\tilde\epsilon=\pm 1$, the metric of the unit sphere $\mathbb S^{2}$ is
\begin{equation}
  \label{eq:132}
  ds^{2}=d\theta^{2}+\sin^{2}\theta d\phi^{2}=4 P^{-2}_{\tilde\epsilon}d\zeta_{\tilde\epsilon} d\widebar \zeta_{\tilde\epsilon}
  ,\quad P_{\tilde\epsilon}=(1+\zeta_{\tilde\epsilon}\widebar\zeta_{\tilde\epsilon}).
\end{equation}
If the K\"ahler potential is defined by
\begin{equation}
  \label{eq:113}
  K^q_{\tilde\epsilon}=2q\ln P_{\tilde\epsilon},
\end{equation}
and $d=\partial+\widebar\partial$ with $\partial =d\zeta_{\tilde\epsilon}\partial_{\zeta_{\tilde\epsilon}}$,
$\widebar \partial =d\widebar \zeta_{\tilde\epsilon}\partial_{\widebar \zeta_{\tilde\epsilon}}$,
\begin{equation}
  \label{eq:110}
  \begin{split}
    & a_{\tilde\epsilon}= \partial(\frac{\imath}{2} K^{q}_{\tilde\epsilon})-\widebar\partial(\frac{\imath}{2} K^{q}_{\tilde\epsilon})
    =-d (\frac \imath 2 K^q_{\tilde\epsilon})+ \partial (\imath K^q_{\tilde\epsilon})=d(\frac \imath 2K^{q}_{\tilde\epsilon})-\widebar\partial
  (\imath K^{q}_{\tilde\epsilon}),\\
    & \sigma_{\tilde\epsilon}=  \bar\partial\partial (\imath K^q_{\tilde\epsilon})=-2\imath q P^{-2}_{\tilde\epsilon}d\zeta_{\tilde\epsilon}\wedge d\widebar\zeta_{\tilde\epsilon}.
  \end{split}
\end{equation}
The associated Dirac brackets are given by
\begin{equation}
  \label{eq:131}
  {\{\zeta_{\tilde\epsilon},\widebar\zeta_{\tilde\epsilon}\}}^{*}=\frac{P^{2}_{\tilde\epsilon}}{2\imath q}.
\end{equation}

For integration on the sphere in these coordinates, note that
\begin{equation}
  \label{eq:56}
  \frac{1}{2\pi\imath}\int_{{\mathbb S^{2}}}\frac{d\zeta_{\tilde\epsilon}\wedge d\widebar\zeta_{\tilde\epsilon}}{P_{\tilde\epsilon}^{2}}=1,
\end{equation}
which can be explicitly checked in this context by (i) integrating over
$U_{\tilde\epsilon}^{\frac{\pi}{2}-\varepsilon}$, that is to say $\mathbb S^{2}$ minus a small cap
$D_{\tilde\epsilon}^{\varepsilon}$ surrounding the point at infinity, which is not covered in the
coordinate neighborhood $U_{\tilde\epsilon}$, (ii) using that the integrand may be
written as $-d\wedge \partial \ln (1+\zeta_{\tilde\epsilon}\widebar\zeta_{\tilde\epsilon})$ and applying Stokes'
theorem, (iii) using that $\zeta_{\tilde\epsilon}=\frac{2}{\varepsilon}e^{- \imath\tilde\epsilon\phi}$ on the
boundary, see~\eqref{eq:66} for more details.

We conclude this section by giving the expressions of the Noether charges. The Noether charges are gauge-invariant and expressed as
\begin{equation}
  \label{eq:41}
  Q_{\pm}=q\sqrt{1-\nu^{2}}e^{{\mp \imath\phi}}=2qP^{-1}_{\tilde\epsilon}{\zeta_{\tilde\epsilon}^{\frac{1\pm \tilde\epsilon}{2}}
    \widebar\zeta_{\tilde\epsilon}^{\frac{1\mp \tilde\epsilon}{2}}},\quad Q_{3}=-q\nu=-\tilde\epsilon  qP^{-1}_{\tilde\epsilon}(1-\zeta_{\tilde\epsilon}\bar\zeta_{\tilde\epsilon}),
\end{equation}
with
\begin{equation}
  \label{eq:42}
  \begin{split}
    -{\{Q_{\pm},\cdot\}}^{*}&=\mp\imath \sqrt{1-\nu^{2}}e^{{\mp \imath\phi}}\partial_{\nu}+\frac{\nu e^{{\mp \imath\phi}}}{\sqrt{1-\nu^{2}}}\partial_{\phi}
                    =\pm\imath\tilde\epsilon\big[\zeta_{\tilde\epsilon}^{{1\pm\tilde\epsilon}}\partial_{\zeta_{\tilde\epsilon}}
                    +\widebar\zeta_{\tilde\epsilon}^{{1\mp\tilde\epsilon}}\partial_{\widebar
    \zeta_{\tilde\epsilon}} \big],\\
    -{\{Q_{3},\cdot\}}^{*}&=\partial_{\phi}=
                    -\imath\tilde\epsilon\big[\zeta_{\tilde\epsilon}\partial_{\zeta_{\tilde\epsilon}}
                    -\widebar\zeta_{\tilde\epsilon}\partial_{\widebar
    \zeta_{\tilde\epsilon}}\big],
  \end{split}
\end{equation}
where we have used~\eqref{eq:63}.

In this case,
\begin{equation}
  \label{eq:49}
  Q^{2}=\delta^{\alpha\beta}Q_{\alpha}Q_{\beta}=q^{2}= \boldsymbol{\mu}^{2},
\end{equation}
and, if $\mathcal L_{\alpha}=-\imath{\{Q_{\alpha},\cdot\}}^{*}$, the Casimir is minus the Laplacian on the sphere,
\begin{multline}
  \label{eq:50}
  \mathcal L^{2}=\delta^{\alpha\beta}\mathcal L_{\alpha}\mathcal L_{\beta}=-\Delta_{\mathbb S^{2}}=-[(1-
  \nu^{2})\partial_{\nu^{2}}-2\nu\partial_{\nu}+\frac{1}{1-\nu^{2}}\partial^{2}_{\phi}]\\
  =-[\partial_{\theta}^{2}+\cot\theta\partial_{\theta}+\frac{1}{\sin^{2}\theta}\partial_{\phi}^{2}]=-P^{2}_{\tilde\epsilon}\partial_{\zeta_{\tilde\epsilon}}\partial_{{\widebar\zeta_{\tilde\epsilon}}}.
\end{multline}

\subsection{Geometric action for $T^*\mathrm{SU}(2)$}\label{sec:tsu2-geom-acti}

If $\pi_{\pm}=\pi_{1}\pm \imath \pi_{2}$ and $Q_{\pm}=Q_{1}\pm \imath Q_{2}$,
the non-vanishing Poisson brackets are
\begin{equation}
  \label{eq:21}
  \begin{split}
  & \{\pi_{+},\pi_{-}\}=-2\imath \pi_{3},\ \{\pi_{3},\pi_{\pm}\}=\mp \imath \pi_{\pm},\\
    & \{\psi,\pi_{3}\}=1,\ \{\psi,\pi_{\pm}\}=-e^{\pm\imath\psi}\cot\theta,\ \{\theta,\pi_{\pm}\}=\pm\imath e^{\pm\imath\psi},\ \{\phi,\pi_{\pm}\}=\frac{e^{\pm\imath\psi}}{\sin\theta}.
  \end{split}
\end{equation}
The geometric action for $T^*\mathrm{SU}(2)$ is
\begin{multline}
  \label{eq:28}
  \boxed{S^{T^{*}\mathrm{SU}(2)}[\psi,\theta,\phi,\pi_{3},\pi_{\pm};\xi]=\int dt\, [\pi_{3}(\dot \psi+\cos\theta\dot \phi)+\frac 12 \pi_{+}e^{-\imath\psi}(-\imath \dot\theta+\sin\theta\dot\phi)}\\ \boxed{+\frac 12 \pi_{-}e^{\imath\psi}(\imath \dot\theta+\sin\theta\dot\phi)- \boldsymbol{\xi} Q_{3}]},
\end{multline}
where the Noether charges are
\begin{equation}
  \label{eq:91}
  \begin{split}
    & Q_{3}=\frac 12 (\pi_{+}e^{-\imath\psi}+\pi_{-}e^{\imath\psi})\sin\theta+\pi_{3}\cos\theta,\\
    & Q_{\pm}= \frac {1}{2}[\pi_{+}e^{-\imath\psi}(\cos\theta \pm 1)e^{\mp \imath\phi}+\pi_{-}e^{\imath\psi}(\cos\theta \mp 1)e^{\mp\imath\phi}]-\pi_{3}\sin\theta e^{\mp\imath\phi}.
  \end{split}
\end{equation}
with
\begin{equation}
  \label{eq:29}
  \{Q_{+},Q_{-}\}=2\imath Q_{3},\ \{Q_{3},Q_{\pm}\}=\pm \imath Q_{\pm},
\end{equation}
and
\begin{equation}
  \label{eq:92}
  Q^{2}=\pi_{+}\pi_{-}+\pi_{3}^{2}+\pi_{3}\cos\theta[\pi_{+}e^{-\imath\psi}(1-\frac 12\sin\theta)+\pi_{-}e^{\imath\psi}(1-\frac 12\sin\theta)].
\end{equation}
Explicit expressions for the left and
right invariant vector fields are
\begin{equation}
  \label{eq:110x}
  \begin{split}
     \mathcal L_{3}=\imath\partial_{\phi},\quad \mathcal L_{\pm}=\mp e^{\mp\imath\phi}(\partial_{\theta}\pm \imath\frac{1}{\sin\theta}\partial_{\psi}\mp\imath\cot\theta\partial_{\phi}),\\
     \mathcal R_{3}=-\imath\partial_{\psi},\quad \mathcal R_{\pm}=\pm e^{\pm \imath \psi}(\partial_{\theta}\mp\imath\frac{1}{\sin\theta}\partial_{\phi}\pm \imath\cot\theta\partial_{\psi}),
  \end{split}
\end{equation}
where
\begin{equation}
  \label{eq:109x}
  \mathcal L_{\alpha}=\imath \vec L_{\alpha},\quad \mathcal R_{\alpha}=-\imath\vec R_{\alpha}.
\end{equation}

In the complex parametrizations~\eqref{eq:99y} adapted to the coordinate patches
$U_{\tilde\epsilon}$, we have instead from Appendix~\ref{sec:compl-param} and~\ref{sec:euler-rodr-param} that the non-vanishing
Poisson brackets involving the coordinates become
\begin{equation}
  \label{eq:62}
  \begin{split}
    & \{\psi,\pi_{3}\}=1,\quad \{\psi,\pi_{+}\}=\tilde\epsilon \zeta_{\tilde\epsilon}e^{\imath\psi},\quad \{\psi,\pi_{-}\}=\tilde\epsilon \widebar\zeta_{\tilde\epsilon}e^{-\imath\psi},\\ & \{\zeta_{\tilde\epsilon},\pi_{-}\}=-\imath\tilde\epsilon e^{-\imath\psi}P_{\tilde\epsilon},\quad \{\widebar\zeta_{\tilde\epsilon},\pi_{+}\}=\imath\tilde\epsilon e^{\imath\psi}P_{\tilde\epsilon},
  \end{split}
\end{equation}
while the geometric action becomes
\begin{multline}
  \label{eq:28a}
  \boxed{S^{T^{*}\mathrm{SU}(2)}[\psi,\zeta_{\tilde\epsilon},\widebar\zeta_{\tilde\epsilon},\pi_{3},\pi_{\pm};\xi]
  =\int dt\, \Big(\pi_{3}[\frac{d\psi}{dt}+\imath P^{-1}_{\tilde\epsilon}(\zeta_{\tilde\epsilon}\frac{d\widebar\zeta_{\tilde\epsilon}}{dt}
  -\widebar\zeta_{\tilde\epsilon}\frac{d\zeta_{\tilde\epsilon}}{dt})]}\\ \boxed{-\imath\tilde\epsilon
  \pi_{+}e^{-\imath\psi}P^{-1}_{\tilde\epsilon}\frac{d\widebar\zeta_{\tilde\epsilon}}{dt} +
  \imath\tilde\epsilon
  \pi_{-}e^{\imath\psi}P^{-1}_{\tilde\epsilon}\frac{d\zeta_{\tilde\epsilon}}{dt}-\boldsymbol{\xi} Q_{3}\Big)},
\end{multline}
where the Noether charges are
\begin{equation}
  \label{eq:91a}
  \begin{split}
    & Q_{3}=P^{-1}_{\tilde\epsilon}\Big[\pi_{+}e^{-\imath\psi}\widebar\zeta_{\tilde\epsilon}+\pi_{-}e^{\imath\psi}\zeta_{\tilde\epsilon}+
      \pi_{3}\tilde\epsilon(1-\zeta_{\tilde\epsilon}\widebar\zeta_{\tilde\epsilon})\Big],\\
    & Q_{\pm}= P^{-1}_{\tilde\epsilon}\Big[\pm \pi_{+}e^{-\imath\psi}\widebar\zeta^{1\mp\tilde\epsilon}_{\tilde\epsilon}\mp \pi_{-}e^{\imath\psi}\zeta^{1\pm\tilde\epsilon}_{\tilde\epsilon}-2\pi_{3}\zeta^{\frac{1\pm\tilde\epsilon}{2}}_{\tilde\epsilon}\widebar\zeta^{\frac{1\mp\tilde\epsilon}{2}}_{\tilde\epsilon}\Big].
  \end{split}
\end{equation}
with
\begin{equation}
  \label{eq:92a}
  Q^{2}=\pi_{+}\pi_{-}+\pi_{3}^{2}.
\end{equation}
Explicit expressions for the left and right invariant vector fields are now
\begin{equation}
  \label{eq:110xa}
  \begin{split}
    & \mathcal L_{3}=\tilde\epsilon[\imath\partial_{\psi}+
    \zeta_{\tilde\epsilon}\partial_{\zeta_{\tilde\epsilon}}-\widebar\zeta_{\tilde\epsilon}
    \partial_{\widebar\zeta_{\tilde\epsilon}}],\quad \mathcal L_{\pm}=-\imath\zeta_{\tilde\epsilon}^{\frac{1\pm\tilde\epsilon}{2}}
    \widebar\zeta_{\tilde\epsilon}^{\frac{1\mp\tilde\epsilon}{2}}\partial_{\psi}\mp\tilde\epsilon[ \zeta^{1\pm\tilde\epsilon}\partial_{\zeta_{\tilde\epsilon}}+\widebar\zeta^{1\mp\tilde\epsilon}
    \partial_{\widebar\zeta_{\tilde\epsilon}}],\\
    & \mathcal R_{3}=-\imath\partial_{\psi},\quad \mathcal R_{+}=\tilde\epsilon e^{\imath \psi}(-\imath\zeta_{\tilde\epsilon}\partial_{\psi}+P_{\tilde\epsilon}
    \partial_{\widebar\zeta_{\tilde\epsilon}})=-\widebar{\mathcal R_{-}}.
  \end{split}
\end{equation}

By construction, when evaluating the observables $Q_{\alpha},Q^{2}$ on the constraint
surface $\pi_{\pm}=0$, $\pi_{3}=-q=\epsilon\boldsymbol{\mu}$, one gets those associated to a single
coadjoint orbit discussed in the previous section.

\subsection{$\mathrm{SU}(2)$ model space}\label{sec:su2-model-space}

The model space for $SU(2)$ is obtained from the previous section by assuming
that $\pi_{3}\neq 0$ and imposing the second class constraints
$\pi_{+}=0=\pi_{-}$. Dirac brackets are given by
\begin{equation}
  \label{eq:108}
  {\{f,g\}}^{*}=\{f,g\}+\frac{1}{2\imath\pi_{3}}[\{f,\pi_{+}\}\{g,\pi_{-}\}-\{f,\pi_{-}\}\{g,\pi_{+}\}].
\end{equation}
More explicitly, on the second class constraint surface, the non-vanishing brackets are given by
\begin{equation}
  \label{eq:141}
  {\{\psi,\pi_{3}\}}^{*}= 1,\quad {\{\psi,\theta\}}^{*}=\frac{\cot\theta}{\pi_{3}},\quad {\{\theta,\phi\}}^{*}=\frac{1}{\pi_{3}\sin\theta},
\end{equation}
respectively by
\begin{equation}
  \label{eq:232}
  {\{\psi,\pi_{3}\}}^{*}= 1,\quad {\{\psi,\zeta_{\tilde\epsilon}\}}^{*}=-\frac{P_{\tilde\epsilon}\zeta_{\tilde\epsilon}}{2\pi_{3}}
  =\overline{{\{\psi,\widebar\zeta_{\tilde\epsilon}\}}^{*}},\quad {\{\zeta_{\tilde\epsilon},\widebar\zeta_{\tilde\epsilon}\}}^{*}=\frac{\imath P^{2}_{\tilde\epsilon}}{2\pi_{3}}.
\end{equation}
The geometric action becomes
\begin{equation}
  \label{eq:219}
  \boxed{S^{\mathcal M_\mu}[\psi,\theta,\phi,\pi_{3};\xi]=\int dt\ [\pi_{3}(\dot \psi+\cos\theta\dot \phi) - \boldsymbol{\xi} \pi_{3}\cos\theta]},
\end{equation}
or,
\begin{equation}
  \label{eq:219a}
  \boxed{S^{\mathcal M_\mu}[\psi,\zeta_{\tilde\epsilon},\widebar\zeta_{\tilde\epsilon},\pi_{3};\xi]=\int dt\, \Big(\pi_{3}[\frac{d\psi}{dt}+\imath P^{-1}_{\tilde\epsilon}(\zeta_{\tilde\epsilon}\frac{d\widebar\zeta_{\tilde\epsilon}}{dt}
  -\widebar\zeta_{\tilde\epsilon}\frac{d\zeta_{\tilde\epsilon}}{dt})]-\boldsymbol{\xi} Q_{3}\Big)},
\end{equation}
with associated symplectic 2-form, 
\begin{equation}
  \label{eq:58}
  \sigma=d\pi_{3}\wedge (d\psi+\cos\theta d\phi)-\pi_{3}\sin\theta d\theta \wedge d\phi,
\end{equation}
respectively
\begin{equation}
  \label{eq:236}
  \sigma_{\tilde\epsilon}=d\pi_{3}\wedge [d\psi+\imath P^{-1}_{\tilde\epsilon}(\zeta_{\tilde\epsilon}{d\widebar\zeta_{\tilde\epsilon}}
  -\widebar\zeta_{\tilde\epsilon}{d\zeta_{\tilde\epsilon}})]+ \pi_{3}2\imath
  P^{-2}_{\tilde\epsilon}d\zeta_{\tilde\epsilon}\wedge d\widebar\zeta_{\tilde\epsilon},
\end{equation}
while the Noether charges reduce to
\begin{equation}
  \label{eq:91b}
  Q_{3}=\pi_{3}\cos\theta,\quad
    Q_{\pm}= -\pi_{3}\sin\theta e^{\mp\imath\phi},
\end{equation}
respectively to
\begin{equation}
  \label{eq:237}
  Q_{3}=\pi_{3}\tilde\epsilon P^{-1}_{\tilde\epsilon}(1-\zeta_{\tilde\epsilon}\widebar\zeta_{\tilde\epsilon}),\quad
  Q_{\pm}= -2\pi_{3}P^{-1}_{\tilde\epsilon}\zeta^{\frac{1\pm\tilde\epsilon}{2}}_{\tilde\epsilon}
  \widebar\zeta^{\frac{1\mp\tilde\epsilon}{2}}_{\tilde\epsilon},
\end{equation}
with
\begin{equation}
  \label{eq:29a}
  {\{Q_{+},Q_{-}\}}^{*}=2\imath Q_{3},\ {\{Q_{3},Q_{\pm}\}}^{*}=\pm \imath Q_{\pm},\quad Q^{2}=\pi_{3}^{2}.
\end{equation}
Finally, the left invariant vector fields are generated by the Noether charges
in the Dirac bracket $X_{Q_{\alpha}}=-{\{Q_{\alpha},\cdot\}}^{*}$ and given by~\eqref{eq:110x} multiplied by
$-\imath$, respectively~\eqref{eq:110xa} multiplied by $-\imath$, where the explicit computation
uses~\eqref{eq:63}. From the right invariant vector fields, only
$X_{\pi_{3}}=-{\{\pi_{3},\cdot\}}^{*}=\partial_{\psi}$ remains.

Alternatively, the model space may be considered as a first class constrained system
associated to $T^{*}\mathrm{SU}(2)$ by dropping either the constraint $\pi_{_+}=0$ and keeping
the constraint $\pi_-=0$, or by dropping the latter and keeping the former. The
associated generator of (abelian) gauge transformations is then $\gamma=\pi_-\varepsilon^-$ or $\gamma=\pi_+\varepsilon^+$.

Similarly, the theory associated to a single coadjoint orbit can be understood
as a non-abelian first class constrained system associated to $T^{*}\mathrm{SU}(2)$ by
imposing in addition $\pi_{3}+q=0$, so that the generator of gauge
transformation becomes
\begin{equation}
  \label{eq:51}
  \gamma=(\pi_3+q)\varepsilon^3+\pi_-\varepsilon^-,\quad {\rm or}\quad \gamma=(\pi_3+q)\varepsilon^3+\pi_+\varepsilon^+.
\end{equation}

\subsection{$\mathrm{SU}(2)$ as a constrained system of nonzero quaternions}\label{sec:unconstr-model-as}

As mentioned in Section \ref{sec:descriptions}, it is also interesting to describe the system associated to $T^{*} \mathrm{SU}(2)$ as a constrained system of a larger ``embedding'' phase space. In order to do so, we consider the implicit Euler-Rodrigues parametrization of $\mathrm{SU}(2)$ as discussed
in equations~\eqref{eq:138},~\eqref{eq:127}. Associated expressions for left and right invariant
vector fields and Maurer-Cartan forms as well as the coadjoint representation
are provided in Appendix~\ref{sec:euler-rodr-param}. In these terms, the geometric action for
$T^*\mathrm{SU}(2)$ becomes
\begin{equation}
  \label{eq:126}
  S^{T^{*}\mathrm{SU}(2)}[\alpha^{A},\pi_{\alpha};\xi]=\int dt\, \pi_{\alpha}R\indices{^\alpha_{A}}
  (\dot \alpha^{A}- \boldsymbol{\xi} L\indices{_{3}^{A}}),
\end{equation}
with Noether charges given by
\begin{equation}
  \label{eq:182}
  Q_{\alpha}= \pi_{\beta}R\indices{^{\beta}_{A}}L\indices{_{\alpha}^{A}}.
\end{equation}

Dropping the constraint~\eqref{eq:127} amounts to considering the Lie group of
(non-unimodular) nonzero quaternions $\mathbb H^{*}$ which can be represented
by
\begin{equation}
  \label{eq:128}
 z_{1}=x^{0}+\imath x^{3},\ z_{2}=\imath(x^{1}+\imath x^{2}),\quad  g=\begin{pmatrix} \bar z_{1} & \bar z_{2} \\ -z_{2}  & z_{1}
  \end{pmatrix}=x^{0}\sigma_{0}+x^{\beta}(-\imath\sigma_{\beta}).
\end{equation}

This corresponds to replacing $a,b$ and $\alpha^{A}=(\alpha^{0},\alpha^{\beta})$ in the
Euler-Rodrigues parametrization by $z_{1},z_{2}$ and $x^{A}=(x^{0},x^{\beta})$ in
order to emphasize that these coordinates are now unconstrained coordinates on
$\mathbb R^{4}-{0}$. Associated expressions for left and right invariant vector
fields $\vec L_{A},\vec R_{A}$ and Maurer-Cartan forms $L^{A},R^{A}$ as well as
the coadjoint representation for the Lie group of nonzero quaternions are
provided in Appendix~\ref{sec:left-right-invariant-h}.

The geometric action for $T^{*}\mathbb H^{*}$ is
\begin{equation}
  \label{eq:158}
  \boxed{S^{T^{*}\mathbb H^{*}}[x^{B}, \pi_{A}; \xi]=\int dt\, \pi_{A}R\indices{^{A}_{B}}[\dot x^{B}-L\indices{_{C}^{B}} \xi^{C}]},
\end{equation}
with
\begin{equation}
  \label{eq:156}
  \{x^{A},x^{B}\}=0=\{\pi_{0}, \pi_{A}\},\quad \{ \pi_{ \alpha}, \pi_{ \beta}\}=\epsilon\indices{^{ \gamma}_{ \alpha \beta}}\pi_{ \gamma},\quad \{x^{A}, \pi_{B}\}=R\indices{_{B}^{A}},
\end{equation}
\begin{equation}
  \label{eq:154}
  \{Q_{0}, Q_{A}\},\quad \{Q_{ \alpha}, Q_{ \beta}\}=-\epsilon\indices{^{ \gamma}_{ \alpha \beta}}Q_{ \gamma},
\end{equation}
and
\begin{equation}
\{ Q_{A}, \pi_{B}\}=0.\label{bis}
\end{equation}
Furthermore, if $\delta_{AB}$ is the non-degenerate invariant metric on the reductive
Lie algebra described by the structure constants~\eqref{eq:164}, used together
with its inverse to lower and raise Lie algebra indices,
\begin{equation}
  \label{eq:191}
  \pi_{A}\pi^{A}=Q_{A}Q^{A}.
\end{equation}

Equivalently, in terms of Darboux coordinates,
\begin{equation}
  \label{eq:99}
   S^{T^{*}\mathbb H^{*}}[x^{B},p_{A};\xi]=\int dt\, p_{B}[\dot x^{B}-L\indices{_{C}^{B}} \xi^{C}],
 \end{equation}
 and, when using the explicit expressions in Appendix~\ref{sec:left-right-invariant-h},
 \begin{equation}
   \label{eq:165}
   \begin{split}
   & \pi_{0}=\frac 12 x^{A}p_{A},\quad  \pi_{\alpha}=\frac 12 (x_{0}p_{\alpha}-x_{\alpha}p_{0}+\epsilon_{\alpha\beta\gamma}x^{\beta}p^{\gamma}),\\
 &   Q_{0}=\frac 12 x^{A}p_{A}, \quad  Q_{\alpha}=\frac 12 (x_{0}p_{\alpha}-x_{\alpha}p_{0}
   -\epsilon_{\alpha\beta\gamma}x^{\beta}p^{\gamma}),
   \end{split}
 \end{equation}
with
\begin{equation}
   \label{eq:146}
   \pi_{A}\pi^{A}=\frac 14 R^{2}S^{2}=Q_{A}Q^{A},\quad R^{2}=x^{A}x_{A},\quad S^{2}=p_{A}p^{A}.
\end{equation}

Consider the second class constraints $\mathcal L=0=\mathcal S$,
\begin{equation}
  \label{eq:226}
  \mathcal L=R-1,\quad \mathcal S= \frac{\pi_{0}}{R}=\frac{x^{A}\pi_{B}R\indices{^{B}_{A}}}{R}=\frac{ x^{A}p_{A}}{2R},\quad \{\mathcal L,\mathcal S\}=\frac 12.
\end{equation}
The geometric action for $T^{*}\mathrm{SU}(2)$ in Euler-Rodrigues parametrization in~\eqref{eq:126}
is obtained from the one for quaternions in~\eqref{eq:158} by imposing the second class
constraints
\begin{equation}
  \label{eq:147}
   S^{T^{*}\mathrm{SU}(2)}[x^{B},\pi_{A},\lambda,\lambda ';\xi]=\int dt\, [\pi_{A}R\indices{^{A}_{B}}(\dot x^{B}-L\indices{_{C}^{B}}\xi^{C})-\lambda\mathcal L-\lambda ' \mathcal S],
\end{equation}
where $\lambda(t),\lambda '(t)$ are Lagrange multipliers, as can directly be seen by solving
the constraints in the action.

One may also convert this set of second class constraints into the single first
class constraint given by $\mathcal S=0$,
\begin{equation}
  \label{eq:147a}
   S^{T^{*}\mathrm{SU}(2)}[x^{B}, \pi_{A},\lambda ';\xi]=\int dt\, [ \pi_{A}R\indices{^{A}_{B}}(\dot x^{B}-L\indices{_{C}^{B}}\xi^{C})-\lambda' \mathcal S].
\end{equation}
This first class constraint generates the gauge symmetry, $\delta_{\varepsilon}\cdot=\{\cdot,\mathcal S\}\varepsilon$,
\begin{equation}
  \label{eq:153}
  \delta_{\epsilon}x^{A}=\varepsilon\frac{x^{A}}{2R},\quad \delta_{\epsilon}  \pi_{0}=\varepsilon\frac{\pi_{0}}{2R},\ \delta_{\varepsilon}\pi_{\alpha}=0 \iff \delta_{\varepsilon}p_{A}=\varepsilon\frac{x_{A}x^{B}p_{B}-R^{2}p_{A}}{2R^{3}}.
\end{equation}

Instead of the real variables $x^{A}$, one may also use the complex variables
$z_{1},z_{2}$ and instead of $ \pi_{A}=(\pi_{0}, \pi_{\alpha})$,
$ \rho_{\pm}, \pi_{\pm}$, where $\pi_{\pm}=\pi_{1}\pm \imath \pi_{2}$,
$\rho_{\pm}=\pi_{0}\pm\imath\pi_{3}$. The relevant expressions in this
parametrization are provided in Appendix~\ref{sec:phase-space-h}.

\subsection{$\mathrm{SU}(2)$ model space from quaternions. Adapted gauge fixing}\label{sec:su2-model-space-2}

It is now possible to obtain the model space from the embedding phase space
$T^{*}\mathbb H^{*}$ described above. Consider the phase space $(x^{A},\pi_{B})$
of dimension $8$ corresponding to the $4$ canonical pairs $(x^{A},p_{B})$. The
model space corresponds to imposing the 4 second class constraints
\begin{equation}
  \label{eq:173a}
  \chi_{A}=(R-1,\frac{\pi_{0}}{R},\frac{\pi_{+}}{R},\frac{\pi_{-}}{R}),
\end{equation}
on the theory described by the geometric action for $\mathbb H^{*}$. One remains
with the coordinates $z_{1},z_{2}$ constrained by
$|z_{1}|^{2}+|z_{2}|^{2}=1$ and $\pi_{3}$. 

However, a better way to prepare the system for quantization is the following. Starting
from the description of $T^{*}SU(2)$ as a first class constraint system of
$T^{*}\mathbb H^{*}$ in~\eqref{eq:147a}, one may choose a more convenient gauge fixing condition
than $R=1$. For instance one may also choose $\sqrt{\pi_{A}\pi^{A}}=R$ or the
symmetric condition $\sqrt{\pi_{A}\pi^{A}}=R^{2}$. When one imposes in addition
$\pi_{+}=0=\pi_{-}$, the latter reduces to $|\pi_{3}|=R^{2}$, so that
\begin{equation}
  \label{eq:202}
  \{R-\frac{|\pi_{3}|}{R},\frac{\pi_{0}}{R}\}=\frac 12 +\frac{|\pi_{3}|}{2R^{2}}\approx 1.
\end{equation}

In this case, the model space is obtained by using the following set of
second class constraints,
\begin{equation}
  \label{eq:200}
  \chi_{A}=(\frac{R^{2}-|\pi_{3}|}{R},\frac{\pi_{0}}{ R},\frac{\pi_{+}}{R},\frac{\pi_{-}}{R}).
\end{equation}
In the following we assume that the second class constraints are imposed
strongly. This implies that $\pi_{0}=0=\pi_{\pm}$, $|\pi_{3}|=R^{2}$ and that
the model space is described by the unconstrained variables
$z_{1},z_{2},\bar z_{1},\bar z_{2}$ (except for the condition that $R\neq 0$).

The observables defined on the constraint surface by
\begin{equation}
  q_+=2\epsilon' \bar z_1 z_2,\quad q_-=2\epsilon' z_1\bar z_2,\quad q_3=\epsilon' (z_1\bar z_1-z_2\bar z_2), \quad \epsilon' =\sign(\pi_{3}),
  \label{c}
\end{equation}
satisfy
\begin{equation}
  \label{eq:193}
  R^{4}=q_{+}q_{-}+q_{3}^{2}.
\end{equation}
Since the Dirac brackets of first class fonctions agree with their Poisson
brackets on the constraint surface, it follows from~\eqref{bis} and~\eqref{eq:181} that
\begin{equation}
  \label{eq:192a}
  {\{R^{2},q_3\}}^{*}=0={\{R^{2},q_{\pm}\}}^{*},\quad {\{q_{+},q_{-}\}}^{*}=2\imath q_{3},
  \quad {\{q_{\pm},q_{3}\}}^{*}=\mp\imath q_{\pm}.
\end{equation}

The explicit expressions for the Dirac brackets are obtained from 
\begin{equation}
  \label{eq:173b}
  C_{AB}\equiv\{\chi_{A},\chi_{B}\}= \begin{pmatrix} 0 & 1  & 0 & 0\\
    -1 & 0 & 0 & 0 \\
    0 & 0 & 0 & -2\imath \epsilon'  \\ 0 & 0 & 2\imath \epsilon'  & 0
  \end{pmatrix},
\end{equation}
\begin{equation}
  \label{eq:175a}
  C^{{-1}AB}= \begin{pmatrix} 0 & -1 & 0 & 0\\
    1 & 0 & 0 & 0 \\
    0 & 0 & 0 & -\frac{\imath \epsilon'}{2}\\ 0 & 0 & \frac{\imath \epsilon'}{2} & 0
  \end{pmatrix}.
\end{equation}
As a consequence, the Dirac brackets of functions of
$z_{1},z_{2},\bar z_{1},\bar z_{2}$ are given by
\begin{multline}
  \label{eq:198}
  {\{f,g\}}^{*}=\frac{\epsilon'}{R^{2}}\big[\{f,\pi_{3}\}\{g,\pi_{0}\}-\{f,\pi_{0}\}\{g,\pi_{3}\}\\
  -\frac{\imath}{2}\{f,\pi_{+}\}\{g,\pi_{-}\}
  +\frac{\imath}{2}\{f,\pi_{-}\}\{g,\pi_{+}\}\big],
\end{multline}
or, equivalently,
\begin{equation}
  \label{eq:180a}
  {\{z_{1},\bar z_{1}\}}^{*}=\frac{\imath\epsilon' }{2},\quad  {\{z_{2},\bar z_{2}\}}^{*}=\frac{\imath\epsilon'}{2},
  \quad {\{z_{1},z_{2}\}}^{*}=0,\quad
  {\{ z_{1},\bar z_{2}\}}^{*}=0.
\end{equation}
It is this simple representation of the Dirac brackets of the fundamental
variables that the adapted gauge fixing allows one to achieve. In particular,
\begin{equation}
  \label{eq:185a}
  \begin{split}
    &{\{z_{1},q_{+}\}}^{*}=\imath \epsilon'  z_{2},\quad {\{z_{1},q_{-}\}}^{*}=0,\quad {\{z_{1},q_{3}\}}^{*}=\frac{\imath \epsilon' }{2} z_{1},\\
    &{\{z_{2},q_{+}\}}^{*}=0,\quad {\{z_{2},q_{-}\}}^{*}=\imath \epsilon' z_{1},\quad {\{z_{2},q_{3}\}}^{*}=-\frac{\imath \epsilon' }{2} z_{2}.
  \end{split}
\end{equation}

Let  $\Xi=(1,2)$. The additional change of variables,
\begin{equation}
  \label{eq:199}
  a_{1}=\frac{1+\epsilon' }{2}{\sqrt{\frac{2}{\hbar}}}\bar z_{1}+\frac{1-\epsilon' }{2}{\sqrt{\frac{2}{\hbar}}} z_{1},
  \quad a_{2}=\frac{1+\epsilon' }{2}{\sqrt{\frac{2}{\hbar}}}\bar z_{2}+\frac{1-\epsilon' }{2}{\sqrt{\frac{2}{\hbar}}} z_{2},
\end{equation}
allows one to write the Dirac brackets~\eqref{eq:180a} as\footnote{Instead of a bar for complex conjugation, we now use a star.}
\begin{equation}
  \label{eq:203}
  {\{a_{\Xi},a^{*}_{\Xi'}\}}^{*}=\frac{1}{\imath\hbar}\delta_{\Xi,\Xi'},\quad  {\{a_{\Xi},a_{\Xi'}\}}^{*}=0= {\{a^{*}_{\Xi},a^{*}_{\Xi'}\}}^{*}.
\end{equation}
The number operators for the two types of oscillators are
\begin{equation}
  \label{eq:90}
  N_{1}=a^{*}_1a_1,\quad N_{2}=a^{*}_2a_2,\quad N=N_{1}+N_{2},\quad M=N_{1}-N_{2}.
\end{equation}
In terms of these creation and destruction operators, the Noether charges in~\eqref{c} and $R^{2}$ in the expression for $Q^{2}$ in~\eqref{eq:193}
become
\begin{equation}
  \begin{split}
  & q_+=\hbar[\frac{1+\epsilon' }{2}a^{*}_{2}a_{1}
  -\frac{1-\epsilon' }{2}a^{*}_{1}a_{2}],\quad q_-= \hbar[\frac{1+\epsilon' }{2}a^{*}_{1}a_{2}
    -\frac{1-\epsilon' }{2}a^{*}_{2}a_{1}],\\
    & q_3=\frac{\hbar\epsilon' }{2}M,\quad R^{2}=\frac{\hbar}{2}N.
      \label{d}
  \end{split}
\end{equation}

The Dirac brackets~\eqref{eq:203} are unchanged under the (canonical)
transformation $a_{\Xi}\to a^{*}_{\Xi}$, $a^{*}_{\Xi}\to -a_{\Xi}$. Performing
this transformation on the Noether charges in~\eqref{d} produces an isomorphic
algebra of charges. If one performs this transformation on only one of the pairs
of oscillators, the second one for definiteness, one finds
\begin{equation}
  \begin{split}
  & q'_+=-\hbar[\frac{1+\epsilon' }{2}a_{1}a_{2}
  +\frac{1-\epsilon' }{2}a^{*}_{1}a^{*}_{2}],\quad q'_-= \hbar[\frac{1+\epsilon' }{2}a^{*}_{1}a^{*}_{2}
    +\frac{1-\epsilon' }{2}a_{2}a_{1}],\\
    & q'_3=\frac{\hbar\epsilon' }{2}N=\epsilon' R^{2},\quad (R^{2})'=\frac{\hbar}{2}M=\epsilon' q_{3},
      \label{e}
  \end{split}
\end{equation}
both the relation~\eqref{eq:193} and the algebra~\eqref{eq:192a} continue to hold
for the primed functions.

If one wants to recover a single coadjoint orbit, one needs to impose the
additional constraint $\pi_{3}=\epsilon\boldsymbol{\mu}$. Together with $\pi_{3}=\epsilon' R^{2}$, and since
both $\boldsymbol{\mu}$ and $R$ are positive, this means that
\begin{equation}
  \label{eq:44}
  \epsilon=\epsilon',\quad \boldsymbol{\mu}= R^{2}=\frac{\hbar}{2}N.
\end{equation}

\section{Application to $\mathrm{SU}(2)$: Quantum theory}\label{sec:quantum-theory}

\subsection{Quantization of a single coadjoint orbit }\label{sec:quant-singke-coadj}

We start our study of the quantum theory by considering the reduced phase space
quantization of a single coadjoint orbit. In order to show that the associated
Hilbert space carries a single unitary irreducible representation of
$\mathrm{SU}(2)$, one needs basic elements from geometric quantization. Details
and an extensive commented list of references on geometric quantization can be
found for instance in~\cite{nlab:geometric_quantization} and constrained systems in this context are
discussed in~\cite{Sniatycki1983,Ashtekar:1986ty,Gotay:1986zy,Blau:1987zt,Blau:1987ay}.

The integrality condition of geometric quantization implies here the
quantization of the radius in half integer units of $\hbar$,
\begin{equation}
  \label{eq:54}
  \int_{\mathbb S^{2}}\sigma=2\pi\hbar n,\ n\in \mathbb Z\iff  q= \frac {\hbar n}{2} 
  \Longrightarrow  \boldsymbol{\mu}=\hbar j,\quad j\in \mathbb N/2, \quad \frac{q}{\hbar}=-\epsilon j.
\end{equation}
The operators on phase space associated to the Noether charges through the
prequantization formula of geometric quatization depend on the potential
one-form. They are obtained by replacing derivatives by covariant derivatives in
the Hamiltonian vectors fields,
\begin{equation}
  \label{eq:60}
  D^{\tilde\epsilon}_{A}=\partial_{A}-\frac{\imath}{\hbar}a^{\tilde\epsilon}_{A},
\end{equation}
multiplying by $-\imath\hbar$ and adding the charge. In spherical coordinates with potential~\eqref{eq:97},
\begin{equation}
  \label{eq:61}
  D^{\tilde\epsilon}_{\phi}=\partial_{\phi}+\imath \epsilon j [\tilde\epsilon-\nu],\quad D^{\tilde\epsilon}_{\nu}=\partial_{\nu},
\end{equation}
More explicitly\footnote{Up to conventions, these operators agree with those
  constructed in~\cite{Wu:1976ge} (see also~\cite{Dray:1984gy} for a discussion close to the
  current context).},
\begin{equation}
  \begin{split}
  \frac{1}{\hbar}\widehat Q^{\tilde\epsilon}_\pm & =e^{\mp\imath\phi}\big[\mp\sqrt{1-\nu^{2}}\partial_{\nu}
  +\frac{\nu}{\sqrt{1-\nu^{2}}}(-\imath\partial_{\phi}+\tilde\epsilon\epsilon j)-\epsilon j\frac{1}{\sqrt{1-\nu^{2}}}\big]=\mathcal J_{\mp},\\
  \frac{1}{\hbar}\widehat Q^{\tilde\epsilon}_{3}  & =-\imath\partial_{\phi}+\tilde\epsilon\epsilon j=\mathcal J_{3},
  \end{split}
\end{equation}
with associated modified Casimir operator
\begin{multline}
  \label{eq:107}
  \frac{1}{\hbar^{2}}{(\widehat Q^{\tilde\epsilon})}^{2} =-\Delta_{\mathbb S^{2}}+\frac{2(1-\mu\tilde\epsilon)}{1-\mu^{2}}\big[j^{2}+\tilde\epsilon \epsilon j(-\imath\partial_{\phi})\big]=\\=-(1-\mu^{2})\partial^{2}_{\mu}+2\mu\partial_{\mu}
  +\frac{1}{1-\mu^{2}}\big[{(-\imath\partial_{\phi}
    +\tilde\epsilon\epsilon j)}^{2}-2\mu\epsilon j
    (-\imath\partial_{\phi}+\tilde\epsilon\epsilon j)+j^{2}\big].
\end{multline}

In complex coordinates, it follows from~\eqref{eq:110} that it is advantageous to perform a
gauge transformation that gets rid of the exact term in the second expression of
the potentials\footnote{If one chooses to get rid of the exact term in the third
  expression, the role of $\zeta_{\tilde\epsilon}$ and $\widebar\zeta_{\tilde\epsilon}$ in the
  considerations below will be exchanged.}. The transformed potential becomes
\begin{equation}
  \label{eq:120}
  \frac{\imath}{\hbar}a^{\prime}_{\tilde\epsilon}=\frac{\imath}{\hbar}a_{\tilde\epsilon}+\epsilon j d(\ln P_{\tilde\epsilon})=2\epsilon jd\zeta_{\tilde\epsilon}\partial_{\zeta_{\tilde\epsilon}} \ln P_{\tilde\epsilon}.
\end{equation}
When taking into account that
$\exp(K^{\frac{\epsilon j}{2}}_{\tilde\epsilon})=P_{\tilde\epsilon}^{\epsilon j}$,
the associated wave functions are
$\psi'=P_{\tilde\epsilon}^{\epsilon j}\psi$. The holomorphic polarization
spanned by $\partial_{\widebar\zeta_{\tilde\epsilon}}$ is preserved by the Hamiltonian vector
fields in~\eqref{eq:42}, it satisfies $D'_{\widebar\zeta_{\tilde\epsilon}}=\partial_{\widebar\zeta_{\tilde\epsilon}}$
so that polarized wave functions $D'_{\widebar\zeta_{\tilde\epsilon}}\psi'=0$ are holomorphic,
$\psi'=\psi'(\zeta_{\tilde\epsilon})$. For polarized wave functions
$\psi=P_{\tilde\epsilon}^{-\epsilon j}\psi'(\zeta_{\tilde\epsilon})$ which are well
defined on $\mathbb S^{2}$ to exist, one needs $\epsilon=1$,
\begin{equation}
  \label{eq:163}
  \psi(\zeta_{\tilde\epsilon},\widebar\zeta_{\tilde\epsilon})
  =P_{\tilde\epsilon}^{-j}\psi'(\zeta_{\tilde\epsilon}),
\end{equation}
and also that $\psi'(\zeta_{\tilde\epsilon})$ are complex linear combinations of
$1,\zeta_{\tilde\epsilon},\dots \zeta^{2j}_{\tilde\epsilon}$.

Furthermore,
\begin{equation}
  \label{eq:124}
  D_{\zeta_{\tilde\epsilon}}'\psi'=[\partial_{\zeta_{\tilde\epsilon}}-{2 j}\partial_{\zeta_{\tilde\epsilon}}\ln P_{\tilde\epsilon}]\psi'.
\end{equation}
The associated quantum operators preserving the chosen polarization are
\footnote{In the explicit computation, the relations in~\eqref{eq:63} may be used in order
  to simplify expressions.}
\begin{equation}
  \label{eq:57}
  \begin{split}
  \frac{1}{\hbar}\widehat Q^{\prime\tilde\epsilon}_{\pm}
  &=\pm\tilde\epsilon\big[\zeta_{\tilde\epsilon}^{{1\pm\tilde\epsilon}}\partial_{\zeta_{\tilde\epsilon}}
                    +\widebar\zeta_{\tilde\epsilon}^{{1\mp\tilde\epsilon}}\partial_{\widebar
    \zeta_{\tilde\epsilon}} \big]- j\zeta_{\tilde\epsilon}^{\frac{1\pm \tilde\epsilon}{2}}(1\pm\tilde\epsilon),\\
    \frac{1}{\hbar}\widehat Q^{\prime\tilde\epsilon}_{3}&=
-\tilde\epsilon\big[\zeta_{\tilde\epsilon}\partial_{\zeta_{\tilde\epsilon}}
                    -\widebar\zeta_{\tilde\epsilon}\partial_{\widebar
    \zeta_{\tilde\epsilon}}                                      \big]+\tilde\epsilon j,
  \end{split}
\end{equation}
with modified Casimir operator,
\begin{equation}
  \label{eq:65}
  \frac{1}{\hbar^{2}}{(\widehat Q^{\prime\tilde\epsilon})}^{2} = -P_{\tilde\epsilon}^{2}\partial_{\zeta_{\tilde\epsilon}}\partial_{{\widebar\zeta_{\tilde\epsilon}}}
  +2\imath jP_{\tilde\epsilon}\widebar\zeta_{\tilde\epsilon}\partial_{\widebar
    \zeta_{\tilde\epsilon}}+j(j+1),
\end{equation}
and where the terms involving $\partial_{\widebar\zeta_{\tilde\epsilon}}$ can be dropped when
acting on the polarized wave functions $\psi'$. With this understanding if
\begin{equation}
  \label{eq:122}
  \widehat Q^{\tilde\epsilon}_{\alpha}\psi=P_{\tilde\epsilon}^{-j}\widehat Q^{\prime\tilde\epsilon}_{\alpha}\psi',
\end{equation}
and
$\frac{1}{\hbar}\widehat Q^{\tilde\epsilon}_{\alpha}=-\mathcal L^{\tilde\epsilon}_{\alpha}$,
the Hilbert space carries a single unitary irreducible representation of $SU(2)$.

The wavefunctions considered above can also be expressed in terms of $SU(2)$ coherent states~\cite{Barut:1970qf,Perelomov:1971bd,GILMORE1972391}. Un-normalized $\mathrm{SU}(2)$ coherent states may be defined as
\begin{equation}
  \label{eq:167}
  |\widebar\zeta_{\tilde\epsilon}\rangle=e^{\widebar\zeta_{\tilde\epsilon} \mathcal J_{-\tilde\epsilon}}|j,\tilde\epsilon j\rangle=\sum_{m=\tilde\epsilon j}^{-\tilde\epsilon j}\frac{\widebar\zeta^{j-\tilde\epsilon m}_{\tilde\epsilon}}{(j-\tilde\epsilon m)!}\mathcal J_{-\tilde\epsilon}^{j-\tilde\epsilon m}|j,\tilde\epsilon j\rangle.
\end{equation}
Furthermore, one has
\begin{equation}
  \label{eq:168}
  |jm\rangle={\Big[\frac{(j+\tilde\epsilon m)!}{(2j)!(j-\tilde\epsilon m)!}\Big]}^{\frac 12}\mathcal J_{-\tilde\epsilon}^{j-\tilde\epsilon m}|j,\tilde\epsilon j\rangle,\
   \langle jm,\widebar\zeta_{\tilde\epsilon}\rangle=
  {\Big[\frac{(2j)!}{(j+\tilde\epsilon m)!(j-\tilde\epsilon m)!}\Big]}^{\frac 12}\widebar\zeta^{j-\tilde\epsilon m}_{\tilde\epsilon},
\end{equation}
and also
\begin{equation}
  \label{eq:169}
  \langle\eta_{\tilde\epsilon},\widebar\zeta_{\tilde\epsilon}\rangle={(1+\eta_{\tilde\epsilon}
  \widebar\zeta_{\tilde\epsilon})}^{2j}=e^{K^{j}_{\tilde\epsilon}(\eta_{\tilde\epsilon,\widebar\zeta_{\tilde\epsilon}})},
\end{equation}
with
\begin{equation}
  \label{eq:16}  K^{j}_{\tilde\epsilon}(\eta_{\tilde\epsilon},\widebar\zeta_{\tilde\epsilon})=2j\ln(1+\eta_{\tilde\epsilon}\widebar\zeta_{\tilde\epsilon}).
\end{equation}

If
\begin{equation}
  \label{eq:157}
  \psi'(\zeta_{\tilde\epsilon})=\langle \zeta_{\tilde\epsilon},\psi\rangle,\quad \overline{\psi'(\zeta_{\tilde\epsilon})}=\langle \psi,\widebar\zeta_{\tilde\epsilon}\rangle,
\end{equation}
the inner product becomes
\begin{equation}
  \label{eq:160}
  \langle \phi,\psi\rangle=\frac{2j+1}{2\pi \imath}\int_{\mathbb S^{2}}\frac{d \zeta_{\tilde\epsilon}\wedge d\widebar\zeta_{\tilde\epsilon}}{{P_{\tilde\epsilon}}^{2}}\,\widebar\phi\, \psi
  =\frac{2j+1}{2\pi \imath}\int_{\mathbb S^{2}}{d \zeta_{\tilde\epsilon}\wedge d\widebar\zeta_{\tilde\epsilon}}\,
  e^{-K^{j+1}_{\tilde\epsilon}}\langle \phi, \widebar \zeta_{\tilde\epsilon}\rangle \langle \zeta_{\tilde\epsilon},\psi\rangle,
\end{equation}
when taking into account the integrals worked out in~\eqref{eq:68}.
In this representation, the angular momentum operators are
\begin{equation}
  \label{eq:166}
  \frac{1}{\hbar}\widehat Q^{\prime\tilde\epsilon}_{\pm}=-\mathcal J_{\mp}^{\prime\tilde\epsilon},\quad \frac{1}{\hbar}\widehat Q^{\prime\tilde\epsilon}_{3}=\mathcal J_{3}^{\prime\tilde\epsilon},
\end{equation}
or, more explicitly,
\begin{equation}
  \label{eq:171}
  \mathcal J_{3}^{\prime\tilde\epsilon} = -\tilde\epsilon\zeta_{\tilde\epsilon}\partial_{\zeta_{\tilde\epsilon}}+\tilde\epsilon j,\quad
  \mathcal J^{\prime\tilde\epsilon}_{\pm}=\pm\tilde\epsilon\zeta^{1\mp\tilde\epsilon}_{\tilde\epsilon}\partial_{\zeta_{\tilde\epsilon}}
  +j(1\mp\tilde\epsilon)\zeta^{\frac{1\mp\tilde\epsilon}{2}}_{\tilde\epsilon}.
\end{equation}

In the operator formalism, the character for an element associated to $e_{3}$
may be evaluated directly since the basis $|jm\rangle$ diagonalizes
$\widehat Q_{3}$ by construction,
\begin{equation}
  \label{eq:123}
  \boxed{\chi(\boldsymbol{\xi})={\rm Tr}\, e^{-\frac{\imath{\boldsymbol{\xi}}}{\hbar} \hat Q^{3}}=\sum_{m=-j}^{j} e^{-\imath \boldsymbol{\xi}m}=
    \frac{\sin{\frac{(2j+1)\boldsymbol{\xi}}{2}}}{\sin\frac{
        \boldsymbol{\xi}
      }{2}}}.
  \end{equation}
This character, or more generally a matrix element,
may also be evaluated using path integrals by repeating the steps used in the
standard holomorphic representation,
\begin{equation}
  \label{eq:133}
  \langle \eta_{\tilde\epsilon},e^{-\frac{\imath}{\hbar} (t_{f}-t_{i})\hat H^{\prime \tilde\epsilon} } \widebar\zeta_{\tilde\epsilon}\rangle=\int^{\zeta_{\tilde\epsilon}(t_{f})=\eta_{\tilde\epsilon}}_{
    \widebar\zeta_{\tilde\epsilon}(t_{i})=\widebar\zeta_{\tilde\epsilon}}
  \prod_{\tau}\frac{d\zeta_{\tilde\epsilon}(\tau) d\widebar\zeta_{\tilde\epsilon}(\tau)}{2\pi\imath}e^{\frac{\imath}{\hbar}S_{I}}.
\end{equation}
Here, the improved action is
\begin{equation}
  \label{eq:162}
  S_{I}= \int_{t_{i}}^{t_{f}}dt\big[\frac{\hbar}{2\imath}(\partial_{\eta_{\tilde \epsilon}}K_{\tilde\epsilon}^{j}\frac{d{\eta_{\tilde\epsilon}}}{dt}-\partial_{\widebar\zeta_{\tilde \epsilon}}K_{\tilde\epsilon}^{j}\frac{d{\widebar\zeta_{\tilde\epsilon}}}{dt})-H^{\tilde\epsilon}_{N}\big]+\frac{\hbar}{2\imath}\big[K_{\tilde\epsilon}^{j}(t_{f})+K_{\tilde\epsilon}^{j}(t_{i})],
\end{equation}
and $H^{\tilde\epsilon}_{N}$ is the analog for $\mathrm{SU}(2)$ coherent states of the normal
symbol for $\hat H^{\prime \tilde\epsilon}$,
\begin{equation}
  \label{eq:7}
  H^{\tilde\epsilon}_{N}(\eta_{\tilde\epsilon},\widebar\zeta_{\tilde\epsilon})=
  e^{-K^{j}_{\tilde\epsilon}}\langle \eta_{\tilde\epsilon}, \hat H^{\prime \tilde\epsilon}\widebar\zeta_{\tilde\epsilon}\rangle.
\end{equation}
In particular,
\begin{equation}
  \label{eq:121}
  \mathcal J^{\prime\tilde\epsilon}_{3 N}=\tilde\epsilon j\frac{1-\eta_{\tilde\epsilon}\widebar\zeta_{\tilde\epsilon}}{1+\eta_{\tilde\epsilon}\widebar\zeta_{\tilde\epsilon}},\quad
  \mathcal J^{\prime\tilde\epsilon}_{\pm N}=2j\frac{\widebar\zeta^{\frac{1\pm\tilde\epsilon}{2}}_{\tilde\epsilon}
    \eta^{\frac{1\mp\tilde\epsilon}{2}}_{\tilde\epsilon}}{1+\eta_{\tilde\epsilon}\widebar\zeta_{\tilde\epsilon}}.
\end{equation}

The Euler-Lagrange equations of motion associated to $S_{I}$ are
\begin{equation}
  \label{eq:229}
  \frac{d \widebar\zeta_{\tilde\epsilon}}{dt}=-\frac{\imath}{2j\hbar}{(1+\eta_{\tilde\epsilon}\widebar\zeta_{\tilde\epsilon })}^{2}\frac{\partial H^{\tilde\epsilon}_{N}}{\partial\eta_{\tilde\epsilon}},\quad
  \frac{d \eta_{\tilde\epsilon}}{dt}=\frac{\imath}{2j\hbar}{(1+\eta_{\tilde\epsilon}\widebar\zeta_{\tilde\epsilon })}^{2}\frac{\partial H^{\tilde\epsilon}_{N}}{\partial\widebar\zeta_{\tilde\epsilon}}.
\end{equation}
For solutions with fixed values for $\eta_{\tilde\epsilon}$ at $t_{f}$ and $\widebar\zeta_{\tilde\epsilon}$ at $t_{i}$, the variational principle has a true extremum on-shell, without any boundary terms,
$\delta S_{I}\approx 0$ if $\delta \eta_{\tilde\epsilon}(t_{f})=0=\delta\widebar\zeta_{\tilde\epsilon}(t_{i})$.
If one takes as quantum Hamiltonian
$ \hat H^{\prime\tilde\epsilon}_{J}=\hat Q^{\prime\tilde\epsilon}_{3}$,
the equations of motion become
\begin{equation}
  \label{eq:230}
  \frac{d \widebar\zeta_{\tilde\epsilon}}{dt}=\imath\tilde\epsilon\widebar\zeta_{\tilde\epsilon},\quad
  \frac{d \eta_{\tilde\epsilon}}{dt}=-\imath\tilde\epsilon\eta_{\tilde\epsilon},
\end{equation}
with unique solution satisfying the boundary conditions given by
\begin{equation}
  \label{eq:233}
  \widebar\zeta_{\tilde\epsilon}(t)=\widebar\zeta_{\epsilon}e^{\imath \tilde\epsilon(t-t_{i})},\quad \eta_{\tilde\epsilon}(t)=\eta_{\tilde\epsilon}e^{\imath \tilde\epsilon(t_{f}-t)}.
\end{equation}
In this case, the path integral reduces to the on-shell action $S_{I}^{\rm cl}$. If $\boldsymbol{\xi}=t_{f}-t_{i}$,
\begin{equation}
  \label{eq:234}
  \langle \eta_{\tilde\epsilon},e^{-\frac{\imath}{\hbar} \boldsymbol{\xi}\hat Q^{\prime \tilde\epsilon}_{3} } \widebar\zeta_{\tilde\epsilon}\rangle=e^{\frac{\imath}{\hbar}S^{\rm cl}_{I}}=
  e^{-\imath\tilde\epsilon \boldsymbol{\xi}Z j}{(1+\widebar\zeta_{\tilde\epsilon}\eta_{\tilde\epsilon}e^{\imath\tilde\epsilon \boldsymbol{\xi} j})}^{2j}.
\end{equation}
When expanding the integrand and using the integrals~\eqref{eq:68}, the trace
\begin{equation}
  \label{eq:235}
  \Tr\, e^{-\frac{\imath}{\hbar} \boldsymbol{\xi}\hat Q^{\prime \tilde\epsilon}_{3} }=\frac{2j+1}{2\pi \imath}\int_{\mathbb S_{2}}d\zeta_{\tilde\epsilon}\wedge d\widebar\zeta_{\tilde\epsilon}\, \langle \zeta_{\tilde\epsilon},e^{-\frac{\imath}{\hbar} \boldsymbol{\xi}\hat Q^{\prime \tilde\epsilon}_{3} } \widebar\zeta_{\tilde\epsilon}\rangle e^{-K^{j+1}},
\end{equation}
yields the expression~\eqref{eq:123} for the character.

\subsection{Quantization of $T^*\mathrm{SU}(2)$}\label{sec:quantization-tsu2}

Due to their canonical symplectic structure, cotangent bundles of Lie groups are
more straightforward to quantize than coadjoint orbits. We will consider here
the quantization of $T^*\mathrm{SU}(2)$. This will allow us, in the following
subsection, to apply the Dirac quantization procedure to reduce to the quantized model space and also to a the unitary irreducible representation corresponding to a single quantized coadjoint orbit.

The quantization of $T^{*}G$ should be such that the classical real observables
\begin{equation}
  \label{eq:30}
  \pi_{\alpha}=R\indices{_{\alpha}^{i}}p_{i},\quad Q_{\alpha}=L\indices{_{\alpha}^{i}}p_{i},
\end{equation}
are represented by hermitian quantum operators. This involves ordering
ambiguities. The prequantization formula of geometric quantization, applied to
the observables $\pi_{\alpha},Q_{\alpha}$, which are linear in the $\pi$'s
reduces to the first term determined by the Hamiltonian vector fields given
in~\eqref{eq:79},
\begin{equation}
  \label{eq:83}
  \hat Q_{\alpha}=-\imath\hbar \vec L_{\alpha},\quad \hat \pi_{\alpha}=-\imath\hbar (\vec R_{\alpha}+f\indices{^{\gamma}_{\beta\alpha}}\pi_{\gamma}\frac{\partial}{\partial \pi_{\beta}}).
\end{equation}
When the Hilbert space consists of scalar densities on $G$ of weight $1/2$,
\begin{equation}
  \label{eq:36}
\tilde\Psi'(g')=|\frac{\partial g^{i}}{\partial g^{\prime j}}|^{\frac 12}\tilde\Psi(g),\quad   \langle \Phi,\Psi\rangle = \int_{G} d^{n}g\, \tilde\Phi^{*}(g)\tilde\Psi(g).
\end{equation}
the associated operators are represented by
\begin{equation}
  \label{eq:76}
  \hat\pi_{\alpha} \tilde \Psi(g)=-\imath\hbar [\vec R_{\alpha}+\frac 12 \partial_{i}R\indices{_{\alpha}^{i}}]\tilde\Psi(g),\quad \hat Q_{\alpha} \tilde\Psi(g)= -\imath\hbar [\vec L_{\alpha}+\frac 12 \partial_{i}L\indices{_{\alpha}^{i}}]\tilde \Psi(g).
\end{equation}
and are hermitian,
\begin{equation}
  \label{eq:38}
  \hat \pi^{\dagger}_{\alpha}=\hat \pi_{\alpha},\quad \hat Q^{\dagger}_{\alpha}=\hat Q_{\alpha}.
\end{equation}

Furthermore, suppose that there is a bi-invariant metric on the group, and let
$\mathbf g=\det g_{{ij}}$. One may then replace scalar densities by scalar
fields
\begin{equation}
  \label{eq:39}
  \tilde\Psi=|\mathbf g|^{\frac 14}\Psi,\quad \Psi'(g')=\Psi(g),\quad \langle \Phi,\Psi\rangle = \int_{G} d^{n}g{|\mathbf g|}^{\frac 12}\Phi^{*}(g)\Psi(g).
\end{equation}
In this case, the hermitian operators simplify to
\begin{equation}
  \label{eq:45}
  \begin{split}
  &\hat\pi_{\alpha} \Psi(g)=-\imath\hbar [\vec R_{\alpha}+\frac 12 {|\mathbf g|^{-\frac 12}}\partial_{i}({|
    \mathbf g|^{\frac 12}}R\indices{_{\alpha}^{i}})]\Psi(g)=-\imath\hbar \vec R_{\alpha}
    \Psi(g)
    ,\\
  &\hat Q_{\alpha} \Psi(g)=-\imath\hbar [\vec L_{\alpha}+\frac 12 {|\mathbf g|^{-\frac 12}}\partial_{i}({|
    \mathbf g|^{\frac 12}}L\indices{_{\alpha}^{i}})]\Psi(g)=-\imath\hbar \vec L_{\alpha}\Psi(g),
  \end{split}
\end{equation}
because
\begin{equation}
  \label{eq:4}
  \partial_{i}({|\mathbf g|^{\frac 12}}R\indices{_{\alpha}^{i}})=0=\partial_{i}({|\mathbf g|^{\frac 12}}L\indices{_{\alpha}^{i}}),
\end{equation}
when using that left and right invariant vector fields commute. Alternatively,
if $D_{i}$ denotes the Christoffel connection associated with $g_{ij}$, this may
also be seen from $\mathcal L_{\xi}g_{ij}=D_{i}\xi_{j}+D_{j}\xi_{i}$,
${|\mathbf g|^{-\frac 12}}\partial_{i}({|\mathbf g|^{\frac 12}}\xi^{i})=D_{i}\xi^{i}$
and~\eqref{eq:15}.

In the case of $T^{*}\mathrm{SU}(2)$, we consider wave functions polarized with respect
to the standard vertical polarization, i.e, wave functions on configuration
space $\mathrm{SU}(2)$. A basis for the Hilbert space $L^{2}(\mathrm{SU}(2))$ that forms a
representation of left and right invariant vector fields is provided by the
Wigner $\mathcal D$ matrices. In particular, the (complex conjugates of the)
Wigner functions simultaneously diagonalize $\mathcal J^{2},\mathcal J_{3}$,
$\mathcal R_{3}$, where the latter is the operator associated to $\pi_{3}$. How to
explicitly construct the Wigner matrices is discussed for instance in section in
section 3.8 of~\cite{biedenharn1981angular}. Here we content ourselves to using standard results from the
literature adapted to our parametrization.

In the expressions below, $j$ is integer ($j=0,1,2,\dots$) and $m',m\in \mathbb Z$,
$|m'|\leq j,|m|\leq j$, or $j$ is half integer
$(j=\frac 12, \frac 12 +1, \frac 12+2,\dots)$ and $m',m\in \mathbb Z+\frac 12$,
$|m'|\leq j,|m|\leq j$. In the parametrization~\eqref{eq:99x}, left and right invariant vector
fields are given in~\eqref{eq:110x}\footnote{ $\mathcal R_{3},\mathcal R_{\pm}$ agree with $\mathcal J_{3},\mathcal J_{\pm}$ in eq.~(3.102), $\mathcal L_{3},\mathcal L_{\pm}$ with $-\mathcal P_{3},-\mathcal P_{\pm}$ in eq~(3.122) of~\cite{biedenharn1981angular}}. The associated Wigner matrices\footnote{cf.~eq.~(3.59), (3.65) of~\cite{biedenharn1981angular}} are given by
\begin{equation}
  \label{eq:107x}
\boxed{\langle jm'm|\psi\theta\phi\rangle\equiv  D^{j}_{m'm}(\psi\theta\phi)=e^{-\imath m'\psi}d^{j}_{m'm}(\theta)e^{-\imath m\phi}},
\end{equation}
\begin{multline}
  \label{eq:111x}
  d^{j}_{m'm}(\theta)=\sqrt{(j+m')!(j-m')!(j+m)!(j-m)!}\\
  \sum_{r={\max (0,m-m')}}^{{\min}(j-m',j+m)}\frac{{(-)}^{m'-m+r}
    \cos^{2j-2r-m'+m}\frac{\theta}{2}\sin^{2r+m'-m}\frac{\theta}{2}}{(j+m-r)!r!(m'-m+r)!(j-m'-r)!},
\end{multline}
and satisfy\footnote{cf.~eq.~(3.82) of~\cite{biedenharn1981angular}},
\begin{equation}
  \label{eq:125}
  d^{j}_{m'm}={(-)}^{m'-m}d^{j}_{-m',-m},
\end{equation}
so that one may also write
\begin{multline}
  \label{eq:111y}
  d^{j}_{m'm}(\theta)=\sqrt{(j+m')!(j-m')!(j+m)!(j-m)!}\\
  \sum_{r={\max (0,m'-m)}}^{{\min}(j+m',j-m)}\frac{{(-)}^{r}
    \cos^{2j-2r+m'-m}\frac{\theta}{2}\sin^{2r-m'+m}\frac{\beta}{2}}{(j-m-r)!r!(m-m'+r)!(j+m'-r)!}.
\end{multline}

If wave functions are defined by
\begin{equation}
  \label{eq:85}
  \langle \psi\theta\phi|jm'm\rangle=D^{j*}_{m'm},
\end{equation}
the action of right invariant vector fields is given by\footnote{cf.~eq.~(3.104)
  of~\cite{biedenharn1981angular}},
\begin{equation}
  \label{eq:112x}
  \mathcal R_{\pm}D^{j*}_{m'm}=\sqrt{(j\mp m')(j\pm m'+1)}D^{j*}_{m'\pm 1m},\quad \mathcal R_{3}D^{j*}_{m'm}=m'D^{j*}_{m'm},
\end{equation}
while the one for left invariant vector fields is\footnote{cf.~eq.~(3.123)
  of~\cite{biedenharn1981angular}} is
\begin{equation}
  \label{eq:113x}
  \mathcal L_{\pm}D^{j*}_{m'm}=-\sqrt{(j\pm m)(j\mp m+1)}D^{j*}_{m'm\mp 1},\quad \mathcal L_{3}D^{j*}_{m'm}=-mD^{j*}_{m'm}.
\end{equation}
In other words, the standard relations are obtained for the operators
\begin{equation}
  \label{eq:129}
  \mathcal J_{3}=-\mathcal L_{3},\quad\mathcal J_{\pm}=-\mathcal L_{\mp}.
\end{equation}
Furthermore,
\begin{equation}
  \label{eq:114x}
  \mathcal R^{2}D^{j*}_{m'm}=\mathcal J^{2}D^{j*}_{m'm}=j(j+1)D^{j*}_{m'm}.
\end{equation}

Spin-weighted spherical harmonics may be defined in
terms of Wigner functions by
\begin{equation}
  \label{eq:125x}
  {}_{s}Y_{jm}={(-)}^{m-s}\sqrt{\frac{2j+1}{4\pi}}D^{j*}_{sm}|_{\psi=0}\quad
  \iff e^{\imath s\psi}{}_{s}Y_{jm}={(-)}^{m-s}\sqrt{\frac{2j+1}{4\pi}}D^{j*}_{sm}.
\end{equation}
Standard spherical harmonics for integer $j$ are given
by\footnote{cf.~(3.138) of~\cite{biedenharn1981angular}}
\begin{equation}
  \label{eq:116}
  {}_{0}Y_{jm}(\theta\phi)=\sqrt{\frac{2j+1}{4\pi}}D^{j*}_{0m}(\phi\theta\psi)=\sqrt{\frac{2j+1}{4\pi}}e^{\imath m\phi}d^{j}_{m0}(\theta).
\end{equation}
This agrees with the above in the parametrization adopted here because\footnote{cf.~eq.~(3.81)
  of~\cite{biedenharn1981angular}} ${(-)}^{m}d^{j}_{0m}(\theta)=d^{j}_{m0}(\theta)$.

Using~\eqref{eq:110x}, this gives
\begin{multline}
  \label{eq:120x}
  \mathcal R_{\pm}D^{j*}_{sm}(\psi\theta\phi)=\pm e^{\imath(s\pm 1)\psi}(\partial_{\theta}\mp \imath\frac{1}{\sin\theta}\partial_{\phi}\mp s\cot\theta)D^{j*}_{sm}|_{\psi=0}\\
  =\pm e^{\imath(s\pm 1)\psi}\sin^{\pm s}\theta(\partial_{\theta}\mp \imath\frac{1}{\sin\theta}\partial_{\phi})(\sin^{\mp s}\theta D^{j*}_{sm}|_{\psi=0}),
\end{multline}
and, if $\eth,\widebar\eth$ are defined as
\begin{equation}
  \label{eq:123x}
  \begin{split}
    &{\big(\mathcal R_{+} D^{j*}_{sm}\big)}|_{\psi=0}=\eth D^{j*}_{sm}|_{\psi=0},\\ & \big(\mathcal R_{-} D^{j*}_{m-s}\big)|_{\psi=0}=                                                                          -\widebar\eth D^{j*}_{sm}|_{\psi=0},\\ & \big(\mathcal R_{3} D^{j*}_{sm}(\psi\theta\phi)\big)|_{\psi=0}=s D^{j*}_{sm}|_{\psi=0},
  \end{split}
\end{equation}
it follows that
\begin{equation}
  \label{eq:121x}
  \begin{split}
    \eth {}_{s}Y_{jm}(\theta\phi)=\sin^{s}\theta(\partial_{\theta}-\imath\frac{1}{\sin\theta}\partial_{\phi})(\sin^{-s}\theta{}_{s}Y_{jm}(\theta\phi)),\\
    \widebar\eth {}_{s}Y_{jm}(\theta\phi)=\sin^{-s}\theta(\partial_{\theta}+\imath\frac{1}{\sin\theta}\partial_{\phi})(\sin^{s}\theta{}_{s}Y_{jm}(\theta\phi)).
  \end{split}
\end{equation}
In terms of spin-weighted spherical harmonics, equation~\eqref{eq:112x} becomes
\begin{equation}
  \label{eq:115x}
  \begin{split}
    \eth {}_{s}Y_{jm}=-\sqrt{(j-s)(j+s+1)}{}_{s+1}Y_{jm},\quad \widebar \eth {}_{s}Y_{jm}=\sqrt{(j+s)(j-s+1)}{}_{s-1}Y_{jm},
  \end{split}
\end{equation}
while $[\mathcal R_{+},\mathcal R_{-}]=2\mathcal R_{3}$ becomes
\begin{equation}
  \label{eq:122x}
  [\widebar\eth,\eth]{}_{s}Y_{jm}=2s{}_{s}Y_{jm}.
\end{equation}
Furthermore,
\begin{equation}
  \label{eq:124x}
  \begin{split}
    &\widebar\eth\eth {}_{s}Y_{jm}=-(j-s)(j+s+1){}_{s}Y_{jm}=[s(s+1)-j(j+1)]{}_{s}Y_{jm},\\
    &\eth\widebar\eth {}_{s}Y_{jm}=-(j+s)(j-s+1){}_{s}Y_{jm}=[s(s-1)-j(j+1)]{}_{s}Y_{jm},\\
    & \frac 12(\widebar\eth\eth+\eth\widebar\eth){}_{s}Y_{jm}=[s^{2}-j(j+1)]{}_{s}Y_{jm}.
  \end{split}
\end{equation}

For comparison, note that the coordinate expressions of $\eth,\bar\eth$ in~\eqref{eq:121x} and the
raising/lowering formulas in~\eqref{eq:115x} agree with the ones in~\cite{Penrose:1984}, eq.~(4.15.122),
(4.15.106) for $R=\frac{1}{\sqrt 2}$ and the ones in~\cite{Beyer:2013loa}, eq.~(2.12),
(2.13). When using~\eqref{eq:111y}, the explicit expression for ${}_{s}Y_{jm}(\theta\phi)$
defined here agrees with eq.~(2.8), (2.10) of~\cite{Beyer:2013loa}.

In the implicit Euler-Rodrigues parametrization~\eqref{eq:138}, one has\footnote{cf.
  eq~(3.89)-(3.90) of~\cite{biedenharn1981angular}},
\begin{multline}
  \label{eq:135}
  D^{j}_{m'm}(\alpha^{0},\alpha^{\alpha})=\sqrt{(j+m')!(j-m')!(j+m)!(j-m)!}\\
  \sum_{r={\rm \max (0,m-m')}}^{{\rm \min}(j-m',j+m)}
  \frac{  \bar a^{j+m-r}\bar b^{m'-m+r}{(-b)}^{r}
    a^{j-m'-r}}{(j+m-r)!r!(m'-m+r)!(j-m'-r)!},
\end{multline}
which upon substitution of~\eqref{eq:99x} yields~\eqref{eq:107x},~\eqref{eq:111x}, and where
\begin{equation}
  \label{eq:136bis}
  D^{j}(\alpha^{\prime0},\alpha^{\prime\beta})D^{j}(\alpha^{0},\alpha^{\beta})=D^{j}(\alpha^{\prime\prime0},\alpha^{\prime\prime\beta}).
\end{equation}
When taking into account the orthogonality conditions discussed in section 3.9
of~\cite{biedenharn1981angular}, this concludes the quantization of the
$T^{*}SU(2)$ in terms of the Hilbert space of wave functions $D^{j*}_{m'm}$
together with the representation of the symmetries and the momenta.

In complex parametrizations~\eqref{eq:99y}, with Wigner functions
$D^{j}_{m'm}(\psi\zeta_{\tilde\epsilon}\widebar\zeta_{\tilde\epsilon})$ obtained
by substituting~\eqref{eq:173} into~\eqref{eq:135}, left and right invariant vector fields are given
in~\eqref{eq:110xa}. In this case, when defining spin-weighted spherical harmonics
${}_{s}Y_{jm}(\zeta_{\tilde\epsilon}\widebar\zeta_{\tilde\epsilon})$ in terms of these Wigner
functions at $\psi=0$ as in~\eqref{eq:125x}, one now gets instead of~\eqref{eq:120x},
\begin{equation}
  \begin{split}
  \label{eq:120xz}
    & \mathcal R_{+}D^{j*}_{sm}=\tilde\epsilon e^{\imath(s+1)\psi}(P_{\tilde\epsilon}\partial_{\widebar\zeta_{\tilde\epsilon}}+s\zeta_{\tilde\epsilon})D^{j*}_{sm}|_{\psi=0}=\tilde\epsilon e^{\imath(s+ 1)\psi}P_{\tilde\epsilon}^{1-s}\partial_{\widebar\zeta_{\tilde\epsilon}}(P_{\tilde\epsilon}^{s} D^{j*}_{sm}|_{\psi=0}),\\
    & \mathcal R_{-}D^{j*}_{sm}=-\tilde\epsilon e^{\imath(s-1)\psi}(P_{\tilde\epsilon}\partial_{\zeta_{\tilde\epsilon}}
      -s\widebar\zeta_{\tilde\epsilon})D^{j*}_{sm}|_{\psi=0}=-\tilde\epsilon e^{\imath(s- 1)\psi}P_{\tilde\epsilon}^{1+s}\partial_{\zeta_{\tilde\epsilon}}(P_{\tilde\epsilon}^{-s} D^{j*}_{sm}|_{\psi=0}).
  \end{split}
\end{equation}
If $\eth,\widebar\eth$ are still defined by~\eqref{eq:123x}, it now follows that
\begin{equation}
  \label{eq:242}
 \begin{split}
    \eth {}_{s}Y_{jm}(\zeta_{\tilde\epsilon}\widebar\zeta_{\tilde\epsilon})=\tilde\epsilon P_{\tilde\epsilon}^{1-s}\partial_{\widebar\zeta_{\tilde\epsilon}}(P_{\tilde\epsilon}^{s} {}_{s}Y_{jm}(\zeta_{\tilde\epsilon}\widebar\zeta_{\tilde\epsilon})),\\
    \widebar\eth {}_{s}Y_{jm}(\zeta_{\tilde\epsilon}\widebar\zeta_{\tilde\epsilon})=\tilde\epsilon P_{\tilde\epsilon}^{1+s}\partial_{\zeta_{\tilde\epsilon}}(P_{\tilde\epsilon}^{-s} {}_{s}Y_{jm}(\zeta_{\tilde\epsilon}\widebar\zeta_{\tilde\epsilon})).
  \end{split}
\end{equation}
Note that the effect of a reparametrization of the type $g'=\bar h g$ with $h=e^{-\frac{\bar\psi}{2}\imath\sigma_{3}}$ and $\bar\psi$ constant, consists in a shift $\psi\to \psi+\bar\psi$ in $g$. There is no inhomogeneous term in the transformation of the associated Maurer-Cartan form, $dg'g^{\prime -1}=\bar h dg g^{-1}\bar h^{-1}$. When choosing $\bar\psi =\pi$ and applying this reparametrization to $g_{-}$ in~\eqref{eq:99y}, one gets
\begin{equation}
  \label{eq:247}
  g'_{-}=\imath P_{-}^{{-\frac 12}}
  \begin{pmatrix} -\widebar\zeta_{-}\, e^{-\imath\frac{\psi}{2}} & e^{-\imath\frac{\psi}{2}} \\
      e^{\imath\frac{\psi}{2}} & \zeta_{-}\,e^{\imath\frac{\psi}{2}}
\end{pmatrix}
\end{equation}
On account of~\eqref{eq:47Bxd}, the associated $\mathcal R_{\pm}$ change by an overall minus sign, with $\mathcal R_{3}$ unchanged. After the change induced by this reparametrization, the coordinate expressions~\eqref{eq:242} for $\eth,\widebar\eth$ in terms of $\zeta_{-},\widebar\zeta_{-}$ agree with those in (4.15.117) of~\cite{Penrose:1984}.

In the Hilbert space associated to the quantization of $T^{*}\mathrm{SU}(2)$, one can compute the partition function for the Casimir operator,
\begin{equation}
  \label{eq:260}
  \boxed{Z(b)=\Tr_{L^{2}(\mathrm{SU}(2))}\, e^{{-b\mathcal R^{2}}}=\sum_{j\in\mathbb N/2}{(2j+1)}^{2}e^{{-bj(j+1)}}=
  e^{\frac b4}\sum_{m=0}^{\infty} m^{2} e^{-\frac b4 m^{2}}}.
\end{equation}
This expression is adapted to a low temperature expansion,  $b\gg 1$:
\begin{equation}
  \label{eq:269}
  Z(b)=1+4 e^{-\frac{3b}{4}}+\mathcal O(e^{-2b}).
\end{equation}
In order to access the high temperature expansion, $b\ll 1$, the partition function may be written as
\begin{equation}
  \label{eq:273}
  Z(b)=-\frac{e^{\frac b4}}{2\pi}\theta'(t)|_{t=\frac{b}{4\pi}},
\end{equation}
where the Jacobi theta function is
\begin{equation}
  \label{eq:274}
  \theta(t)=\vartheta_{3}(0|\imath t)=\sum_{n\in \mathbb Z}e^{-t\pi n^{2}}.
\end{equation}
Using the modular transformation $\theta(t)=\frac{1}{\sqrt t}\theta(\frac 1t)$ and the expansion $\theta(t)=1+2e^{-\pi t}+2e^{-4\pi t}+2e^{-9\pi t}+\dots $ for $t\gg 1$, it follows that for $t\ll 1$,
\begin{equation}
  \label{eq:275}
  \theta(t)=\frac{1}{\sqrt t}[1+2e^{-\frac{\pi}{t}}+2e^{-\frac{4\pi}{t}}+2e^{-9\frac{\pi}{t}}+\dots ],
\end{equation}
so that
\begin{equation}
  \label{eq:276}
  \theta'(t)=-\frac{1}{2 t^{\frac 32}}+e^{-\frac{\pi}{t}}(-\frac{1}{t^{\frac 32}}+2\frac{\pi}{t^{\frac 52}})+
  e^{-\frac{4\pi}{t}}(-\frac{1}{t^{\frac 32}}+2\frac{4\pi}{t^{\frac 52}})+\dots,
\end{equation}
and
\begin{equation}
  \label{eq:277}
  Z(b)=\frac{2\sqrt \pi e^{\frac b4}}{b^{\frac 32}}\big[1+e^{-\frac{4\pi^{2}}{b}}(-\frac{16 \pi^{2}}{b}+2)+\mathcal O(\frac{e^{-\frac{16\pi^{2}}{b}}}{b})\big].
\end{equation}

\subsection{Quantum model space and coadjoint orbit from Dirac quantization}\label{sec:reduct-after-quan}

When applying Dirac's procedure to quantize first class constrained system, the reduction
to the quantized model space follows by imposing a single first class
constraint, either $\hat\pi_{-}=0$ or $\hat\pi_{+}=0$, on the Hilbert space of wave functions,
\begin{equation}
  \label{eq:130}
  \hat \pi_{-}\Psi(\psi\theta\phi)=0\quad{\rm or}\quad \hat \pi_{+}\Psi(\psi\theta\phi)=0,\quad
  \Psi(\psi\theta\phi)=\Psi_{j}^{m' m}D^{j*}_{m'm},
\end{equation}
where $\Psi_{j}^{m' m}\in \mathbb C$ and the summation convention applies.
This is equivalent to
\begin{equation}
  \label{eq:86}
  \widebar\eth {}_{s}Y_{{jm}}=0 \quad{\rm or}\quad \eth {}_{s}Y_{{jm}}=0.
\end{equation}
The general solution\footnote{see e.g.~\cite{Penrose:1984}, (4.15.60)
  or~\cite{A.S.Galperin13674} section 4.3} to this equation is generated by the
sum over $j$ of the $2j+1$ vectors ${}_{-j}Y_{jm}$ or equivalently the
$D^{j*}_{-jm}$ in the former case, and by ${}_{j}Y_{jm}$ or equivalently the
$D^{j*}_{jm}$ in the latter.

In other words, in this approach, the Hilbert space of the quantized model space
consists of vectors of the form
\begin{equation}
  \label{eq:88}
\Psi_{j}^{m} {}_{-j}Y_{jm},\quad{\rm or}\quad
\Psi_{j}^{m} {}_{j}Y_{jm}
\end{equation}
Since for each $j$, these subspaces do carry a unitary irreducible
representation of $\mathrm{SU}(2)$ provided by the $\mathcal J_{3},\mathcal J_{\pm}$,
this proves the proposal for the model space in the simplest case of $\mathrm{SU}(2)$.

When imposing the first class constraint
\begin{equation}
  \label{eq:89}
  \hat \pi_{3}-\epsilon\boldsymbol{\mu}=0,\quad \hat\pi_{3}=\hbar \mathcal R_{3}=-\imath\hbar\partial_{\psi},
\end{equation}
on the Hilbert space of wave functions $D^{j}_{sm}$, it follows that non trivial
solutions exists if and only if $\boldsymbol{\mu}$ is quantized according to
\begin{equation}
  \label{eq:43}
  \epsilon\boldsymbol{\mu}=\hbar s,
\end{equation}
with $s$ half-integer and one thus remains with the wave functions at this fixed
value of $s$, which is positive or negative depending on whether the coadjoint
representative was chosen along the positive or negative $z$-axis.

In order to have a non-trivial intersection between this and the previous
condition $\hat \pi_{-}=0$ or $\hat \pi_{+}=0$, one has to choose $\hat\pi_{\epsilon}=0$ and
thus to correlate the choice of the first class constraint (in the conversion
from a second to a first class system) with the orientation of the coadjoint
representative along the $z$-axis. It then follows that
\begin{equation}
  \label{eq:93}
  \boldsymbol{\mu}=\hbar j,
\end{equation}
so that one indeed finds a single unitary irreducible representation carried by
$ \mathcal H^{G_{\mu^{\epsilon\hbar j}}\backslash SU(2)}$, the Hilbert space generated by the
$2j+1$ vectors ${}_{\epsilon j}Y_{jm}$ at fixed $j$,
\begin{equation}
  \label{eq:94}
 \sum_{m=-j}^{j}\Psi_{(j)}^{m}{}_{\epsilon j}Y_{jm}.
\end{equation}

\subsection{Quantum model space from gauge fixed quaternions}\label{sec:quant-extend-space}

Starting from the alternative description of the model space developed in
Section \ref{sec:su2-model-space-2}, we are instructed to find a quantization in a Hilbert space that
realizes the Dirac brackets~\eqref{eq:192a},~\eqref{eq:180a} and~\eqref{eq:185a} in terms of operators and their
commutators.

If $\Xi=(1,2)$, we get for the quantum version of~\eqref{eq:180a},
\begin{equation}
  \label{eq:201}
  [\hat a_{\Xi},\hat a^{\dagger}_{\Xi'}]=\delta_{\Xi,\Xi'},\quad [\hat a_{\Xi},\hat a_{\Xi'}]=0=[\hat a^{\dagger}_{\Xi},\hat a^{\dagger}_{\Xi'}].
\end{equation}

If we define
\begin{equation}
  \label{eq:155}
  \mathcal O_{+}=\frac{1}{\hbar}\hat q_{-},\quad \mathcal O_{-}=\frac{1}{\hbar}\hat q_{+},\quad \mathcal O_{3}=\frac{1}{\hbar} \hat q_{3},
\end{equation}
the correspondence rule and the Dirac brackets~\eqref{eq:192a} imply that these
operators satisfy the standard algebra given in~\eqref{eq:100}.

Defining
\begin{equation}
  \label{eq:95}
  \hat N_{1}=\hat a^{\dagger}_{1}\hat a_{1},\quad \hat N_{2}=\hat a^{\dagger}_{2}\hat a_{2},\quad \hat N=\hat N_{1}+\hat N_{2},\quad \hat M=\hat N_{1}-\hat N_{2},
\end{equation}
this also follows directly from
\begin{equation}
  \label{eq:204}
  \begin{split}
    &  \mathcal O_{+}=\frac{1+\epsilon' }{2}\hat a^{\dagger}_{1}\hat a_{2}-\frac{1-\epsilon' }{2}\hat a^{\dagger}_{2}\hat a_{1},
      \quad \mathcal O_{-}=\frac{1+\epsilon' }{2}\hat a^{\dagger}_{2}\hat a_{1}-\frac{1-\epsilon' }{2}\hat a^{\dagger}_{1}\hat a_{2},\\
    & \mathcal O_{3}=\frac{\epsilon' }{2} \hat M,
  \end{split}
\end{equation}
and the commutation relations~\eqref{eq:201}. The Casimir operator is
\begin{equation}
  \label{eq:44a}
  \mathcal O^{2}=\frac 14 {(\hat N)}^{2}+\frac 12 \hat N,
\end{equation}
while additional quadratic observables are
\begin{equation}
  \label{eq:208}
  \frac{\hat R^{2}}{\hbar}=\frac 12 \hat N,\quad \frac{\hat q'_+}{\hbar}=-[\frac{1+\epsilon' }{2}\hat a_{1}\hat a_{2}
  +\frac{1-\epsilon' }{2}\hat a^{\dagger}_{1}\hat a^{\dagger}_{2}],\quad \frac{\hat q'_-}{\hbar}=\frac{1+\epsilon' }{2}\hat a^{\dagger}_{1}\hat a^{\dagger}_{2}
  +\frac{1-\epsilon' }{2}\hat a_{2}\hat a_{1}.
\end{equation}
Note that, when $\epsilon'=1$, the Noether charges
can be written in terms of the ``Jordan map''
\begin{equation}
  \label{eq:205}
  \mathcal O_{\alpha}=\frac{1}{2}\hat a^{\dagger}_{\Xi}\sigma_{\alpha}^{\Xi\Xi'}\hat a_{\Xi'},
\end{equation}
which is the starting point of the analysis in~\cite{osti_4389568} (see
e.g.~\cite{biedenharn1981angular} chapter 5 or~\cite{Wheeler2000} for reviews).

In the standard orthonormal basis,
\begin{equation}
  \label{eq:90a}
  \begin{split}
    & |n_{1},n_{2})=\frac{1}{\sqrt{n_{1}!n_{2}!}}{(a_{1}^{\dagger})}^{n_{1}}{(a_{2}^{\dagger})}^{n_{2}}|0),
    \\
    & \mathcal O^{2}|n_{1},n_{2})=j(j+1)|n_{1},n_{2}),\quad j=\frac 12(n_{1}+n_{2}),\\ & \mathcal O_{3}|n_{1},n_{2})=m|n_{1},n_{2}),\quad m=\frac 12(n_{1}-n_{2}),
  \end{split}
\end{equation}
so that
\begin{equation}
  \label{eq:96}
  |jm\rangle=|j+m,j-m),
\end{equation}
and the Hilbert space decomposes into the sum of unitary irreducible
representations of $SU(2)$ with multiplicity one, as it should for the quantized
model space.

In holomorphic representation, this Hilbert space may also be described in terms of coherent states as
\begin{equation}
  \label{eq:206}
  |a_{\xi}\rangle=e^{a^{\xi} \hat a_{\xi}^{\dagger}}|0\rangle,\quad \psi(a^{*}_{\xi})=\langle a^{*}_{\xi}|\psi\rangle,
\end{equation}
with inner product
\begin{equation}
  \label{eq:207}
  \langle\phi|\psi\rangle=\int \prod_{\xi}\frac{da^{*}_{\xi}da^{\xi}}{2\pi\imath}e^{-a^*_{\xi}a^{\xi}}\phi^{*}(a^{\xi})\psi(a^{*}_{\xi}).
\end{equation}

If one now restricts to a single coadjoint orbit, the quantum version of~\eqref{eq:44}
implies that $\boldsymbol{\mu}$ is quantized in half-integer units of $\hbar$,
\begin{equation}
  \label{eq:96a}
  \boldsymbol{\mu}= \frac{\hbar}{2} (n_{1}+n_{2})=\hbar j,\quad \hat N=2j\mathbb 1,
\end{equation}
with the associated unitary irreducible representation described by states that
are created by monomials of order $2j$ in $\hat a^{\dagger}_{\Xi}$ acting on the
vacuum $|0\rangle$.

How this understanding may be used to efficiently construct the Wigner functions
$D^{j}_{m' m}$ is explained for instance in section 5.4 of~\cite{biedenharn1981angular}. One thus
recovers from the current perspective of constrained systems and Dirac brackets
the solution to quantum angular momentum found by Schwinger~\cite{osti_4389568}.

Since the Casimir is a dynamical operator in the quantum model space, one can compute
\begin{equation}
  \label{eq:257}
  \boxed{Z(b)=\Tr\, e^{-b\mathcal O^{2}}=\sum_{j\in \mathbb N/2}(2j+1)e^{-bj(j+1)}=e^{\frac b4}\sum_{m=0}^{\infty}me^{-\frac b4 m^{2}}}.
\end{equation}
This expression is adapted to a low temperature expansion:  if $b\gg 1$,
\begin{equation}
  \label{eq:259}
  Z(b)=1+2e^{-\frac{3b}{4}}+\mathcal O(e^{{-2b}}).
\end{equation}
Defining $F(m)=e^{\frac b4}me^{-\frac{b}{4}m^{2}}$, an expression adapted to a
high temperature expansion $b\ll 1$, can be obtained from the Euler-Maclaurin
formula. When using that $B_{2}=\frac{1}{6}$, $B_{4}=-\frac{1}{30}$,
\begin{multline}
  \label{eq:258}
  Z(b)=\frac 12 F(0)+\sum_{m\in \mathbb N^{*}} F(m)=\int^{\infty}_{0}dx F(x)-\frac 12 B_{2}F'(0)-\frac{1}{4!}F'''(0)+\dots
  \\=e^{\frac b4}[\frac{2}{b}-\frac{1}{12}-\frac{1}{480}b+\dots].
\end{multline}

\section{Conclusions}\label{sec:conclusions}

Based on the considerations on model spaces in~\cite{bernstein1976models} and in~\cite{Alekseev:1990mp,La:1990ty}, we
propose that the symplectic space defined in~\eqref{eq:12} is a crucial
ingredient in the construction of the model space.

The proposal has been tested in detail in the case of $\mathrm{SU}(2)$, where
standard techniques from the theory of constrained systems allow one to recover
various results on ``the quantum theory of angular momentum'' (see e.g.~\cite{biedenharn1981angular} for a
review) from a unified perspective. This gives rise in particular to a natural
understanding of spin-weighted/monopole spherical harmonics~\cite{Newman1966,Goldberg1967,Wu:1976ge,Wu:1977qk,Eastwood:1982aa}
and their relation to Wigner functions along the lines
of~\cite{Dray:1984gy,URBANTKE2003125,nabertopology,Beyer:2013loa,Straumann:2014uka,Boyle:2016tjj}. It also allows one to establish a direct
connection to the construction of the model space of $\mathrm{SU}(2)$ by
Schwinger~\cite{osti_4389568}, and thus also to original formulation in terms of holomorphic
functions in~\cite{bernstein1976models}.

Closely related considerations on $T^{*}\mathrm{SU}(2)$ and its quantizations
can be found for instance in~\cite{Freidel:2010aq,Freidel:2010bw,Calcinari:2020bft}, and also in~\cite{AndradeeSilva:2020ofl} in the
context of Isham's group quantization scheme~\cite{Isham:1983zr}.

In another related recent paper~\cite{Basile:2023vyg},
following~\cite{Barnich2022} in using constrained Hamiltonian techniques to
study dynamical systems associated to coadjoint orbits, an understanding of Howe
duality is presented by starting from $\mathrm{GL}(n,\mathbb C)$.

The case of more general groups will be addressed elsewhere. In particular, we
plan to study the case of non-compact groups such as
$\mathrm{SL}(2,\mathbb R)$, of semi-direct product groups such as the Euclidean
or the Poincar\'e groups, and of infinite-dimensional groups such as Kac-Moody,
Virasoro or the (centrally extended) $\mathrm{BMS}_3$ and $\mathrm{BMS}_4$
groups that arise in three-dimensional and four-dimensional gravity, and for
which geometric actions have been constructed for example in~\cite{Barnich:2017jgw,Barnich2022,Beauvillain:2024dou}.

\section*{Acknowledgments}\label{sec:acknowledgements}

\addcontentsline{toc}{section}{Acknowledgments}

G.~B.~thanks A.~Kleinschmidt, M.~Grigoriev, B.~Oblak, M.~Beauvillain,
S.~Speziale, S.~Vitouladitis and A.~Alekseev, while T.~S.~thanks G.~Kozyreff for
useful discussions. This research is supported by the F.R.S.-FNRS Belgium
through an aspirant fellowship for Thomas Smoes and through convention IISN
4.4514.08.

\appendix

\section{Path integral quantization}\label{sec:path-integr-meas}

We collect here some results on path integrals for symplectic manifolds that we
will use by following chapter 15 of~\cite{Henneaux:1992ig} (and in particular,
Exercises 15.5 and 15.8). Consider a symplectic manifold $M$ with local
coordinates $z^A$. The symplectic potential and two form are denoted by
\begin{equation}
  \label{eq:197}
  a=a_{\Delta}dz^\Delta,\quad \sigma=da=  \frac{1}{2} \sigma_{\Delta\Gamma}dz^\Delta dz^\Gamma,
\end{equation}
while the first order action
\begin{equation}
  \label{eq:209}
  S_{H}=\int_{{t_{i}}}^{t_{f}}dt\,[a_{\Delta}\dot z^{\Delta}-H].
\end{equation}
The Liouville measure to be used in the Hamiltonian path integral in non-Darboux
coordinates is
\begin{equation}
  \label{eq:8}
  \prod_{t}{\sqrt{\det \sigma_{\Delta\Gamma}}} \frac{d^{2n}z}{{(2\pi\hbar)}^n},\quad d^{2n}z=\frac{1}{(2n)!}\epsilon_{\Delta_1\dots\Delta_{2n}}dz^{\Delta_1}\dots dz^{\Delta_{2n}}.
\end{equation}

For instance, in the case of $T^{*}G$ in non-Darboux coordinates $z^{\Delta}=(g^{i},\pi_{\alpha})$, explicit expressions are given in~\eqref{eq:40},
\begin{equation}
  \label{eq:17}
  \sigma_{\Delta\Gamma}=\begin{pmatrix} \pi_\delta f^\delta_{\rho\sigma} R\indices{^\rho_i} R\indices{^\sigma_j} & -R\indices{^\beta_j}\\
           R\indices{^\alpha_j} & 0
         \end{pmatrix},\quad \sqrt{\det \sigma_{\Delta\Gamma}}=|\det R\indices{^\alpha_i}|,
\end{equation}
If the Lie algebra basis is taken as $e_j=\frac{\partial}{\partial g^j}|_{g^k=0}$, with
$g^k=0$ corresponding to the identity element in the group,
$R\indices{^j_i}(g^k=0)=\delta^j_i$, so that $\det R\indices{^j_i}>0$ by continuity.
It follows that $\det R\indices{^\alpha_i}>0$ if $e_\alpha$ is a Lie algebra basis with
the same orientation than $e_j$.

The phase space dependent measure factor $\sqrt{\det \sigma_{\Delta\Gamma}}$ may
be exponentiated into the path integral action to yield
\begin{equation}
  \label{eq:172}
  W=\int_{t_{i}}^{t_{f}}dt\, [a_{\Delta}\dot z^{\Delta}-H-\frac{\hbar}{2\imath}\delta(0)\Tr\ln \sigma_{\Delta\Gamma}],
\end{equation}
Consider
\begin{equation}
  \label{eq:196}
  \langle F\rangle\equiv\langle z_{t_f},T\widehat F\, z_{t_{i}}\rangle=\int \prod_{t}\frac{d^{2n}z}{{(2\pi\hbar)}^n}\,e^{\frac{\imath}{\hbar}W}.
\end{equation}
When using the Schwinger-Dyson equations in the form
\begin{equation}
  \label{eq:194}
  0=\langle \sigma^{\Delta\Gamma}(t)\frac{\delta W}{\delta z^{\Gamma}(t)}\rangle +\frac{\hbar}{\imath}\langle \frac{\partial\sigma^{\Delta\Gamma}}{\partial z^{\Gamma}}(t)\rangle \delta(0),
\end{equation}
and the closure of the symplectic form,
$\partial_{{[\Delta_{1}}}\sigma_{\Delta_{2}\Delta_{3}]}=0$, the singular terms proportional
to $\delta(0)$ cancel out. In summary, in path integral computations, one may
use the Hamiltonian equations of motion
\begin{equation}
  \label{eq:195}
  \dot z^{\Gamma}=\sigma^{\Gamma\Delta}\frac{\partial H}{\partial z^{\Delta}},
\end{equation}
while disregarding the singular term in~\eqref{eq:172}. Note however that one has to use
both (i) the correct symbol for the Hamiltonian adapted to the ordering at hand,
and (ii) the improved kinetic term with suitable boundary terms that guarantee
that the action has a true extremum for solutions of the equations of
motion~\eqref{eq:195} satisfying the boundary conditions adapted to the external
states. Finally, if the symbol for the Hamiltonian is such that the right hand
sides of the equations of motion in~\eqref{eq:195} are at most linear in $z^{\Delta}$, one may
expect the result to be simply given by the value of the improved action at the
extremum.

\section{Parametrizations of $\mathrm{SU}(2)$}\label{sec:parametrizations-su2}

\subsection{Generalities and notation}\label{sec:gener-notat}

We will use as Lie algebra basis $e_{\alpha}=-\frac{\imath}{2}\sigma_{\alpha}$,
$\alpha=(1,2,3)$ where the Pauli matrices are
\begin{equation}
  \label{eq:37}
  \sigma_0=\begin{pmatrix}
    1 & 0 \\
    0 & 1
  \end{pmatrix},\
  \sigma_1=\begin{pmatrix}
    0 & 1 \\
    1 & 0
  \end{pmatrix},\ \sigma_2=\begin{pmatrix}
    0 & -\imath\\
    \imath & 0
  \end{pmatrix},\ \sigma_3=\begin{pmatrix}
    1& 0\\
    0& -1
  \end{pmatrix},
\end{equation}
\begin{equation}
  \label{eq:33} \sigma_{\alpha}\sigma_{\beta}=\delta_{\alpha\beta}{\sigma_{0}}+\epsilon\indices{^{\gamma}_{\alpha\beta}}\imath\sigma_{\gamma},
\end{equation}
so that the structure constants are
\begin{equation}
  \label{eq:101}
  [e_{\alpha},e_{\beta}]=\epsilon\indices{^{\gamma}_{\alpha\beta}}e_{\gamma}.
\end{equation}

Consider the left and right invariant vector fields on $\mathrm{SU}(2)$,
$\vec L_{\alpha},\vec R_{\alpha}$ and let
\begin{equation}
  \label{eq:102}
  \vec L_{\pm}=\vec L_{1}\pm \imath\vec L_{2},\quad \vec R_{\pm}=\vec R_{1}\pm \imath\vec R_{2},\quad \mathcal L_{\alpha}=\imath\vec L_{\alpha},\quad \mathcal R_{\alpha}=-\imath \vec R_{\alpha},
\end{equation}
If $\mathcal O_{\alpha}$ is either $\mathcal L_{\alpha}$ or
$\mathcal R_{\alpha}$, it follows that
\begin{equation}
  \label{eq:100}
    [\mathcal O_{\alpha},\mathcal O_{\beta}]=\imath \epsilon^{\gamma}_{\alpha\beta}\mathcal O_{\gamma},\quad
     [\mathcal O_{3},\mathcal O_{\pm}]=\pm\mathcal O_{\pm},\quad [\mathcal O_{+},\mathcal O_{-}]=2\mathcal O_{3},\quad
      [\mathcal L_{\alpha},\mathcal R_{\beta}]=0.
\end{equation}
For the Casimir operator, we have
\begin{equation}
  \label{eq:67}
  \begin{split}
  &\mathcal O^{2}\equiv \mathcal O_{1}^{2}+\mathcal O_{2}^{2}+\mathcal O_{3}^{2},\quad
  [\mathcal O^{2},\mathcal O_{\alpha}]=0,\quad
  \mathcal O_{-}\mathcal O_{+}=\mathcal O^{2}-\mathcal O_{3}^{2}-\mathcal O_{3},\\ & \mathcal O_{+}\mathcal O_{-}=\mathcal O^{2}-\mathcal O_{3}^{2}+O_{3},\quad \mathcal O^{2}=\frac 12(\mathcal O_{-}\mathcal O_{+}+\mathcal O_{+}\mathcal O_{-})+\mathcal O_{3}^{2}.
\end{split}
\end{equation}

Lie algebra indices $\alpha,\beta,\dots$ are raised and lowered with the invariant tensors
$\delta^{\alpha\beta},\delta_{\alpha\beta}$, so that adjoint and coadjoint representation can be identified,
$e_{*}^{\alpha}=-\frac{\imath}{2}\sigma^{\alpha}$, and the pairing between $\mathfrak{su}(2)$ and
$\mathfrak{su}{(2)}^{*}$ can be written as
\begin{equation}
  \label{eq:103}
  \langle \mu,\xi\rangle=\langle \mu_{\alpha}e_{*}^{\alpha},\xi^{\beta}e_{\beta}\rangle=\mu_{\alpha}\xi^{\alpha}=-2{\rm Tr}(\mu\xi)=\frac 12 {\rm Tr}(\mu_{\alpha}\sigma^{\alpha}\xi^{\beta}\sigma_{\beta}).
\end{equation}
The associated bi-invariant metric on the group is
\begin{equation}
  \label{eq:55}
ds^{2}=-2{\rm Tr}(g^{{-1}}dg g^{-1}dg)=g_{ij}dg^{i}dg^{j},\quad  g_{ij}=\delta_{\alpha\beta}R\indices{^{\alpha}_{i}}R\indices{^{\beta}_{j}}=\delta_{\alpha\beta}L\indices{^{\alpha}_{i}}L\indices{^{\beta}_{j}}.
\end{equation}
For later use, the norm and unit vector associated to a Lie algebra (co-)~vector
$\mu$ are denoted by
\begin{equation}
  \label{eq:212}
  \boldsymbol{\mu}=\sqrt{\mu_{\alpha}\mu^{\alpha}},\quad \hat \mu^{\alpha}=\frac{\mu^{\alpha}}{\boldsymbol{\mu}}.
\end{equation}

\subsection{Exponential parametrization}\label{sec:expon-param}

Coordinates on $\mathrm{SU}(2)$ are $g^{\alpha}\equiv\omega^{\alpha}$, with
\begin{equation}
  \label{eq:25}
  \begin{split}
&  \omega=\sqrt{\omega^{\alpha}\omega_{\alpha}},\quad
\hat \omega^{\alpha}=\frac{\omega^{\alpha}}{\omega},\quad d\omega=\hat\omega_{\alpha}d\omega^{\alpha},\quad d\hat \omega^{\alpha}=(\delta^{\alpha}_{\beta}-\hat \omega^{\alpha}\hat \omega_{\beta})\frac{d \omega^{\beta}}{\omega},\\
&  0\leq \omega< 4\pi,\ \hat \omega^{\alpha}\in \mathbb S^{2}: \hat\omega^{1}+\imath\hat\omega^{2}=\sin\theta e^{\imath\varphi},\ \hat \omega^{3}=\cos\theta,\ 0< \theta < \pi,\ 0\leq \varphi<2\pi,\\
&    g=e^{\omega^{\alpha}e_{\alpha}}=\cos{\frac{\omega}{2}}\sigma_{0} +2\sin{\frac{\omega}{2}}\hat\omega^{\alpha}e_{\alpha},\\ & g^{-1}=g^{\dagger}=e^{-\omega^{\alpha}e_{\alpha}}=\cos{\frac{\omega}{2}}\sigma_{0} -2\sin{\frac{\omega}{2}}\hat\omega^{\alpha}e_{\alpha}.
  \end{split}
\end{equation}
It follows that
\begin{equation}
  \begin{split}
  \label{eq:23} & dg=-\frac 12 \sin\frac{\omega}{2}\hat\omega_{\beta}d\omega^{\beta}\sigma_{0}+\big[\cos\frac{\omega}{2}\hat\omega^{\alpha}\hat\omega_{\beta}d\omega^{\beta}+\frac{2\sin\frac{\omega}{2}}{\omega}(\delta^{\alpha}_{\beta}- \hat \omega^{\alpha}\hat \omega_{\beta})d\omega^{\beta}\big]e_{\alpha},\\
& g^{-1}dg=\big[ \hat\omega^{\gamma}\hat\omega_{\beta}d\omega^{\beta}+\frac{2\cos\frac{\omega}{2}\sin\frac{\omega}{2}}{\omega}(\delta^{\gamma}_{\beta}-\hat\omega^{\gamma}\hat\omega_{\beta})d\omega^{\beta} - \frac{2\sin^{2}\frac{\omega}{2}}{\omega}\epsilon\indices{^{\gamma}_{\alpha\beta}}\hat\omega^{\alpha}d\omega^{\beta}\big]e_{\gamma},\\
&  dg g^{-1}=\big[ \hat\omega^{\gamma}\hat\omega_{\beta}d\omega^{\beta}+\frac{2\cos\frac{\omega}{2}\sin\frac{\omega}{2}}{\omega}(\delta^{\gamma}_{\beta}-\hat\omega^{\gamma}\hat\omega_{\beta})d\omega^{\beta}
  +\frac{2\sin^{2}\frac{\omega}{2}}{\omega}\epsilon\indices{^{\gamma}_{\alpha\beta}}\hat\omega^{\alpha}d\omega^{\beta}\big]e_{\gamma},
  \end{split}
\end{equation}
and thus
\begin{equation}
  \label{eq:24}
  \begin{split}
    & L\indices{^{\gamma}_{\beta}}=\hat\omega^{\gamma}\hat\omega_{\beta}
      +\frac{2\cos\frac{\omega}{2}\sin\frac{\omega}{2}}{\omega}
      (\delta^{\gamma}_{\beta}-\hat\omega^{\gamma}\hat\omega_{\beta})
      -\frac{2\sin^{2}\frac{\omega}{2}}{\omega}\epsilon\indices{^{\gamma}_{\delta\beta}}
      \hat\omega^{\delta},\\
    & L\indices{_{\alpha}^{\beta}}=\hat\omega^{\beta}\hat\omega_{\alpha}
      +\frac{\omega\cos\frac{\omega}{2}}{2\sin\frac{\omega}{2}}
      (\delta^{\beta}_{\alpha}-\hat\omega^{\beta}\hat\omega_{\alpha})-
      \frac{\omega}{2}\epsilon\indices{^{\beta}_{\alpha\sigma}}\hat\omega^{\sigma},\\
    & R\indices{^{\gamma}_{\beta}}=\hat\omega^{\gamma}\hat\omega_{\beta}
      +\frac{2\cos\frac{\omega}{2}\sin\frac{\omega}{2}}{\omega}
      (\delta^{\gamma}_{\beta}-\hat\omega^{\gamma}\hat\omega_{\beta})
      +\frac{2\sin^{2}\frac{\omega}{2}}{\omega}
      \epsilon\indices{^{\gamma}_{\delta\beta}}\hat\omega^{\delta},\\
    &  R\indices{_{\alpha}^{\beta}}
      =\hat\omega^{\beta}\hat\omega_{\alpha}
      +\frac{\omega\cos\frac{\omega}{2}}{2\sin\frac{\omega}{2}}
      (\delta^{\beta}_{\alpha}-\hat\omega^{\beta}\hat\omega_{\alpha})+
      \frac{\omega}{2}\epsilon\indices{^{\beta}_{\alpha\sigma}}\hat\omega^{\sigma},\\
    & \mathcal R(\hat \omega,\omega)\indices{^{\gamma}_{\alpha}}
      =R\indices{^{\gamma}_{\beta}}L\indices{_{\alpha}^{\beta}}=
                  \hat\omega^{\gamma}\hat \omega_{\alpha}+\cos{\omega}
                  (\delta^{\gamma}_{\alpha}-\hat\omega^{\gamma}\hat\omega_{\alpha})
                  -\sin{\omega}\epsilon\indices{^{\gamma}_{\alpha\delta}}\hat\omega^{\delta}.
 \end{split}
\end{equation}
with
$\mathcal R(\hat \omega,\omega)\indices{_{\alpha}^{\gamma}}=L\indices{^{\gamma}_{\beta}}R\indices{_{\alpha}^{\beta}}
=R\indices{_{\alpha\beta}}L\indices{^{\gamma\beta}}$.

\subsection{Adapted Euler angles}\label{sec:adapted-euler-angles-1}

In this parametrization $g^{i}=\psi,\theta,\phi$, with
$g=e^{-\frac{\psi}{2}\imath\sigma_{3}}e^{-\frac{\theta}{2}\imath\sigma_{2}}e^{-\frac{\phi}{2}\imath\sigma_{3}}$, and
$0\leq \psi<4\pi$, $0\leq \phi<2\pi$,  $0< \theta< \pi$. It follows that
\begin{multline}
  \label{eq:46Bx}
    dg g^{-1}=-\imath\sigma_{3}(d\frac{\psi}{2}+\cos\theta d\frac{\phi}{2})-\imath \sigma_{2}(\cos\psi d\frac{\theta}{2}+\sin\psi\sin\theta d\frac{\phi}{2})\\-\imath\sigma_{1}(-\sin\psi d\frac{\theta}{2}+\cos\psi\sin \theta d\frac{\phi}{2}),
\end{multline}
\begin{multline}
    g^{-1}dg=-\imath\sigma_{3}(d\frac{\phi}{2}+\cos\theta d\frac{\psi}{2})-\imath \sigma_{2}(\cos\phi d\frac{\theta}{2}+\sin\theta\sin\phi d\frac{\psi}{2})\\-\imath\sigma_{1}(\sin\phi d\frac{\theta}{2}-\cos\phi\sin \theta d\frac{\psi}{2}),
\end{multline}
\begin{equation}
  \label{eq:47Bx}
  R\indices{^{\alpha}_{i}}=\begin{pmatrix} 0 & -\sin \psi & \cos \psi\sin \theta
    \\ 0& \cos \psi & \sin \psi \sin \theta\\ 1 & 0 & \cos \theta\\
  \end{pmatrix},\quad R\indices{_{\beta}^{i}}=\begin{pmatrix} -\frac{\cos \psi\cos \theta}{\sin \theta} & -\sin \psi & \frac{\cos \psi}{\sin \theta}
    \\ -\frac{\sin \psi\cos \theta}{\sin \theta}& \cos \psi & \frac{\sin \psi}{\sin \theta}\\ 1 & 0 & 0\\
  \end{pmatrix},
\end{equation}
\begin{equation}
L\indices{^{\alpha}_{i}}=\begin{pmatrix} -\cos \phi\sin \theta & \sin \phi & 0
    \\ \sin \phi \sin \theta& \cos \phi & 0 \\ \cos \theta & 0 & 1\\
  \end{pmatrix},\quad L\indices{_{\beta}^{i}}=\begin{pmatrix} -\frac{\cos \phi}{\sin \theta} & \sin \phi & \frac{\cos \phi\cos \theta}{\sin \theta}
    \\ \frac{\sin \phi}{\sin \theta} & \cos \phi & -\frac{\sin \phi\cos \theta}{\sin \theta}\\ 0 & 0 & 1\\
  \end{pmatrix},
\end{equation}
\begin{multline}
  \label{eq:49Bx}
  R\indices{^{\alpha}_{i}}L\indices{_{\beta}^{i}}=\\
\begin{pmatrix}
 \cos \psi \cos \theta \cos \phi-\sin \psi \sin \phi & -\cos \psi \cos \theta \sin \phi-\sin \psi \cos \phi & \cos \psi \sin \theta \\
 \sin \psi\cos \theta \cos \phi+\cos \psi \sin \phi & \cos \psi \cos \phi-\sin \psi \cos \theta \sin \phi & \sin \psi \sin \theta \\
 -\sin \theta \cos \phi & \sin \theta \sin \phi & \cos \theta \\
\end{pmatrix},
\end{multline}
with $L\indices{^{\alpha}_{i}}R\indices{_{\beta}^{i}}=R\indices{_{\beta i}}L\indices{^{\alpha i}}$.

The metric and its inverse are
\begin{equation}
  \label{eq:72}
  ds^{2}=(d\psi^{2}+d\theta^{2}+d\phi^{2}+2\cos\theta d\psi d\phi ),\quad g^{ij}=\frac{1}{\sin^{2}\theta}\begin{pmatrix} 1 & 0 & -\cos\theta \\ 0 & \sin ^2 \theta & 0\\ -\cos\theta & 0 & 1
  \end{pmatrix}.
\end{equation}
and $\mathbf g=\sin^{2}\theta$,
and the Laplacian is
\begin{equation}
  \label{eq:82}
  \Delta_{SU(2)}=\frac{1}{\sin^{2}\theta}[\partial^{2}_{\psi}-2\cos\theta \partial_{\psi}\partial_{\phi}+\sin ^2 \theta \partial ^2 _\theta + \sin \theta \cos \theta \partial _\theta +\partial_{\phi}^{2}].
\end{equation}

In order to relate to the discussion in Section~\ref{sec:reduced-phase-space},
where well-defined expressions for the potential one-forms on the coadjoint
orbits in the patches containing the north and south poles have been
constructed, one considers a gauge transformation by the element
\begin{equation}
  \label{eq:231}
  h_{\tilde\epsilon}=e^{\frac{\tilde\epsilon\phi}{2}\imath\sigma_{3}},
\end{equation}
whose effect on the group element~\eqref{eq:99x} is the shift $\psi\to \psi-\tilde\epsilon\phi$. It follows that the gauge conditions $\chi_{\tilde\epsilon}=\psi+\tilde\epsilon\phi=0$ that avoid the Gribov ambiguities before the gauge transformation, are replaced by $\psi=0$ after the gauge transformations.
The effect of this gauge transformation is to replace $a$ in~\eqref{eq:26} by $a_{\tilde\epsilon}$ in~\eqref{eq:97} while the Hamiltonian in~\eqref{eq:52} is unchanged. More generally, left and right invariant Maurer-Cartan forms and vector fields have now to be computed from
\begin{equation}
  \label{eq:99g}
  g_{\tilde\epsilon}=e^{-\frac{\psi-\tilde\epsilon\phi}{2}\imath\sigma_{3}}
  e^{-\frac{\theta}{2}\imath\sigma_{2}}e^{-\frac{\phi}{2}\imath\sigma_{3}}
  =\begin{pmatrix} \cos\frac{\theta}{2}e^{-\imath\frac{\psi+\phi(1-\tilde\epsilon)}{2}} & -\sin\frac{\theta}{2}e^{-\imath\frac{\psi-\phi(1+\tilde\epsilon)}{2}} \\
    \sin\frac{\theta}{2}e^{\imath\frac{\psi-\phi(1+\tilde\epsilon)}{2}} & \cos\frac{\theta}{2}e^{\imath\frac{\psi+\phi(1-\tilde\epsilon)}{2}}
\end{pmatrix}.
\end{equation}
We will do so in complex parametrization in Section~\ref{sec:compl-param}.

\subsection{Euler-Rodrigues parametrization}\label{sec:euler-rodr-param}

The implicit Euler-Rodrigues parametrization of $\mathrm{SU}(2)$ given in~\eqref{eq:138}
is related to the explicit parametrization in terms of adapted Euler angles in~\eqref{eq:99x}
through
\begin{equation}
  \label{eq:140}
  \begin{split}
    & \alpha^{0}=\cos\frac{\theta}{2}\cos\frac{\psi+\phi}{2},\quad \alpha^{3}=\cos\frac{\theta}{2}\sin\frac{\psi+\phi}{2},
    \\ & \alpha^{2}=\sin\frac{\theta}{2}\cos\frac{\psi-\phi}{2},\quad \alpha^{1}=-\sin\frac{\theta}{2}\sin\frac{\psi-\phi}{2}.
  \end{split}
\end{equation}
In the Euler-Rodrigues parametrization, the $SU(2)$ group law becomes the
quaterionic composition rule,
\begin{equation}
  \label{eq:136}
  g(\alpha^{\prime0},\alpha^{\prime\beta})g(\alpha^{0},\alpha^{\beta})=g(\alpha^{\prime\prime0},\alpha^{\prime\prime\beta}),\quad
  \bar g(\alpha^{\prime0},\alpha^{\prime\beta})\bar g(\alpha^{0},\alpha^{\beta})=\bar g(\alpha^{\prime\prime0},\alpha^{\prime\prime\beta}),
\end{equation}
where\footnote{cf.~equation (2.42) of~\cite{biedenharn1981angular}}
\begin{equation}
  \label{eq:137a}
  \alpha^{\prime\prime0}=\alpha^{\prime0}\alpha^{0}-\alpha^{\prime\beta}\alpha_{\beta},\quad  \alpha^{\prime\prime\beta}=\alpha^{0}\alpha^{\prime \beta} +\alpha^{\prime 0}\alpha^{\beta}+\epsilon\indices{^{\beta}_{\gamma\delta}}\alpha^{\prime\gamma}\alpha^{\delta}.
\end{equation}

For any explicit parametrization $\alpha^{A}=\alpha^{A}(g^{i})$ of $\mathbb S^{3}$, let $g_{ij}=
\frac{\partial \alpha^{A}}{\partial g^{i}}\frac{\partial \alpha_{}{A}}{\partial g^{j}}$ be the induced metric with inverse $g^{jk}$. We have
\begin{equation}
  \label{eq:149}
  \delta^{A}_{B}=\alpha^{A}\alpha_{B}+\frac{\partial\alpha^{A}}{\partial g^{i}}g^{ij}\frac{\partial\alpha_{B}}{\partial g^{j}},
\end{equation}
and, as a consequence of the constraint~\eqref{eq:127},
\begin{equation}
  \label{eq:148}
  \alpha_{A}d\alpha^{A}=0=ad\bar a+\bar a da+bd\bar b+\bar bdb.
\end{equation}

When using that the inverse group element is
\begin{equation}
  \label{eq:142a}
  g^{-1}=\begin{pmatrix} a & -\bar b\\ b & \bar a
  \end{pmatrix},
\end{equation}
and choosing as Lie algebra basis $e_{\alpha}=-\frac 12\imath\sigma_{\alpha}$,
we get
\begin{equation}
  \label{eq:143a}
  \begin{split}
    & dg g^{-1}=
      \begin{pmatrix} d\bar a a+d\bar b b & d\bar b\bar a-d\bar a\bar b
        \\ da b-db a & da\bar a+db\bar b
  \end{pmatrix}=e_{\alpha}R\indices{^{\alpha}_{i}}dg^{i},\\
    & g^{-1}dg =
      \begin{pmatrix} ad\bar a+\bar b
        d b & ad\bar b-\bar bda
        \\ bd\bar a-\bar a db & \bar ada+bd\bar b
  \end{pmatrix}=e_{\alpha}L\indices{^{\alpha}_{i}}dg^{i},
  \end{split}
\end{equation}
with
\begin{equation}
  \label{eq:152}
  R\indices{^{\alpha}_{i}}=R\indices{^{\alpha}_{B}}\frac{\partial \alpha^{B}}{\partial g^{i}},\quad L\indices{^{\alpha}_{i}}=L\indices{^{\alpha}_{B}}\frac{\partial \alpha^{B}}{\partial g^{i}}
\end{equation}
\begin{equation}
  \label{eq:144a}
  R\indices{^{\alpha}_{B}}=2
  \begin{pmatrix} -\alpha^{1} & \alpha^{0} & -\alpha^{3} & \alpha^{2}\\
    -\alpha^{2} & \alpha^{3} & \alpha^{0} & -\alpha^{1} \\
    -\alpha^{3} & -\alpha^{2} & \alpha^{1} & \alpha^{0}
  \end{pmatrix},\quad L\indices{^{\alpha}_{B}}=2
  \begin{pmatrix} -\alpha^{1} & \alpha^{0} & \alpha^{3} & -\alpha^{2}\\
    -\alpha^{2} & -\alpha^{3} & \alpha^{0} & \alpha^{1} \\
    -\alpha^{3} & \alpha^{2} & -\alpha^{1} & \alpha^{0}
  \end{pmatrix},
\end{equation}
Let $L\indices{^{B}_{\beta}}\equiv \frac 14 {(L\indices{^{\beta}_{B}})}^{T}$. We have $L\indices{^{\alpha}_{B}}L\indices{^{B}_{\beta}}=\delta^{\alpha}_{\beta}$. Note that
\begin{equation}
  \label{eq:150}
  L\indices{^{\alpha}_{B}}\alpha^{B}=0=\alpha_{B}L\indices{^{B}_{\beta}}.
\end{equation}
It follows that
\begin{equation}
  \label{eq:151}
  L\indices{_{\beta}^{i}}=g^{ij}\frac{\partial \alpha_{B}}{\partial g^{j}}L\indices{^{B}_{\beta}}.
\end{equation}
and\footnote{cf.~equation (2.22) of~\cite{biedenharn1981angular}}
\begin{multline*}
  R\indices{^{\alpha}_{i}}L\indices{_{\beta}^{i}}=\\
  \begin{pmatrix}
    {(\alpha^{0})}^{2}+{(\alpha^{1})}^{2}-{(\alpha^{2})}^{2}-{(\alpha^{3})}^{2} & 2(-\alpha^{0}\alpha^{3}+\alpha^{1}\alpha^{2}) & 2(\alpha^{0}\alpha^{2}+\alpha^{1}\alpha^{3})\\
    2(\alpha^{0}\alpha^{3}+\alpha^{1}\alpha^{2}) & {(\alpha^{0})}^{2}-{(\alpha^{1})}^{2}+{(\alpha^{2})}^{2}-{(\alpha^{3})}^{2} & 2(-\alpha^{0}\alpha^{1}+\alpha^{2}\alpha^{3}) \\
    2(-\alpha^{0}\alpha^{2}+\alpha^{1}\alpha^{3}) & 2(\alpha^{0}\alpha^{1}+\alpha^{2}\alpha^{3}) & {(\alpha^{0})}^{2}-{(\alpha^{1})}^{2}-{(\alpha^{2})}^{2}+{(\alpha^{3})}^{2}
  \end{pmatrix}\\=
  \begin{pmatrix}
    \frac 12 (a^{2}+\bar a^{2} -b^{2}-b^{2})& \frac{\imath}{2}(a^{2}-\bar a^{2}+b^{2}-\bar b^{2}) & -(ab+\bar a\bar b)\\
    -\frac{\imath}{2}(a^{2}-\bar a^{2}-b^{2}+\bar b^{2}) & \frac 12 (a^{2}+\bar a^{2} +b^{2}+\bar b^{2}) & \imath (ab-\bar a\bar b) \\
     a\bar b+\bar ab & \imath(a\bar b-\bar ab) & a\bar a-b\bar b
  \end{pmatrix}.
\end{multline*}

\subsection{Complex parametrizations}\label{sec:compl-param}

In these parametrizations, one uses
$g^{i}=\psi,\zeta_{\tilde\epsilon},\widebar\zeta_{\tilde\epsilon}$ in terms of
the complex coordinates of~\eqref{eq:77}. The group
elements~\eqref{eq:99g} become
\begin{equation}
  \label{eq:99y}
  g_{-}=P_{-}^{{-\frac 12}}
  \begin{pmatrix} \widebar\zeta_{-}\, e^{-\imath\frac{\psi}{2}} & - e^{-\imath\frac{\psi}{2}} \\
      e^{\imath\frac{\psi}{2}} & \zeta_{-}\,e^{\imath\frac{\psi}{2}}
\end{pmatrix},\quad
  g_{+}=P_{+}^{{-\frac 12}}\begin{pmatrix} e^{-\imath\frac{\psi}{2}} &
      -\widebar\zeta_{+}\, e^{-\imath\frac{\psi}{2}}\\
    \zeta_{+}\,e^{\imath\frac{\psi}{2}}  &
                                        e^{\imath\frac{\psi}{2}}
\end{pmatrix}.
\end{equation}
In terms of the Euler-Rodrigues parametrization, we have
\begin{equation}
  \label{eq:173}
  a_{-}=P_{-}^{{-\frac 12}}{\zeta_{-}e^{\imath\frac{\psi}{2}}},\quad
  b_{-}=-P_{-}^{{-\frac 12}}{e^{\imath\frac{\psi}{2}}},\quad
  a_{+}=P_{+}^{{-\frac 12}}{e^{\imath\frac{\psi}{2}}},\quad b_{+}
  =-P_{+}^{{-\frac 12}}{\zeta_{+}e^{\imath\frac{\psi}{2}}}.
\end{equation}
All formulas can be obtained from the previous section by substitution. For
later use, we provide the explicit expressions,
\begin{multline}
  \label{eq:47Bxd}
  R\indices{^{\alpha}_{i}}=P^{-1}_{\tilde\epsilon}\begin{pmatrix} 0 &
    {\imath \tilde\epsilon e^{\imath\psi}} & {-\imath \tilde\epsilon e^{-\imath\psi}}
    \\ 0& \tilde\epsilon{e^{\imath\psi}} & \tilde\epsilon {e^{-\imath\psi}}\\ P_{\tilde\epsilon} & {-\imath \widebar\zeta_{\tilde\epsilon}} & {\imath \zeta_{\tilde\epsilon}}\\
  \end{pmatrix},\\ R\indices{_{\beta}^{i}}
  =\frac 12 \begin{pmatrix}\tilde\epsilon\widebar\zeta_{\tilde\epsilon}e^{-\imath\psi}+\tilde\epsilon\zeta_{\tilde\epsilon}e^{{\imath\psi}} & -{\imath \tilde\epsilon e^{-\imath\psi}}P_{\tilde\epsilon} & {\imath \tilde\epsilon e^{\imath\psi}P_{\tilde\epsilon}} \\
   \imath \tilde\epsilon\widebar\zeta_{\tilde\epsilon}e^{-\imath\psi} -\imath \tilde\epsilon\zeta_{\tilde\epsilon}e^{\imath\psi}& \tilde\epsilon{e^{-\imath\psi}P_{\tilde\epsilon}} &
    \tilde\epsilon{e^{\imath\psi}P_{\tilde\epsilon}} \\ 2 & 0 & 0
  \end{pmatrix},
\end{multline}
\begin{multline}
L\indices{^{\alpha}_{i}}=P_{\tilde\epsilon}^{-1}\begin{pmatrix} -(\zeta_{\tilde\epsilon}+\widebar\zeta_{\tilde\epsilon}) & \imath & -\imath   \\
  -\imath \tilde\epsilon(\widebar\zeta_{\tilde\epsilon}-\zeta_{\tilde\epsilon}) & \tilde\epsilon & \tilde\epsilon\\
  \tilde\epsilon(1-\zeta_{\tilde\epsilon}\widebar\zeta_{\tilde\epsilon}) & \imath\tilde\epsilon\widebar\zeta_{\tilde\epsilon} & -\imath\tilde\epsilon\zeta_{\tilde\epsilon}
\end{pmatrix},\\
L\indices{_{\beta}^{i}}=\frac 12 \begin{pmatrix}
  -(\zeta_{\tilde\epsilon}+\widebar\zeta_{\tilde\epsilon}) & -\imath(1-\zeta_{\tilde\epsilon}^{2}) & \imath(1-\widebar\zeta^{2}_{\tilde\epsilon}) \\
  -\imath \tilde\epsilon(\widebar\zeta_{\tilde\epsilon}-\zeta_{\tilde\epsilon})& \tilde\epsilon(1+\zeta^{2}_{\tilde\epsilon}) & \tilde\epsilon(1+\widebar\zeta^{2}_{\tilde\epsilon}) \\
  \tilde\epsilon 2 & -2\imath\tilde\epsilon\zeta_{\tilde\epsilon} & 2\imath\tilde\epsilon\widebar\zeta_{\tilde\epsilon}
  \end{pmatrix}.
\end{multline}
\begin{equation}
  \label{eq:161}
  R\indices{^{\alpha}_{i}}L\indices{_{\beta}^{i}}
  =P_{\tilde\epsilon}^{-1}
  \begin{pmatrix}\frac{\tilde\epsilon e^{{\imath\psi}}(1-\zeta_{\tilde\epsilon}^{2})+\tilde\epsilon e^{{-\imath\psi}}(1-\widebar\zeta_{\tilde\epsilon}^{2})}{2}
    & \frac{\imath e^{{\imath\psi}}(1+\zeta_{\tilde\epsilon}^{2})-\imath e^{{-\imath\psi}}(1+\widebar\zeta_{\tilde\epsilon}^{2}) }{2}     & {\zeta_{\tilde\epsilon}e^{{\imath\psi}}+\widebar\zeta_{\tilde\epsilon}e^{{-\imath\psi}}}\\
      \frac{-\imath {\tilde\epsilon} e^{{\imath\psi}}(1-\zeta_{\tilde\epsilon}^{2})+\imath{\tilde\epsilon} e^{{-\imath\psi}}(1-\widebar\zeta_{\tilde\epsilon}^{2})}{2}
    & \frac{e^{{\imath\psi}}(1+\zeta_{\tilde\epsilon}^{2})+ e^{{-\imath\psi}}(1+\widebar\zeta_{\tilde\epsilon}^{2}) }{2}     & {\imath\widebar\zeta_{\tilde\epsilon}e^{{-\imath\psi}}-\imath \zeta_{\tilde\epsilon}e^{{\imath\psi}}}\\
    {-(\zeta_{\tilde\epsilon}+\widebar\zeta_{\tilde\epsilon})}  & {-\imath\tilde\epsilon (\widebar\zeta_{\tilde\epsilon}-\zeta_{\tilde\epsilon})} &
                                                                                              {\tilde\epsilon(1-\zeta_{\tilde\epsilon}\widebar\zeta_{\tilde\epsilon})}
  \end{pmatrix},
\end{equation}
with $L\indices{^{\alpha}_{i}}R\indices{_{\beta}^{i}}=R\indices{_{\beta i}}L\indices{^{\alpha i}}$.

Useful identities in complex parametrization are
\begin{equation}
  \label{eq:63}
  \begin{split}
    & P_{\tilde\epsilon}^{2}\partial_{\zeta_{\tilde\epsilon}}\big[P_{\tilde\epsilon}^{{-1}}{\zeta_{\tilde\epsilon}^{\frac{1\pm\tilde\epsilon}{2}}
      \widebar\zeta_{\tilde\epsilon}^{\frac{1\mp\tilde\epsilon}{2}}}\big]=  \frac{1\pm\tilde\epsilon}{2}\zeta_{\tilde\epsilon}^{\frac{-1\pm\tilde\epsilon}{2}}\widebar\zeta_{\tilde\epsilon}^{\frac{1\mp\tilde\epsilon}{2}}
  -\frac{1\mp\tilde\epsilon}{2}\zeta_{\tilde\epsilon}^{\frac{1\pm\tilde\epsilon}{2}}\widebar\zeta_{\tilde\epsilon}^{\frac{3\mp\tilde\epsilon}{2}}
    ={\pm\tilde\epsilon}\widebar\zeta_{\tilde\epsilon}^{{1\mp\tilde\epsilon}},\\
    & P_{\tilde\epsilon}^{2}\partial_{\widebar\zeta_{\tilde\epsilon}}\big[P_{\tilde\epsilon}^{{-1}}
      {\zeta_{\tilde\epsilon}^{\frac{1\pm\tilde\epsilon}{2}}\widebar\zeta_{\tilde\epsilon}^{\frac{1\mp\tilde\epsilon}{2}}}\big]=    \frac{1\mp\tilde\epsilon}{2}\zeta_{\tilde\epsilon}^{\frac{1\pm\tilde\epsilon}{2}}\widebar\zeta_{\tilde\epsilon}^{\frac{-1\mp\tilde\epsilon}{2}}
  -\frac{1\pm\tilde\epsilon}{2}\zeta_{\tilde\epsilon}^{\frac{3\pm\tilde\epsilon}{2}}\widebar\zeta_{\tilde\epsilon}^{\frac{1\mp\tilde\epsilon}{2}}
    ={\mp\tilde\epsilon}\zeta_{\tilde\epsilon}^{{1\pm\tilde\epsilon}},\\
& P_{\tilde\epsilon}^{{-1}}[\zeta_{\tilde\epsilon}^{\frac{1\pm \tilde\epsilon}{2}}
    \widebar\zeta_{\tilde\epsilon}^{\frac{1\mp \tilde\epsilon}{2}}\mp \tilde\epsilon \zeta_{\tilde\epsilon}\widebar\zeta_{\tilde\epsilon}^{1\mp \tilde\epsilon}]=\widebar\zeta^{\frac{1\mp\tilde\epsilon}{2}}_{\tilde\epsilon}\frac{1\mp\tilde\epsilon}{2},\quad   P_{\tilde\epsilon}^{{-1}}[\zeta_{\tilde\epsilon}^{\frac{1\pm \tilde\epsilon}{2}}
  \widebar\zeta_{\tilde\epsilon}^{\frac{1\mp \tilde\epsilon}{2}}\pm \tilde\epsilon \widebar\zeta_{\tilde\epsilon}\zeta_{\tilde\epsilon}^{1\pm \tilde\epsilon}]=\zeta^{\frac{1\pm\tilde\epsilon}{2}}_{\tilde\epsilon}\frac{1\pm\tilde\epsilon}{2}.
  \end{split}
\end{equation}

For integration on the sphere, with the notation as in the discussion
after~\eqref{eq:56}, we have
\begin{multline}
  \label{eq:66}
  \frac{1}{2\pi\imath}\int_{{\mathbb S^{2}}}\frac{d\zeta_{\tilde\epsilon}\wedge d\widebar\zeta_{\tilde\epsilon}}{P_{\tilde\epsilon}^{2}}=\lim_{\varepsilon\to 0^{+}}-\frac{1}{2\pi\imath}
  \int_{U_{\tilde\epsilon}^{\frac{\pi}{2}-\varepsilon}}d\wedge \partial \ln P_{\tilde\epsilon}\\=\lim_{\varepsilon\to 0^{+}}\frac{1}{2\pi\imath}\oint_{\partial U_{\tilde\epsilon}^{\frac{\pi}{2}-\varepsilon}}d\zeta_{\tilde\epsilon}\frac{\bar\zeta_{\tilde\epsilon}}{P_{\tilde\epsilon}}=
  \frac{1}{2\pi\imath}\int^{2\pi}_{0}  \imath e^{-\tilde\epsilon \imath\phi}d\phi e^{\tilde\epsilon\imath\phi}=1.
\end{multline}
Note that the orientation of curve is decreasing $\phi$ if the cap that is
excluded involves the north pole, $\tilde\epsilon =-1$, but increasing $\phi$ if
the cap that is excluded involves the south pole, $\tilde\epsilon=+1$. More
generally, for $-j\leq m,m'\leq j$ and $m,m',j$ integer or half-integer,
\begin{equation}
  \label{eq:68}
  \boxed{\frac{1}{2\pi\imath}\int_{{\mathbb S^{2}}}\frac{d\zeta_{\tilde\epsilon}\wedge d\widebar\zeta_{\tilde\epsilon}\, \widebar\zeta_{\tilde\epsilon}^{j-\tilde\epsilon m'}\zeta_{\tilde\epsilon}^{j-\tilde\epsilon m}}{P_{\tilde\epsilon}^{2j+2}}
  =\delta_{m,m'}\frac{(j-\tilde\epsilon m)!(j+\tilde\epsilon m)!}{(2j+1)!}}.
\end{equation}
Indeed, when going back to spherical coordinates, it follows directly that the
result vanishes unless $m=m'$. Furthermore, for $j-\tilde\epsilon m>0$,
\begin{multline}
  \label{eq:64}
  \frac{d\zeta_{\tilde\epsilon}\wedge d\widebar\zeta_{\tilde\epsilon}\,{(\zeta_{\tilde\epsilon}\widebar\zeta_{\tilde\epsilon})}^{j-\tilde\epsilon m}}{P_{\tilde\epsilon}^{2j+2}}
  =  -\frac{1}{2j+1}d\wedge \big[d\widebar\zeta_{\tilde\epsilon} \frac{\widebar\zeta^{{-1}}_{\tilde\epsilon}{(\zeta_{\tilde\epsilon}\widebar\zeta_{\tilde\epsilon})}^{j-\tilde\epsilon m}}{P_{\tilde\epsilon}^{2j+1}}\big]\\
  +\frac{j-\tilde\epsilon m}{2j+1}\frac{d\zeta_{\tilde\epsilon}\wedge d\widebar\zeta_{\tilde\epsilon}
    \,{(\zeta_{\tilde\epsilon}\widebar\zeta_{\tilde\epsilon})}^{j-\tilde\epsilon m-1}}{P_{\tilde\epsilon}^{2j+1}}.
\end{multline}
When integrating the first term over the sphere and using Stokes' theorem as
before, we now get $-\frac{1}{2j+1}\lim_{\varepsilon\to 0^{+}}{(\frac{\varepsilon}{2})}^{2j+2\tilde\epsilon m+2}=0$.
When repeating the reasoning for the last term until $j-\tilde\epsilon m$ drops to zero, one ends up with
\begin{equation}
  \label{eq:69}
  \delta_{m,m'}\frac{(j-\tilde\epsilon m)!}{(2j+1)(2j)\dots(j+\tilde\epsilon m+2)}\frac{1}{2\pi\imath}\int_{{\mathbb S^{2}}}\frac{d\zeta_{\tilde\epsilon}\wedge d\widebar\zeta_{\tilde\epsilon}}{P_{\tilde\epsilon}^{j+\tilde\epsilon m+2}}.
\end{equation}
The result~\eqref{eq:68} follows because
\begin{equation}
  \label{eq:70}
  \boxed{\frac{1}{2\pi\imath}\int_{{\mathbb S^{2}}}\frac{d\zeta_{\tilde\epsilon}\wedge d\widebar\zeta_{\tilde\epsilon}}{P_{\tilde\epsilon}^{j+\tilde\epsilon m+2}}=\frac{1}{j+\tilde\epsilon m+1}},
\end{equation}
which in turn is shown by induction: it holds for $j+\tilde\epsilon m=0$ on account of~\eqref{eq:66},
while
\begin{equation}
  \label{eq:104}
  \frac{1}{P_{\tilde\epsilon}^{j+\tilde\epsilon m+3}}=
  \frac{1}{P_{\tilde\epsilon}^{j+\tilde\epsilon m+1}}\partial_{\zeta_{\tilde\epsilon}}\frac{\zeta_{\tilde\epsilon}}{1+\zeta_{\tilde\epsilon}\widebar\zeta_{\tilde\epsilon}}=\partial_{\zeta_{\tilde\epsilon}}
  \frac{\zeta_{\tilde\epsilon}}{P_{\tilde\epsilon}^{j+\tilde\epsilon m+2}}+(j+\tilde\epsilon m+1)\frac{\zeta_{\tilde\epsilon}\widebar\zeta_{\tilde\epsilon}}{P_{\tilde\epsilon}^{j+\tilde\epsilon m+3}}.
\end{equation}
When writing the numerator of the last term as $\zeta_{\tilde\epsilon}\widebar\zeta_{\tilde\epsilon}=1+\zeta_{\tilde\epsilon}\widebar\zeta_{\tilde\epsilon}-1$, and moving the last term to the left hand side, this implies
\begin{equation}
  \label{eq:118}
  (j+\tilde\epsilon m+2)\frac{1}{P_{\tilde\epsilon}^{j+\tilde\epsilon m+3}}
  =\partial_{\zeta_{\tilde\epsilon}}\frac{\zeta_{\tilde\epsilon}}{P_{\tilde\epsilon}^{j+\tilde\epsilon m+2}}+(j+\tilde\epsilon m+1)\frac{1}{P_{\tilde\epsilon}^{j+\tilde\epsilon m+2}}.
\end{equation}
Integrating over the sphere, the boundary term again vanishes by using Stokes' theorem, while the integral on the right hand side gives $1$ by induction, so that the result holds for $j+\tilde\epsilon m+1$.

\section{Nonzero quaternions}\label{sec:non-unim-quat}

\subsection{Invariant vector fields, Maurer-Cartan forms, (co) adjoint
  representation}\label{sec:left-right-invariant-h}

We consider the Lie group of (non-unimodular) nonzero quaternions
$\mathbb H^{*}$ represented by \eqref{eq:128}. Indices $A,B,\dots$ are lowered and
raised with $\delta_{AB},\delta^{AB}$. We have
\begin{equation}
  \label{eq:142}
  g^{-1}=\frac{1}{R^{2}}\begin{pmatrix} z_{1} & -\bar z_{2}\\ z_{2} & \bar z_{1}
  \end{pmatrix},\quad R^{2}=x^{A}x_{A}=|z_{1}|^{2}+|z_{2}|^{2}.
\end{equation}
Take as a basis for the reductive Lie algebra $e_{A}=(\frac 12\sigma_{0},-\frac 12 \imath\sigma_{\alpha})$, with structure constants
\begin{equation}
  \label{eq:164}
  [e_{0},e_{\alpha}]=0,\quad [e_{ \alpha},e_{\beta}]=\epsilon\indices{^{ \gamma}_{ \alpha\beta}}e_{\gamma}.
\end{equation}
From
\begin{equation}
  \label{eq:143}
  \begin{split}
    & dg g^{-1}=\frac{1}{R^{2}}
      \begin{pmatrix} d\bar z_{1}z_{1}+d\bar z_{2}z_{2} & d\bar z_{2}\bar z_{1}-d\bar z_{1}\bar z_{2}
        \\ dz_{1}z_{2}-dz_{2}z_{1} & dz_{1}\bar z_{1}+dz_{2}\bar z_{2}
  \end{pmatrix}=e_{A}R\indices{^{A}_{B}}dx^{B},\\
    & g^{-1}dg = \frac{1}{R^{2}}
      \begin{pmatrix} z_{1}d\bar z_{1}+\bar z_{2}
        d z_{2} & z_{1}d\bar z_{2}-\bar z_{2}dz_{1}
        \\ z_{2}d\bar z_{1}-\bar z_{1} dz_{2} & \bar z_{1}dz_{1}+z_{2}d\bar z_{2}
  \end{pmatrix}=e_{A}L\indices{^{A}_{B}}dx^{B},
  \end{split}
\end{equation}
it follows that
\begin{equation}
  \label{eq:144}
  R\indices{^{A}_{B}}=\frac{2}{R^{2}}
  \begin{pmatrix} x^{0} & x^{1} & x^{2} & x^{3} \\ -x^{1} & x^{0} & -x^{3} & x^{2}\\
    -x^{2} & x^{3} & x^{0} & -x^{1} \\
    -x^{3} & -x^{2} & x^{1} & x^{0}
  \end{pmatrix},\quad L\indices{^{A}_{B}}=\frac{2}{R^{2}}
  \begin{pmatrix} x^{0} & x^{1} & x^{2} & x^{3} \\ -x^{1} & x^{0} & x^{3} & -x^{2}\\
    -x^{2} & -x^{3} & x^{0} & x^{1} \\
    -x^{3} & x^{2} & -x^{1} & x^{0}
  \end{pmatrix},
\end{equation}
We have
\begin{equation}
  \label{eq:159}
  R\indices{_{A}^{B}}=\frac 12
  \begin{pmatrix} x^{0} & x^{1} & x^{2} & x^{3} \\ -x^{1} & x^{0} & -x^{3} & x^{2}\\
    -x^{2} & x^{3} & x^{0} & -x^{1} \\
    -x^{3} & -x^{2} & x^{1} & x^{0}
  \end{pmatrix},\quad L\indices{_{A}^{B}}=\frac 12
  \begin{pmatrix} x^{0} & x^{1} & x^{2} & x^{3} \\ -x^{1} & x^{0} & x^{3} & -x^{2}\\
    -x^{2} & -x^{3} & x^{0} & x^{1} \\
    -x^{3} & x^{2} & -x^{1} & x^{0}
  \end{pmatrix}.
\end{equation}
In particular,
\begin{equation}
  \label{eq:177}
  \begin{split}
    R\indices{_{\pm}^{A}}=\frac 12(-x^{1}\mp \imath x^{2},x^{0}\pm \imath x_{3},-x^{3}\pm\imath x^{0},x^{2}\mp\imath x^{1}),\\
    R\indices{_{+}^{A}}=\frac 12(\imath z_{2},z_{1},\imath z_{1},-z_{2}),\quad R\indices{_{-}^{A}}=\frac 12(-\imath \bar z_{2},\bar z_{1},-\imath\bar z_{1},-\bar z_{2}),\\
    L\indices{_{+}^{A}}=\frac 12(\imath z_{2},\bar z_{1},\imath \bar z_{1},z_{2}),\quad L\indices{_{-}^{A}}=\frac 12 (-\imath \bar z_{2}, z_{1},-\imath  z_{1},\bar z_{2}).
  \end{split}
\end{equation}
\begin{multline}
  \label{eq:145}
  R\indices{^{A}_{C}}L\indices{_{B}^{C}}=\frac{1}{R^{2}}\\
  \begin{pmatrix} R^{2} & 0 & 0 & 0 \\
    0 & {(x^{0})}^{2}+{(x^{1})}^{2}-{(x^{2})}^{2}-{(x^{3})}^{2} & 2(-x^{0}x^{3}+x^{1}x^{2}) & 2(x^{0}x^{2}+x^{1}x^{3})\\
    0 & 2(x^{0}x^{3}+x^{1}x^{2}) & {(x^{0})}^{2}-{(x^{1})}^{2}+{(x^{2})}^{2}-{(x^{3})}^{2} & 2(-x^{0}x^{1}+x^{2}x^{3}) \\
    0 & 2(-x^{0}x^{2}+x^{1}x^{3}) & 2(x^{0}x^{1}+x^{2}x^{3}) & {(x^{0})}^{2}-{(x^{1})}^{2}-{(x^{2})}^{2}+{(x^{3})}^{2}
  \end{pmatrix}\\=\frac{1}{R^{2}}
  \begin{pmatrix} R^{2} & 0 & 0 & 0 \\
    0 & \frac 12 (z_{1}^{2}+\bar z_{1}^{2} -z_{2}^{2}-\bar z_{2}^{2})& \frac{\imath}{2}(z_{1}^{2}-\bar z_{1}^{2}+z_{2}^{2}-\bar z_{2}^{2}) & -(z_{1}z_{2}+\bar z_{1}\bar z_{2})\\
    0 & -\frac{\imath}{2}(z_{1}^{2}-\bar z_{1}^{2}-z_{2}^{2}+\bar z_{2}^{2}) & \frac 12 (z_{1}^{2}+\bar z_{1}^{2} +z_{2}^{2}+\bar z_{2}^{2}) & \imath (z_{1}z_{2}-\bar z_{1}\bar z_{2}) \\
    0 & z_{1}\bar z_{2}+\bar z_{1}z_{2} & \imath(z_{1}\bar z_{2}-\bar z_{1}z_{2}) & z_{1}\bar z_{1}-z_{2}\bar z_{2}
  \end{pmatrix}.
\end{multline}

\subsection{Phase space of nonzero quaternions in complex variables}\label{sec:phase-space-h}

If one uses instead of $x^{A}$ the complex variables
$z_{1},z_{2}$ and instead of $ \pi_{A}=(\pi_{0}, \pi_{\alpha})$, $ \pi_{0}, \pi_{+}, \pi_{-}, \pi_{3}$, where $\pi_{\pm}=\pi_{1}\pm \imath \pi_{2}$, with
$ \pi_{0}, \pi_{3}$ are real and $\pi_{-}= \pi_{+}^{*}$,
one gets
\begin{equation}
  \label{eq:174}
  \begin{split}
    &\{ \pi_{A}, \pi_{0}\}=0,\quad  \{ \pi_{+}, \pi_{-}\}=-2\imath  \pi_{3},\quad \{\pi_{\pm},\pi_{3}\}=\pm\imath\pi_{\pm},\\
    & \{z_{1},\pi_{0}\}=\frac 12 z_{1},\quad \{z_{2},\pi_{0}\}=\frac 12 z_{2},\quad\{z_{1},\pi_{3}\}=\frac 12 \imath z_{1},\quad \{z_{2},\pi_{3}\}=\frac 12 \imath z_{2},\\
    &\{z_{1},\pi_{+}\}=0=\{z_{2},\pi_{+}\},\quad \{z_{1},\pi_{-}\}=-\imath \bar z_{2},\quad \{z_{2}, \pi_{-}\}=\imath \bar z_{1},
  \end{split}
\end{equation}
with other brackets obtained by using complex conjugation and the fact that the
Poisson bracket is real. One may also introduce
\begin{equation}
  \label{eq:186}
  \rho_{\pm}= \pi_{0}\pm \imath\pi_{3},\quad \rho_{-}=\bar \rho_{+},
\end{equation}
so that the Poisson brackets with $ \pi_{3}, \pi_{0}$ become
\begin{equation}
  \label{eq:187}
  \{z_{1},\rho_{+}\}=0,\quad \{z_{1},\rho_{-}\}=z_{1},\quad \{z_{2},\rho_{+}\}=0,
  \quad \{z_{2},\rho_{-}\}=z_{2},
\end{equation}
which implies in particular
\begin{equation}
  \label{eq:227}
  \{R,\rho_{+}\}=\frac 12 R=\{R,\rho_{-}\}.
\end{equation}
Furthermore,
\begin{equation}
  \label{eq:190}
  \{\rho_{+},\rho_{-}\}=0,\ \{ \pi_{+}, \pi_{-}\}=\rho_{-}-\rho_{+},\ \{ \pi_{\pm},\rho_{+}\}=\mp \pi_{\pm},\ \{ \pi_{\pm},\rho_{-}\}=\pm  \pi_{\pm}.
\end{equation}
Similarly, if $Q_{\pm}=Q_{1}\pm\imath Q_{2}$,
\begin{equation}
  \label{eq:181}
  \{ Q_{0}, Q_{A}\}=0,\quad \{ Q_{+}, Q_{-}\}=2\imath  Q_{3},\quad \{ Q_{\pm}, Q_{3}\}=\mp\imath  Q_{\pm}.
\end{equation}
with $Q_{0},Q_{3}$ real and $Q_{-}=\bar Q_{+}$,
\begin{equation}
  \label{eq:183}
  \begin{split}
    \{z_{1},Q_{0}\}=\frac 12 z_{1},\quad \{z_{2},Q_{0}\}=\frac 12 z_{2},\quad\{z_{1},Q_{3}\}=\frac 12\imath z_{1},\quad \{z_{2},Q_{3}\}=-\frac 12\imath z_{2},\\
    \{z_{1},Q_{+}\}=\imath z_{2},\quad \{z_{2},Q_{+}\}=0,\quad \{z_{1},Q_{-}\}=0,\quad \{z_{2},Q_{-}\}=\imath z_{1}.
  \end{split}
\end{equation}
and other brackets obtained by using complex conjugation. In terms of
(non-canonical) complex variables
\begin{equation}
  \label{eq:184}
  \begin{split}
  & Q_{0}=\pi_{0},\quad Q_{3}=\frac{\pi_{3}}{R^{2}}(z_{1}\bar z_{1}-z_{2}\bar z_{2})-\frac{\pi_{+}}{R^{2}}\bar z_{1}\bar z_{2}
  -\frac{\pi_{-}}{R^{2}}z_{1} z_{2},\\ & Q_{+}=2\frac{\pi_{3}}{R^{2}}\bar z_{1} z_{2}+\frac{\pi_{+}}{R^{2}}\bar z_{1}^{2}
  -\frac{\pi_{-}}{R^{2}}z_{2}^{2},\\
  & Q_{-}=2\frac{\pi_{3}}{R^{2}} z_{1} \bar z_{2}+\frac{\pi_{-}}{R^{2}} z_{1}^{2}
    -\frac{\pi_{+}}{R^{2}} \bar z_{2}^{2}.
  \end{split}
\end{equation}
If one introduces
\begin{equation}
  \label{eq:188}
  P_{\pm}=Q_{0}\pm \imath Q_{3},\quad P_{-}=\bar P_{+},
\end{equation}
the Poisson brackets with $Q_{3},Q_{0}$ become
\begin{equation}
  \label{eq:189}
  \{z_{1},P_{+}\}=0,\quad \{z_{1},P_{-}\}=z_{1},\quad \{z_{2},P_{+}\}=z_{2},\quad \{z_{2},P_{-}\}=0.
\end{equation}
and
\begin{equation}
  \label{eq:190a}
  \{P_{+},P_{-}\}=0,\ \{Q_{+},Q_{-}\}=P_{+}+P_{-},\ \{Q_{\pm},P_{+}\}=\pm Q_{\pm},\ \{Q_{\pm},P_{-}\}=\mp Q_{\pm}.
\end{equation}

\vfill
\pagebreak

\addcontentsline{toc}{section}{References}

\printbibliography%

\end{document}